\documentclass[12pt, letterpaper]{article}
\usepackage[utf8]{inputenc}
\usepackage[margin=1in]{geometry}
\usepackage{setspace}
\usepackage[bottom]{footmisc}
\usepackage[english]{babel} 
\usepackage{minted}
\usepackage{indentfirst}
\usepackage[format=hang]{caption}
\usepackage{placeins} 
\usepackage{algorithm}

\usepackage{stackengine} 
\usepackage{algpascal}
\usepackage{algc}
\usepackage{algcompatible}
\usepackage[noend]{algpseudocode}
\algnewcommand\algorithmicinput{\textbf{Input:}}
\algnewcommand\Input{\item[\algorithmicinput]}
\algnewcommand\algorithmicgoal{\textbf{Goal:}}
\algnewcommand\Goal{\item[\algorithmicgoal]}

\usepackage{tabularx}
\usepackage{booktabs}

\usepackage[version=3]{mhchem}
\usepackage{siunitx}
\FloatBarrier 

\usepackage{mathtools}
\usepackage{amsthm, amsmath, amssymb} 

\DeclareMathOperator*{\argsup}{arg\,sup}
\DeclareMathOperator*{\argmin}{arg\,min}
\usepackage{listings}
\theoremstyle{definition}
\newtheorem{definition}{Definition}[subsection]
\newtheorem{theorem}{Theorem}[subsection]
\newtheorem{assumption}{Assumption}[subsection]
\newtheorem{lemma}{Lemma}[section]

\usepackage{float} 
\usepackage{blindtext} 
\usepackage{titlesec}
\usepackage[autostyle]{csquotes}
\usepackage[font=small]{caption}    

\usepackage{booktabs}
\usepackage[flushleft]{threeparttable}
\usepackage[round]{natbib} 
\usepackage{adjustbox}
\usepackage{graphicx, multicol} 
\usepackage{xcolor} 
\usepackage{tikz}
\usepackage{circuitikz}
\usetikzlibrary{positioning}
\usepackage{tikz-qtree} 
\tikzset{every tree node/.style={align=center,anchor=north},
         level distance=2cm} 
\usepackage{subfig}
\usepackage{subfloat}

\usepackage[final, linkcolor=blue, colorlinks = true, citecolor = black, urlcolor=blue]{hyperref}

\usepackage{enumitem} 
\usepackage[yyyymmdd]{datetime} 

\usepackage{subcaption}
\usepackage[capitalise,noabbrev,nameinlink]{cleveref}
\usepackage{fancyhdr} 
\renewcommand\footnoterule{\kern -3pt \hrule width 6.7in \kern 2.6pt} 

\date{}
\title{Variational Regularized Bilevel Estimation \\
for Exponential Random Graph Models} 

\author{Yoon Choi\thanks{Department of Economics, University of Washington. Email: \url{lemineml@uw.edu}. I thank my advisor, Professor Yanqin Fan for her advice, support and fruitful discussions on this project. I also thank Professor Hyeonseok Park for providing guidance on the construction of the proposed algorithm programming, and  Professor Jing Tao and Professor Jackson Bunting at the UW Econometrics Seminar for helpful feedback. All errors are my own.}}

\begin{document}

\begin{titlepage}

\maketitle

\begin{abstract}
     I propose an estimation algorithm for Exponential Random Graph Models (ERGM), a popular statistical network model for estimating the structural parameters of strategic network formation in economics and finance. Existing methods often produce unreliable estimates of parameters for the triangle, a key network structure that captures the tendency of two individuals with friends in common to connect. Such unreliable estimates may lead to untrustworthy policy recommendations for networks with triangles. Through a variational mean-field approach, my algorithm addresses the two well-known difficulties when estimating the ERGM, the intractability of its normalizing constant and model degeneracy. In addition, I introduce $\ell_2$ regularization that ensures a unique solution to the mean-field approximation problem under suitable conditions. I provide a non-asymptotic optimization convergence rate analysis for my proposed algorithm under mild regularity conditions. Through Monte Carlo simulations, I demonstrate that my method achieves a perfect sign recovery rate for triangle parameters for small and mid-sized networks under perturbed initialization, compared to a 50\% rate for existing algorithms. I provide the sensitivity analysis of estimates of ERGM parameters to hyperparameter choices, offering  practical insights for implementation.

    \vspace{2mm}
\vspace{0in}\\
\end{abstract}

\end{titlepage}
\setcounter{page}{0}
\thispagestyle{empty}
\doublespacing

\section{Introduction}
    
     Understanding what determines social connections, how they influence agents' choices, and how their choices result in outcomes in society plays a crucial role in social science areas. As \cite{de2023dependence} warns, ``not considering network structure when it is present may result in unreliable estimates and wrong association among observations."  In particular, considering endogenous network formation when analyzing social networks helps us to better understand how an observed network structure is formed and what network-based policy should be implemented. 
    
    The Exponential Random Graph Model (ERGM) is a well-suited modeling strategy for this purpose\footnote{It is widely used in many social science areas such as economics of education (\cite{mele2022structural}, \cite{badev2017discrete}), urban economics (\cite{liu2015regional}), finance (\cite{wong2015board}) and organizational management (\cite{kim2016understanding}, \cite{gaonkar2018structural}).}. It demonstrates the flexibility and generality of network modeling, as it can accommodate network configurations with complex dependence, such as transitivity as well as agents' attributes.

    Despite its ability to model complex network topologies, estimation of ERGM is notorious for two major obstacles. One is the computationally intractable normalizing constant.\footnote{It is the sum over all $2^{n\choose 2}$ possible networks with $n$ nodes; if there are 10 nodes, the sum involves the computation of $2^{45}$ potential functions, which is infeasible (\cite{dini2021notes}).} Although using a Markov Chain Monte Carlo (MCMC) sampling to approximate the normalizing constant avoids the intractability, it encounters the second difficulty of ERGM, model degeneracy.\footnote{A model degeneracy is a probability distribution that puts most of its mass on a small set of all possible networks with size $n$, either empty networks without any edge, or complete networks with all edges connected. For more discussion, see \cite{caimo2011bayesian} and \cite{snijders2002markov}.} 
    This is critical because an MCMC-based algorithm will generate networks from only small sets of its support, thus leading to 
    unreliable estimates of ERGM parameters when unstable network sufficient statistics are included in the model (\cite{schweinberger2011instability}, \cite{caimo2011bayesian}). Maximum Pseudo-Likelihood Estimation (MPLE) (\cite{besag1974spatial}) is an alternative that shows fast estimation. However, it relies crucially on the weak dependence assumption, leading to an inappropriate approach when a given network has strong global dependence (\cite{caimo2011bayesian}). 

     \cite{mele2023approximate} develops a pioneering variational approximate algorithm that seeks a likelihood closest to the likelihood function of ERGM with respect to the Kullback-Leibler (KL) divergence. They show that their approach is deterministic, thereby avoiding sampling networks. They conclude that the variational approach can be a viable alternative to the MCMC-MLE and MPLE, with competitive performance in mean absolute deviation (MAD) and its estimation runtime. Through a careful implementation of their algorithm and its extension to edge-triangle models, I observe a premature convergence of the algorithm, leading to unreliable estimates of parameters of ERGM.  This finding motivates the development of a systematic algorithm that addresses the convergence issues in \cite{mele2023approximate} while maintaining the computational advantages of their variational approach. 

     Hence, this paper proposes an estimation algorithm for the ERGM, the Variational Regularized Bilevel Estimation Algorithm via a value function approach (VRBEA). Let $F_n(\theta; g_n, \\ \{X_i\}_{i=1}^n, \mu^{*}(\theta)) $ be the upper-level objective function which is the negative log-likelihood function of ERGM, $f_n^{\epsilon}(\theta, \mu'; \{X_i\}_{i=1}^n)$ be the lower-level objective function which is a regularized mean-field approximation to the log-normalizing constant of ERGM.
     Then the proposed algorithm solves the following bilevel optimization problem: 
    \begin{align*}
        \min_{\theta \in \Theta}F_n(\theta; g_n, \{X_i\}_{i=1}^n, \mu^{*}(\theta))& \hspace{5mm} \tag{Upper-level objective}\\ \text{subject to  } \mu^{*}(\theta) \in \argmin_{\mu' \in \mathcal{U}_{\zeta}} f_n^{\epsilon}(\theta, \mu'; \{X_i\}_{i=1}^n). \tag{Lower-level objective}
    \end{align*}
    
     I start the paper by summarizing the contribution of my research and reviewing the literature in \hyperref[sec2]{Section 2}. \hyperref[sec3]{Section 3} reviews the definition of ERGM and bilevel optimization programming as preliminaries. \hyperref[sec4]{Section 4} introduces the log-likelihood of ERGM, its variational estimation approach by \cite{mele2023approximate} and my approach as its extension. \hyperref[sec5]{Section 5} 
    provides the VRBEA in detail. \hyperref[sec6]{Section 6} establishes theoretical non-asymptotic analysis of stationary points obtained by the VRBEA. \hyperref[sec7]{Section 7} demonstrates Monte Carlo simulation results of the VRBEA compared to the existing ones and provides a sensitivity analysis of hyperparameters in my algorithm. \hyperref[sec8]{Section 8} concludes.
    \section{Contributions and literature review}\label{sec2}

    \subsection{Contributions}
    My main contributions are as follows. First, 
    I explicitly formulate the maximum likelihood estimation of ERGM as a bilevel optimization problem. This viewpoint extends the variational approximate algorithm by \cite{mele2023approximate}.
     The bilevel optimization perspective allows the algorithm to solve the two objectives tailored to the specific properties of each objective function, such as convexity or smoothness. Indeed, the lower-level objective function of mean-field MLE of ERGM is nonconvex in the lower-level variable. The bilevel optimization approach I consider in this paper builds on the algorithm in \cite{liu2022bome} and introduce $\ell_2$ regularization. My method hence enables us to solve the lower-level objective despite the presence of nonconvexity that leads to multiple solutions. This approach differs from \cite{mele2023approximate}. They acknowledge that the mean-field problem is generally nonconvex in its variable. They suggest using a global optimization such as simulated annealing in order to find a global solution. One of the drawbacks of using global optimization methods is the prohibitive computation cost. By using the bilevel optimization framework, the VRBEA explicitly addresses the nonconvexity resulting from the variational approximation and reduces the expensive computation cost. 
     
    Second, the VRBEA uses a first-order (gradient descent) method. In contrast, \cite{mele2023approximate} employs a fixed-point algorithm in order to update the lower-level variable that approximates the log-normalizing constant. In their algorithm, the sigmoid function appears as a closed-form solution to the mean-field problem. However, I observe two technical challenges in this algorithm. First, due to the nonconvexity of the lower-level objective function, the fixed-point algorithm may search for a suboptimal stationary point such as local maximum or a saddle point.\footnote{For more details, see \cite{lee2016gradient}. }  Second, the insensitivity of the sigmoid function and the stated convergence criterion can cause early inner-loop termination. The derivative of the sigmoid function is bounded by 0.25. If the change in the upper-level variable is small, the updates may not proceed, leading to premature termination of the inner loop. In addition, the convergence criterion is based on the $1/n^2-$scaled absolute difference between successive mean-field approximation values to the normalizing constant, where $n$ is the number of nodes in an observed network. As $n$ grows, the mean-field approximation can easily satisfy this criterion, even with any random initial choice of the lower-level variable to start the inner loop. As evidence, 
    the Monte Carlo simulation shows the absolute difference between the mean-field approximation values is nearly zero even at the first iteration of the inner loop. Also, I record optimizer messages such as ``convergence due to precision error" or ``abnormal termination in line search." This implies the progress made by the solvers is numerically indistinguishable from zero under the default precision setting because the difference is already small in any direction they search to optimize the log-likelihood function of ERGM. By using a first-order gradient descent method in my proposed algorithm while fixing the number of inner iterations to achieve a desired precision, the VRBEA avoids the above issues and exhibits reliable convergence. 
    
    Third, 
    I introduce $\ell_2$ regularization to the lower-level objective. This strategy guarantees that the lower-level objective function satisfies the Polyak-\L ojasiewicz (PL) inequality (\cite{polyak1963gradient}). The PL inequality is a fundamental condition that enables gradient descent methods in machine learning to achieve a linear convergence rate for nonconvex optimization problems (\cite{karimi2016linear}). The lower-level objective function itself does not meet the global PL inequality. Moreover, it is challenging to show whether it satisfies the local PL inequality. Theoretically, adding the $\ell_2$ regularization term and setting the regularization parameter greater than the minimum eigenvalue of the Hessian matrix of the lower-level objective function converts the lower-level objective function into a strongly convex function of the lower-level variable for any given upper-level variable, leading to the satisfaction of the PL inequality. In practice, one can choose a regularization parameter to reduce the degree of nonconvexity of the lower-level objective function.

     Fourth, 
     I establish a non-asymptotic optimization convergence rate analysis of my algorithm. To my knowledge, this is the first analysis of non-asymptotic optimization convergence rate in the literature of ERGM estimation with variational approach. Two theorems constitute the analysis. The first theorem shows a theoretical blueprint on my proposed algorithm. The theorem establishes that a pre-specified Lyapunov-type energy function $\Phi(\theta, \mu; \gamma) $ as the sum of the upper-level objective function $F_n(\theta)$ and the product of a positive constant $\gamma$ and the constraint of the optimization problem, $q^{\epsilon}(\theta, \mu) = f_n^{\epsilon}(\theta, \mu) - \inf_{\mu' \in \mathcal{U}} f_n^{\epsilon}(\theta, \mu')$, decreases linearly until some outer iteration $t_0$. After $t_0$, the difference between two successive $\Phi$s will be $O(\xi_t^{1.5}),$ where $\xi_t$ is the outer step size at iteration $t$. This theoretically reveals the mechanism of bilevel optimization with the nonconvex lower-level objetive function. The second theorem is about the overall non-asymptotic optimization convergence rate of my algorithm. It proves that the average of a measure of stationarity $\mathcal{K}(\theta,\mu)$ over the outer iteration $T$ is $O(T^{-1/4})$, the same rate \cite{liu2022bome} proved. This rate is optimal in bilevel optimization with the nonconvex lower-level objective function. The measure is defined as the squared magnitude of gradient updating both lower- and upper-level variables and the feasibility of the variables as a solution obtained by the algorithm. Specifically, the measure of stationarity in this paper is the following: 
     \[\mathcal{K}(\theta, \mu) := \ ||\nabla F_n(\theta) +  \lambda^{*}(\theta, \mu) \nabla q^{\epsilon}(\theta, \mu) ||^2 + q^{\epsilon}(\theta, \mu). \tag{stationarity}\label{st} \] 

     This analysis differs from the theoretical analysis on the convergence of mean-field approximation to the log-normalizing constant to the truth and of the log-likelihood of ERGM by providing their lower and upper bound in \cite{mele2023approximate}. 

     Fifth, I demonstrate the performance of my algorithm through numerical simulation. The simulation using a simple model with the number of edges and the triangle reveals that the conventional algorithms, MCMC-MLE and MPLE, suffer from bias when estimating the coefficient of the number of triangles, as shown in \cite{schweinberger2011instability}. Moreover, the algorithm by \cite{mele2023approximate} shows early convergence in many runs due to the insensitivity of the sigmoid function, and its convergence criterion on the variational approximation surrogate. On the other hand, the simulation results illustrate that my method outperforms the existing algorithms with respect to various summary statistics such as bias, mean, and mean absolute deviation (MAD). In addition, I provide sensitivity analysis of the estimates and the objective values to two hyperparameters, the regularization parameter $\epsilon$ and the parameter that controls the speed of constraint satisfaction, $\eta$ (\cite{gong2021bi}). The regularization ($\epsilon$) paths in \hyperref[sec7]{Section 7} illustrate the effect of regularization on the estimates of edge-triangle parameters. The constraint satisfaction ($\eta$) paths show the effect of $\eta$ on the upper-level function value $F_n(\theta)$ and the constraint function value $q^{\epsilon}(\theta, \mu)$. These paths provide practical insights on judicious choices of two hyperparameters $\epsilon$ and $\eta$.

\subsection{Related literature}

\subsubsection{Application of Exponential Random Graph Models}
    The ERGM is widely used in sociology and statistics. However, it is difficult to draw economic interpretation from estimated parameters (\cite{gaonkar2018structural}). A recent study in the econometrics of networks (\cite{mele2017structural}, \cite{badev2017discrete}) has shown that the network formation game (\cite{monderer1996potential}) under mild conditions converges to a unique stationary distribution. The theoretical foundation that the likelihood of observing a network data corresponds to the canonical ERGM enables network scientists to view observed the network data as a draw from the ERGM. Under these assumptions, we need only a single network data set to estimate the structural parameters from the strategic network formation model. The bridge from economic network formation model to the ERGM enables economists to develop a structural model of network formation to study the incentives of social connections among agents. For example, \cite{mele2022structural}
     uses ERGM to study friendship formation in schools, showing that students' preferences depend not only on similarities in their attributes (homophily) but also on the number of common friends that agents have (transitivity\footnote{The transitivity is the tendency of two persons with shared friends to connect (\cite{goodreau2009birds}).}). Accurately estimating the triangle parameter is crucial for distinguishing between these mechanisms and evaluating desegregation policies. Similarly, \cite{gaonkar2018structural} study venture capital networks, showing that the triangle coefficient captures firms' reliance on joint partners, which suggests the observed network structure of venture capital firms is generated by their preference for transitivity as well as their homophily. These examples illustrate that incorporating endogenous network formation is essential for reliable policy recommendations.
\subsubsection{Estimation algorithms for the ERGM}

The commonly used algorithm for the ERGM is the Markov Chain Monte Carlo Maximum Likelihood Estimation (MCMC-MLE), suggested by \cite{geyer1991markov}, further developed by \cite{geyer1992constrained, dahmstrom1993ml, corander1998maximum}. The algorithm suggests that the intractable normalizing constant of ERGM can be approximated by a series of networks generated by the Markov chain. Then by iterating the procedure of finding the parameter vector that maximizes the log-likelihood of ERGM with the approximated log-normalizing constant, one can obtain the parameter estimates. One of the problems in the MCMC-MLE is slow convergence due to the local MCMC sampler used to approximate to the normalizing constant through MCMC. \cite{mele2017structural} shows that the standard local MCMC sampler\footnote{It requires long enough burn-in and thinning. The burn-in is a process of throwing away a pre-determined number of initial samples generated by the Markov Chain Monte Carlo in order to reduce the dependence of samples on the initial parameter set-up, and the thinning is a process of keeping every $k$th sample after the burn-in to reduce high autocorrelation between samples.(\cite{owen2017statistically}). Moreover, the proposal used in the sampler is 1/(n(n-1)), which takes $Cn^2 \log n$ steps in usual cases, $\exp (Cn^2)$ steps in some parameter regions. For more discussion, see \cite{mele2017structural}.} used in the ERGM literature exhibits exponentially slow convergence. 
 A well-known phenomenon in ERGM, model degeneracy, can deteriorate the slow convergence because the performance of MCMC-MLE depends hugely on the choice of initial parameters of ERGM if they are from the extreme basins  -- either empty networks or complete networks (\cite{caimo2011bayesian}). To avoid this trap, \cite{caimo2011bayesian} and \cite{mele2017structural} estimate ERGM parameters using a Bayesian method. They apply the exchange algorithm (\cite{murray2012mcmc}) to overcome the double intractability of posterior and likelihood normalization. However, 
this algorithm still requires sufficient amount of time to generate graph samples.\footnote{\cite{caimo2011bayesian} state in their discussion section that the estimation time takes less than 2 hours for 104 nodes.} The VRBEA addresses the slow convergence issue by taking a deterministic approach as an extension of variational approximate algorithm by \cite{mele2023approximate}.

    Another approach is the Maximum Pseudolikelihood Estimation (MPLE), proposed in  \cite{besag1974spatial}, further developed by \cite{strauss1990pseudolikelihood}. The algorithm maximizes the pseudolikelihood given parameters of interest, the product of the parametric conditional probabilities of forming a link between a pair of two nodes given the rest of the dyads. One of the drawbacks of MPLE is that the estimates of parameters are not accurate in the presence of strong dependence among nodes, despite its fast computation time (\cite{geyer1991markov}). Moreover, confidence intervals computed from the inverse of Fisher information matrix in MPLE are known to be biased (\cite{cranmer2011inferential}), leading to problematic inference on a given network data. 

    To overcome the limitations that the two preceding approaches have, \cite{mele2023approximate} proposes a variational mean-field estimation algorithm that maximizes the log-likelihood function of ERGM with approximation to the log-normalizing constant using a mean-field approximation (\cite{wainwright2008graphical}). The paper shows the bounds on the approximation error of mean-field approximation to the log-normalizing constant and the mean-field likelihood function without the limitation to the size of network, adapting nonlinear large deviation results.

\subsubsection{Bilevel optimization}
Bilevel optimization programming is a special case of multilayer optimization problems, where an optimization problem functions has another optimization problem as its constraint (\cite{sinha2017review}). It is rooted in economics, also known as Stackelberg model (\cite{beck2022brief}), but has been widely applied in many research areas such as machine learning (\cite{hospedales2021meta}, \cite{ustun2024hyperparameter}), environmental economics (\cite{caselli2024bilevel}).
 The definition of bilevel optimization function is as follows:
\begin{definition}(Bilevel Optimization, \cite{liu2021towards}) \hspace{1mm}\\
    For a upper-level objective function $F:\mathbb{R}^n \times \mathbb{R}^{m} \rightarrow \mathbb{R}$ and a lower-level objective function $f:\mathbb{R}^n \times \mathbb{R}^{m} \rightarrow \mathbb{R}$, the bilevel optimization problem is 
    \begin{align*}\label{bi-level}
        &\min_{x\in \mathcal{X},\hspace{1mm} y }F(x,y) \hspace{4mm}\text{subject to}\hspace{2mm} y\in \Psi(x) :=\argmin_{y' \in \mathcal{Y}} f(x,y') \tag{Bilevel}
    \end{align*}
    where $\mathcal{X} \subseteq \mathbb{R}^m$ and 
 $\mathcal{Y} \subseteq \mathbb{R}^n $  are constrained sets satisfying the upper-level constraints $G_p:\mathcal{X} \times \mathcal{Y}\rightarrow \mathbb{R},\hspace{1mm} p \in [P]$ and the lower-level constraints $g_j: \mathcal{X} \times \mathcal{Y}\rightarrow \mathbb{R},\hspace{1mm} j\in[J]$. $\Psi(x):\mathcal{X} \rightarrow \mathbb{R}^m$ is a set-valued function so that $\Psi(x) \subseteq \mathcal{Y}$ for every $x \in \mathcal{X}$.
\end{definition}
Many existing methods have been developed under several assumptions that the lower-level objective function is (strongly) convex or the solution set of lower-level decision variable given a upper-level variable is convex, or even the upper-level objective function is convex. 
When the lower-level objective function is nonconvex, it is unclear about which lower-level solutions should be used to evaluate the upper-level objective function. I adopt a value-function approach to handle the nonconvexity of the lower-level objective function, because it reformulates a given bilevel optimization problem into a single-level optimization algorithm by constructing a value function using the lower-level objective function.

\begin{definition}(A value-function approach bilevel optimization, \cite{liu2022bome}) \hspace{1mm} \\Consider a bilevel optimization problem \ref{bi-level}. The value-funciton approach to solve a given bilevel problem \ref{bi-level} reformulates \ref{bi-level} into the following: 
    \begin{align*}
        &\min_{x\in \mathcal{X},\hspace{1mm} y \in \mathcal{Y}}F(x,y) \hspace{4mm}\text{subject to}\hspace{2mm} q(x,y) = f(x,y) - f(x,y^{*}(x)) \leq 0
    \end{align*}
    where $y^{*}(x) \in \Psi(x)$ and  $\Psi(x)$ is the set of solutions to the lower-level objective function for given $x\in\mathcal{X}$, as defined in \ref{bi-level}.
\end{definition}

\section{ERGM, its log-likelihood, and mean-field approximate MLE}\label{sec3}
\subsection{Exponential Random Graph Models (ERGM)}

Let $[n] = \{1,2,3,...,n\}$ is the set of units in a given cluster or network. A network is represented by an $n \times n$ adjacency matrix $g_n \in \{0,1\}^{n \times n}$. Any $g_n$ is in $\mathcal{G}_n$, where \[\mathcal{G}_n=\{\omega=(\omega_{ij})\hspace{1mm}\big|\hspace{1mm} \omega_{ij}=\omega_{ji}\in\{0,1\},\omega_{ii}=0,  \hspace{1mm} i,j\in [n] 
\}\]
is the set of all binary matrices with $n$ nodes. If unit $j $ and $k$ are connected, $g_{jk} = 1$, and 0 otherwise. $X_i\in\mathbb{R}^{d_x}$ is unit $i$'s covariate in a network. I introduce the formal definition of ERGM. 
\begin{definition} (Exponential random graph models, \cite{chatterjee2013estimating})\hspace{1mm}\\
    Let $\mathcal{G}_n$ be the space of all simple graphs\footnote{Here a simple graph means a undirected, no self-loop or multiple-edge graph (\cite{chatterjee2013estimating}).} on $n$ labeled nodes. An exponential random graph model (ERGM) can be expressed in exponential form 
    \begin{align*}
        \text{Pr}(G=g_n;\theta) = \frac{\text{exp}\big(\sum_{k=1}^K \langle \theta_k, T_k(g_n) \rangle \big)}{\sum_{w\in \mathcal{G}_n}\text{exp}\big(\sum_{k=1}^K \langle\theta_k, T_k(w)\rangle\big)},
    \end{align*}
    where $\theta \in \mathbb{R}^K$ is a real-valued vector of parameters, and $\{T_k\}_{k=1}^K$ are real-valued functions of elements of $\mathcal{G}_n$. Typically, $T_k$ is a function of count of subgraphs of graph from $\mathcal{G}_n$. For instance, $T_1(g_n)$ is the number of edges, $T_2(g_n)$ is the number of 2-stars. $\langle\cdot, \cdot  \rangle : \mathbb{R}^k \times \mathbb{R}^k \rightarrow \mathbb{R}$ denotes the inner product on $\mathbb{R}^k$.
\end{definition}

The ERGM can incorporate observed attributes of each node, where each node is characterized by an $d_x$-dimensional vector of observed attributes $X_i\in \mathcal{X}\subset \mathbb{R}^{d_x}$, $i=1,...,n$. with locally dependent network topologies such as $k-$stars and triangles as in \cite{mele2023approximate}. 
Then the likelihood function of ERGM observing an adjacency matrix $g_n$ with attributes $\{X_i\}_{i=1}^n$ and parameters $\theta$ is 
\begin{align*}
    &\pi_n(\theta\hspace{1mm}| g_n, \{X_i\}_{i=1}^n) = \frac{\text{exp}(Q_n(\theta\hspace{1mm}| g_n, \{X_i\}_{i=1}^n))}{\sum_{w \in \mathcal{G}_n} \text{exp}(Q_n(\theta\hspace{1mm}| w,\{X_i\}_{i=1}^n))}.\tag{$\pi_n$}\label{likelihood}
\end{align*}
 $Q_n$ is a function called a potential that takes parameter $\theta$ as input conditional on the network configuration of $g_n$ as sufficient statistics such as the number of $k-$stars and the number of triangles and a set of covariates $\{X_i\}_{i=1}^n$.
In other words, 
\begin{align}
  Q_n(\theta\hspace{1mm}| g_n,\{X_i\}_{i=1}^n)= \langle \theta, T(g_n) \rangle,\label{equ(1)}  
\end{align}
where $T: \mathcal{G}_n \rightarrow \mathbb{R}^d$, a vector of network statistics as a functions of $g_n.$ 
As an example, \cite{mele2023approximate} defines the potential $Q_n$, based on \cite{chatterjee2013estimating} to multiply a scalar of 2 to the number of edges in the first term, on \cite{wasserman1996logit} for the second term, and on \cite{easley2010networks} for the third term with rescaling by $1/n$, in order for the second and third terms not to blow as $n$ grows : 
\begin{align}\label{melepotential}
    Q_n(\theta\hspace{1mm}| g_n, &\{X_i\}_{i=1}^n) =\underbrace{\sum_{i=1}^n\sum_{j= 1}^n\nu_{ij}g_{ij}}_{\text{Number of direct links}} +\underbrace{\frac{\beta}{n}\sum_{i=1}^n \sum_{j=1}^n \sum_{k=j+1}^n g_{ij}g_{ik}}_{\text{Number of two-stars}}+\underbrace{\frac{\gamma}{6n}  \sum_{i=1}^n\sum_{j=1}^n\sum_{k=1}^ng_{ij}g_{jk}g_{ki}}_{\text{Number of triangles}}.
\end{align}

\subsection{Log-likelihood of ERGM}\label{sec4}
The log-likelihood function of the ERGM is given by
\[l_n(\theta\hspace{1mm}|g_n, \{X_i\}_{i=1}^n):=n^{-2}\log(\pi_n( \theta \hspace{1mm}| g_n, \{X_i\}_{i=1}^n))=T_n(\theta\hspace{1mm}|g_n, \{X_i\}_{i=1}^n)-\psi_n(\theta),\]
where $T_n$ is the potential  $Q_n(\theta\hspace{1mm}| g_n, \{X_i\}_{i=1}^n)$, scaled by $n^{-2}$,  
\begin{align*}
     T_n(\theta\hspace{1mm}|g_n, \{X_i\}_{i=1}^n) &= n^{-2} Q_n(\theta\hspace{1mm}|  g_n, \{X_i\}_{i=1}^n)\tag{$T_n$}\label{T}. 
\end{align*}

The last term of the log-likelihood function of ERGM is the log-normalizing constant of the likelihood of ERGM:
\begin{align*}
    &\psi_n(\theta) = n^{-2}\text{log} \big(\sum_{w\in \mathcal{G}_n} \text{exp}[Q_n(\theta\hspace{1mm}|  w, \{X_i\}_{i=1}^n)]\big) = n^{-2}\text{log} \big(\sum_{w\in \mathcal{G}_n} \text{exp}[n^2 T_n(\theta\hspace{1mm}| w, \{X_i\}_{i=1}^n)]\big) \tag{$\psi_n$}\label{lognormalizing}.
\end{align*} 
 
\subsection{Mean-field approximate MLE}

It is well known that computing the normalizing constant in ERGM is infeasible. In fact, $\mathcal{G}_n$ contains $2^{\binom{n}{2}}$. This indicates that when the size of a network is over 
 20, the cardinality of set containing all possible simple graphs exceeds the number of atoms existing in the Earth ($2^{170}$, \cite{dini2021notes}). To overcome the problem of the intractable log-normalizing constant $\psi_n(\theta)$ in the log likelihood function $l_n(\theta|g_n, \{X_i\}_{i=1}^n)$, \cite{mele2023approximate} propose a variational approximate algorithm. It approximates the log-normalizing constant $\psi_n$ by finding a likelihood function closest to the likelihood function of ERGM with respect to the Kullback-Leibler (KL) divergence, $\psi_n^{MF}$: 
 \begin{align*}\label{mf-ergm}
     l_n^{MF}(\theta\hspace{1mm}|g_n, \{X_i\}_{i=1}^n):=T_n(\theta\hspace{1mm}| g_n, \{X_i\}_{i=1}^n)-\psi_n^{MF}(\theta)\tag{MF}
 \end{align*}
where 
\begin{align*}\label{lognormalizingmf}
    \psi_n^{MF}(\theta)
= \sup_{\substack{\mu\in[0,1]^{n^2},\\ \mu_{ij}=\mu_{ji},\forall i,j}}\Gamma_n(\theta,\mu\hspace{1mm}|\{X_i\}_{i=1}^n):=T_n(\theta | \mu, \{X_i\}_{i=1}^n)-H_n(\mu),\tag{$\psi_n^{MF}$}
\end{align*}
 
where
\[H_n(\mu) = \frac{1}{2n^2}\sum_{i=1}^n\sum_{j= 1}^n[\mu_{ij}\log\mu_{ij}+(1-\mu_{ij})\log(1-\mu_{ij})]\]
is the average entropy of the product of Bernoulli distributions with parameter $\mu_{ij}$, $i,j \in[n]$, $\mu_{ij}=\Pr(g_{ij}=1)$ is the unconditional probability that nodes $i$ and $j$ form a link. 
 Hence, the variational approximate algorithm in \cite{mele2023approximate} solves 
\begin{align*}\label{MF}
    &\max_{\theta} \hspace{1mm}l_n^{MF}(\theta\hspace{1mm}|g_n, \{X_i\}_{i=1}^n):=T_n(\theta\hspace{1mm}|g_n, \{X_i\}_{i=1}^n)-\psi_n^{MF}(\theta)\tag{MF-MLE}\\
    &\text{subject to } \nabla_{\mu}\Gamma_n(\theta,\mu\hspace{1mm}|\{X_i\}_{i=1}^n) = 0. 
\end{align*}

\section{Variational regularized bilevel MLE}
\subsection{Limitations of mean-field approximate MLE}
The optimization procedure in \cite{mele2023approximate} for maximizing the ERGM likelihood involves the following steps. 
For a fixed estimate of the parameter of ERGM, $\theta$, first it solves for an optimal symmetric matrix $\mu$ by maximizing $\Gamma_n(\theta, \mu)$. Finding the optimal matrix  $\mu$ involves determining a solution to the first-order condition (FOC) of $\Gamma_n( \theta, \mu)$.\footnote{This is called a stationary seeking method (\cite{mehra2021penalty}) in the bilevel optimization literature.} Then the sigmoid function, $\sigma(z) = 1/(1+\text{exp}(-z))$, appears as a closed-form solution to the FOC. 

Their update rule for the lower-level variable $\mu$ is to use a fixed point algorithm such that for each inner iteration $k$, each element of $\mu$, $\mu_{ij}$ is updated by $\mu_{ij,k+1}= \sigma(h(\mu_k))$, as a function of $\mu$ at the previous inner step $k$. The iteration continues until the difference between two successive mean-field approximations to the constant, $\psi_n^{MF}(\theta)_{k+1} - \psi_n^{MF}(\theta)_{k}= \Gamma_n(\theta,\mu_{k+1}) - \Gamma_n(\theta,\mu_{k})$, becomes less or equal to a pre-specified threshold. The parameter $\theta$ updates by solving the mean-field approximated log-likelihood function, $l_n^{MF}$, with an updated $\psi_n^{MF}(\theta)_{k+1}$ using a built-in optimization solver in statistical programs such as \texttt{R} or \texttt{Python}. However, I find two technical issues in this algorithm, which yield a premature convergence of the algorithm. 

On one hand, the update of $\mu$ can be small due to the insensitivity of the sigmoid function. The derivative of the sigmoid function is bounded by 0.25.\footnote{The derivative of the sigmoid function, $\sigma'(z)$ is $\sigma(z)(1-\sigma(z))$. It attains its maximum at 0.5, leading to the maximum of 0.25.} So if the change in the argument of the sigmoid function is small,\footnote{For more detailed explanation, see \hyperref[appendix:E]{Appendix E}.} the updates can be small, leading to premature termination of inner loop. In fact, I observed that the absolute differences between $\mu_{k+1}$ and $\mu_{k}$ are significantly small. They range from  $10^{-4}$ to even $10^{-12}$ in the Monte Carlo simulations. 

On the other hand, the convergence criterion is based on the absolute difference between successive mean-field approximation values to the normalizing constant, where $n$ is the number of nodes in an observed network. However, the mean-field approximation to the constant is $1/n^2-$scaled. Hence as $n$ grows, the difference between the two successive mean-field approximations can easily satisfy the threshold, even with any random initial choice of lower-level variable to start the inner loop. As evidence, 
    the Monte Carlo simulation shows the absolute difference between the mean-field approximation values is nearly zero even at the first iteration of inner loop. Also, I record optimizer messages such as ``convergence due to precision error" or ``abnormal termination in line search." This implies the progress made by the solvers is numerically indistinguishable from zero under the default precision setting because the difference is already small in any direction they search to optimize the log-likelihood function of ERGM.

\subsection{Variational regularized bilevel MLE}
Alternatively, I propose the following estimation algorithm for the log likelihood of the ERGM, using (\ref{equ(1)}) as a potential function. My proposed algorithm adopts a value-function approach from \cite{liu2022bome}. However, I extend their algorithm tailored to the estimation of ERGM in two ways. First, since it does not have constraints on both upper- and lower-variables, I modify the algorithm to update the lower-level variable $\mu$ using a projected gradient descent because $\mu$ has constraints $\mu \in \mathcal{U} = \{M\in[0,1]^{n^2}\hspace{1mm}|\hspace{1mm} M_{ij}= M_{ji}, \hspace{1mm}M_{ii}=0 \hspace{2mm}\forall i,j\in [n]\}$. Second, since it is difficult for the lower-level objective function, $\Gamma_n$, to satisfy the PL inequality condition,  I add the $\ell_2 $ regularization term to $\Gamma_n$ to characterize the strong convexity. It enables  $\Gamma_n$ to satisfy the PL inequality condition. 

From the \hyperref[mf-ergm]{mean-field approximation}, I construct a value function. For given $(\theta, \mu)$, let

    \[q^{\epsilon}(\theta, \mu) = f_n^{\epsilon}(\theta, \mu|\{X_i\}_{i=1}^n) - f_n^{\epsilon*}(\theta|\{X_i\}_{i=1}^n),\]

where 

    \[f_n^{\epsilon}(\theta, \mu|\{X_i\}_{i=1}^n)=-\Gamma_n(\theta, \mu|\{X_i\}_{i=1}^n) + \frac{\epsilon}{2n^2}||\mu||_F^2\] and 
    \[f_n^{\epsilon*}(\theta|\{X_i\}_{i=1}^n) = \inf_{\mu\in \mathcal{U}} f_n^{\epsilon}(\theta, \mu|\{X_i\}_{i=1}^n) = f_n^{\epsilon}(\theta, \mu^{*}(\theta)|\{X_i\}_{i=1}^n).\]

$f_n^{\epsilon*}$ is known as the value function (\cite{liu2022bome}). 
Let \[F_n(\theta | g_n, \{X_i\}_{i=1}^n) = -\ell_n^{MF}(\theta| g_n, \{X_i\}_{i=1}^n). \] Then the bilevel optimization of log-likelihood of the ERGM becomes 
\begin{align*}\label{vf}
    \min_{
    \theta, \mu } F_n(\theta| g_n, \{X_i\}_{i=1}^n) :=\big\{ -T_n(\theta|g_n, \{X_i\}_{i=1}^n)-f_n^{\epsilon*}(\theta |\{X_i\}_{i=1}^n)\big\},\hspace{2mm}
    q^{\epsilon}(\theta, \mu)\leq 0 .\tag{Objective}
\end{align*}

\cite{liu2022bome} employs a dynamic barrier gradient descent method proposed by \cite{gong2021bi}. Intuitively, this method seeks  a direction to update $(\theta, \mu)$, which minimizes the upper-level objective function value while keeping the direction to decrease the constraint $q^{\epsilon}( \theta, \mu) \leq 0.$
That is, \cite{liu2022bome} updates the variable at each iteration $t \in [[T]]$, by solving the following: \begin{align*}\label{direction}
(\theta_{t+1}, \mu_{t+1}) =& (\theta_{t}, \mu_{t}) - \xi_t \delta_t,\\
    \delta_t := \argmin_{\delta} ||\nabla F_n(\theta_t\hspace{1mm}|g_n, \{X_i\}_{i=1}^n)-\delta||^2,& \hspace{1mm}\text{subject to }\langle \nabla q^{\epsilon}(\theta_t, \mu_t), \delta \rangle \geq \phi_t, \tag{Update}
\end{align*}
where $\phi_t = \eta ||\nabla q^{\epsilon}(\theta_t, \mu_t)||^2$ is a dynamic barrier with $\eta >0$. One can view $\eta$ as the similarity between the direction of minimizing the value function or constraint $q^{\epsilon}$ and the direction of minimizing the negative log-likelihood function, $F_n$. That is, if $\eta$ is close to 0, then the search direction for the parameter of ERGM, $\theta$, reconciles more on the direction of minimizing $F_n$, while compromising to satisfy the constraint $q^{\epsilon}$. If $\eta$ becomes close to 1, then the search direction for $\theta$ becomes more inclined to meet $q^{\epsilon}$, but sacrificing the purpose of minimizing $F_n$. 
Different from usual bilevel optimization problems, the upper-level objective has only the upper-level variable because the minimum value $f_n^{\epsilon*}(\theta |\{X_i\}_{i=1}^n)$ absorbs the lower-level variable. Hence, in practice, I make the gradient $\nabla F_n$ with respect to the lower-level variable $\mu$ the zero vector in order to match the dimension of gradient of $F_n$.

\section{Description of algorithm}\label{sec5}

The following section describes my bilevel optimization algorithm\footnote{Code is available upon request.} in detail.
The algorithm starts with an initial value for the parameters $\theta_0:= [\theta_{1,0}, \theta_{2,0}, ..., \theta_{d,0}]^{\top}$. For each outer iteration $t\in[[T]]$, the algorithm updates the lower-level variable $\mu_t^{(k)} $ using a projected gradient descent algorithm over $K$ inner iterations, with step size $\alpha^{(k)}$.  
After saving the $K$th lower-level variable $\mu_t^{(K)}$, the algorithm computes the value function $\widehat{q}^{\epsilon}(\theta_t,\mu_t) = f_n^{\epsilon}(\theta_t,\mu_t) - f_n^{\epsilon}(\theta_t,\mu_t^{(K)})$. Then it updates all the lower- and upper-level variables $(\theta_t,\mu_t)$ through a gradient descent with step size $\xi_t$ and the direction or gradient, $\delta_t$, satisfying the minimization constraint $\delta_t = \argmin_{\delta} \frac{1}{2}||\nabla F_n(\theta_t) - \delta||_2^2$ subject to $\langle \nabla F_n(\theta_t), \nabla \widehat{q}^{\epsilon}(\theta_t,\mu_t) \rangle \geq \phi_t$.\footnote{For more information, see \cite{liu2022bome}.}

\small\begin{algorithm}
    \caption{Variational Regularized Bilevel Estimation Algorithm} \label{algo:bi}
    \begin{algorithmic}
    \Goal Solve (\hyperref[vf]{Objective}) for $\theta_1, \theta_2,...\theta_d, \mu_{ij}$ for $i\neq j \in[n]$.
    \Input Initialize $\theta_0:= [\theta_{1,0}, \theta_{2,0}, ..., \theta_{d,0}]^{\top}$, and  $\mu^{(0)}_0$ component-wise randomly drawn from $U[0,1]$ and $\mu_{ii}=0$ for all $i\in[n]$.

    \For{Iteration \texttt{$t=0$ to $T-1$}}
    \State \textbf{Step 1.} Get $\mu_t^{(K)}$ after the following $K$ inner iterations:

    \For{Iteration \texttt{$k=0$ to $K-1$}}
        \For {\texttt{$i=1$ to $n$}}
            \For {\texttt{$j=i+1$ to $n$}}

                    \State $\mu_{ij,t}^{(k+1)} = 
                    \texttt{Proj}[\mu_{ij,t}^{(k)}-\alpha_{\mu}^{(k)} \nabla_{\mu_{ij}} f_n^{\epsilon}(\theta_t, \mu_{ij,t}^{(k)})]$
    
            \EndFor
        \EndFor
        \EndFor
        
        \State \textbf{Step 2.} Set $\widehat{q}^{\epsilon}_t = \widehat{q}(\theta_t,\mu_t) = f_n^{\epsilon}(\theta_t,\mu_t) - f_n^{\epsilon}(\theta_t,\mu_t^{(K)})$
        \State  \textbf{Step 3.} Update $(\theta_t,\mu_t)$: 
        \begin{align*}
            (\theta_{t+1},\mu_{t+1}) = (\theta_t,\mu_t) - \xi_t\delta_t
        \end{align*}
       where 
       \begin{align*}
           \delta_t = \nabla F_n(\theta_t) + \lambda_t \nabla \widehat{q}^{\epsilon}_t,\hspace{4mm}\lambda_t = \max \{\frac{\phi_t - \langle \nabla F_{n}(\theta_t), \widehat{q}^{\epsilon}_t\rangle}{||\widehat{q}_t||^2},0\}, \hspace{2mm}\phi_t = \eta||\widehat{q}^{\epsilon}_t||^2, \hspace{1mm}\eta >0.
       \end{align*}
    \EndFor
    \end{algorithmic}
    \end{algorithm}
\FloatBarrier
\section{Theoretical analysis of algorithm}\label{sec6}
In this section, I present a non-asymptotic analysis of optimization convergence rate of my proposed algorithm built on the algorithm by \cite{liu2022bome}.  Two theorems constitute the analysis. 

The first theorem shows a theoretical blueprint on my proposed algorithm. I define a pre-specified Lyapunov-type energy function $\Phi(\theta, \mu; \gamma) $ as the sum of the upper-level objective function and the product of a positive constant $\gamma$ and the constraint of the optimization problem, $q^{\epsilon}(\theta, \mu) = f_n^{\epsilon}(\theta, \mu) - \inf_{\mu' \in \mathcal{U}} f_n^{\epsilon}(\theta, \mu')$ as follows:  
\begin{align*}\label{energy}
    \Phi(\theta, \mu;\gamma) := F_n(\theta) + \gamma q^{\epsilon}(\theta, \mu)\tag{energy function}
\end{align*}
The theorem states that $\Phi$ decreases linearly in outer step size $\xi_t$ until some outer iteration $t_0$. Here, $t_0$ is the outer step at which the constraint $q^{\epsilon}(\theta_t, \mu_t)$ is smaller than some threshold $b$, a positive value as a function of $L_{n,\epsilon}, M_{n,\epsilon}, \eta$ and $\kappa$. After $t_0$, the difference between two successive $\Phi$s will be $O(\xi_t^{1.5})$. This theoretically reveals the mechanism of bilevel optimization with the nonconvex lower-level objective function.

The second theorem develops a theoretical bound on the overall optimization convergence rate of the algorithm. In other words, this theorem tells how stable and feasible my proposed algorithm can be with respect to a measure of stationarity in \cite{liu2022bome}. To measure the stationarity of iterates provided by the algorithm, \cite{liu2022bome} proposes a measure of stationarity as follows: 
\[\mathcal{K}(\theta, \mu) := \ ||\nabla F_n(\theta) +  \lambda^{*}(\theta, \mu) \nabla q^{\epsilon}(\theta, \mu) ||^2 + q^{\epsilon}(\theta, \mu). \tag{stationarity}\label{st} \]
The square term in the \hyperref[st]{stationarity}\footnote{ By the proposition in \cite{gong2021bi}, my goal is to show the algorithm obtains a sequence $\{(\theta_t, \mu_t)\}_{t=1}^{\infty}$ such that $\mathcal{K}(\theta_t, \mu_t)$ converges to zero as $t \rightarrow \infty.$ } measures the squared $\ell_2$ norm of $\delta_t:= \nabla F_n(\theta_t) + \lambda^{*}(\theta_t, \mu_t) \nabla q^{\epsilon}(\theta_t, \mu_t)$, as the solution to the \hyperref[direction]{problem} in Section \ref{sec3}. The Lagrange multiplier $\lambda^{*}$ is defined as:
\begin{align*}
     \lambda^{*}(\theta, \mu) = \begin{cases}
    \max \bigg\{0, \eta - \frac{\langle \nabla F_n(\theta), \nabla q^{\epsilon}(\theta, \mu)\rangle}{||\nabla q^{\epsilon}(\theta, \mu)||^2}\bigg\}, & \text{for } ||\nabla q^{\epsilon}(\theta, \mu)|| >0 \\
    0 & \text{for } ||\nabla q^{\epsilon}(\theta, \mu)|| = 0.
  \end{cases} 
\end{align*}
$\eta >0$ is a hyper-parameter that controls the speed of constraint satisfaction in the \hyperref[direction]{problem}. Additional $q^{\epsilon}$ shows the feasibility of the solution $(\theta, \mu)$. 
For simplicity I suppress the expression for the dependence of $F_n, f_n, q^{\epsilon}$ on the data $\{X_i\}_{i=1}^n$ and $g_n$. Moreover, I vectorize $\mu$, $\nabla_{\mu} f_n^{\epsilon}(\theta, \mu)$, $\nabla_{\mu} q^{\epsilon}(\theta, \mu))$, and $\nabla_{\mu \mu}^2 f_n^{\epsilon}(\theta, \mu)$ to use $||\cdot||_2$ instead of the Frobenius norm $||\cdot||_F$.

\subsection{Assumptions}

 First of all, I need to assume that the domains of objective functions are nonempty, closed and convex. Nonemptiness guarantees the existence of projection onto $\mathcal{U}$, $\Pi_{\mathcal{U}}(\nu) = \argmin_{\mu' \in \mathcal{U}}\frac{1}{2}||\mu' - \nu||_2^2$. Closedness ensures the well-definedness of $\Pi_{\mathcal{U}}$. 
Convexity of the domain ensures
the uniqueness of the projection onto $\mathcal{U}$ because $\mu' \mapsto 1/2 ||\mu' - \nu||_2^2$ is convex. Also it implies the non-expansivity of projection. All together, the first assumption ensures that the projection onto the constrained domain $\mathcal{U}$ is well-defined and facilitates the projected gradient descent, which will be used to update the lower-level variable $\mu$. For simplicity, I also assume that the domain of upper-level objective function, $\Theta$, is nonempty, compact and convex.
\begin{assumption}\hspace{1mm}(Nonempty, closed and convex domains)\label{closedcvx}\\
    I assume that the ERGM parameter space $\Theta \subseteq \mathbb{R}^{ d_{\theta}}$ and symmetric matrix space $\mathcal{U}=\{M\in [0,1]^{n \times n},M_{ij} = M_{ji},\hspace{1mm} M_{ii}=0\hspace{1mm}\forall i,j\in[n]\}$ are nonempty, closed and convex.
\end{assumption}

The second assumption states that the gradients of the upper- and lower-level objective functions are Lipschitz continuous with respect to the Euclidean norm $||\cdot||_2$.  
\begin{assumption}\hspace{1mm}(Smoothness)\label{smooth}
    For any $(\theta, \mu), ( \theta', \mu') \in \Theta \times \mathcal{U}$, there exists a positive real-valued constant $L_{n,\epsilon}>0$, such that the gradients of lower- and upper-level objective functions are Lipschitz continuous: 
    \begin{align*}
        & ||\nabla F_n(\theta) - \nabla F_n(\theta')||_2\leq L_{n,\epsilon}||\theta- \theta'||_2\\ 
        &||\nabla f_n^{\epsilon}(\theta, \mu)-\nabla f_n^{\epsilon}(\theta', \mu')||_2\leq L_{n,\epsilon}||(\theta, \mu) - (\theta', \mu')||_2.
    \end{align*}
    
\end{assumption}

The third assumption guarantees that the objective functions $F_n$ and $f_n^{\epsilon}$ as well as their gradients are bounded. 
\begin{assumption}\hspace{1mm}(Boundedness)\label{bound}
    There exists a positive $M_{n,\epsilon}>0$ such that $|F_n|$, $||\nabla F_n||_2$, $|f_n^{\epsilon}|$, $||\nabla f_n^{\epsilon}||_2 \leq M_{n,\epsilon}$ for all  $(\theta, \mu) \in \Theta \times \mathcal{U}$ given $\epsilon >0$.
\end{assumption}

The assumptions listed above are standard in bilevel optimization settings with convex lower-level objective function. However, it is unlikely for the lower-level objective function to be (strongly) convex in general. . 
A number of cases in machine learning literature have multiple solutions to the lower-level 
objective function, such as non-convex regularization term and neural network architectures, which
refer to few-shot classification and data hyper-cleaning tasks, respectively(\cite{liu2021value, liu2021towards, liu2024moreau}, \cite{liu2022bome}), due to its nonconvexity. Hence, I need a weaker version of convexity that allows for multiple solutions to an objective function.

In much of the machine learning literature, the Polyak-Lojasiewicz (PL) inequality is assumed on the loss function in optimization problems. It is weaker than convexity, but guarantee a linear convergence rate with \hyperref[smooth]{$L-$smoothness} assumption . 
\begin{assumption}\hspace{1mm}(The PL inequality, \cite{liu2022bome})\label{pl}\\
     Given any $\theta \in \Theta$, we assume that the lower-level objective function $f_n^{\epsilon}(\theta, \mu)$ has a unique minimizer $\mu^{*}(\theta)$. Then there exists a $\kappa >0$ such that 
    \[||\nabla_{\mu}f_n^{\epsilon}(\theta, \mu)||_2^2 \geq \kappa \big[f_n^{\epsilon}(\theta, \mu) - f_n^{\epsilon}(\theta,\mu^{*}(\theta))\big].\]
\end{assumption}

However, it is difficult to directly apply the PL inequality assumption on the original lower-level objective function $-\Gamma_n(\theta, \mu),$ to approximate to the log-normalizing constant of ERGM, \ref{lognormalizingmf} due to the following reason. Finding a PL constant of $-\Gamma_n(\theta,\mu)$ is analytically impossible due to the complexity of function. Even though we try to exploit the equivalence of PL inequality to the error bound (EB) assumption (\cite{karimi2016linear}), we encounter the same issue: we need to find a EB constant. This necessitates use of $\ell_2$ regularization for the lower-level variable $\mu$, which guarantees the global strong convexity of lower-level objective function $f_n^{\epsilon}(\theta, \mu) = -\Gamma_n(\theta,\mu) + \frac{\epsilon}{2n^2}||\mu||_F^2$ under suitable choice of regularization parameter $\epsilon.$

For given $\theta \in \Theta$, 
I assume that
there exists a positive constant $\rho(\theta) >0$ such that the smallest eigenvalue of the Hessian matrix, $\rho(\theta) = \inf_{\mu \in \mathcal{U}}\lambda_{\text{min}}(\nabla_{\mu \mu}^2 f_n(\theta, \mu)) + \epsilon/n^2$ over $\mathcal{U}$. A judicious choice of $\epsilon$ that satisfies $\rho(\theta) >0$ guarantees the $\rho(\theta)-$strong convexity of lower-level objective function, $f_n^{\epsilon}(\theta, \mu)$, leading it to obtain a unique minimizer $\mu^{*}(\theta)$ given any $\theta \in \Theta.$ Letting $\kappa(\theta) = 2\rho(\theta) >0$, the global strong convexity of lower-level objective function $f_n^{\epsilon}(\theta, \mu)$ with parameter $\rho(\theta)>0$.
I also modify \hyperref[pl]{the PL inequality} to the projected gradient setting since the lower-level objective is to minimize $f_n^{\epsilon}(\theta, \mu)$ over a set of constraints $\mathcal{U}$ on the lower-level variable $\mu$. Hence, the assumption becomes the following: 
\begin{assumption}\hspace{1mm}(The projected PL inequality\label{ppl})\\
     Given any $\theta \in \Theta$, we assume that the lower-level objective function $f_n^{\epsilon}(\theta, \mu)$ has a unique minimizer $\mu^{*}(\theta)$. Moreover, with inner learning rate $\alpha \in (0, 1/L_{n,\epsilon}]$, $\mu$ is updated by the following rule: For each $k \in [[K]]:= \{0, 
     1,2,...,K-1\}$, \[\mu^{(k+1)} = \Pi_{\mathcal{U}}(\mu^{(k)} - \alpha\nabla_{\mu}f_n^{\epsilon}(\theta, \mu)) =\mu^{(k)} - \alpha G_{\alpha}^{\epsilon}(\mu^{(k)}; \theta), \tag{Update}\]\label{pgd} with the projected gradient mapping
     \[G_{\alpha}^{\epsilon}(\mu;\theta) = \frac{1}{\alpha} (\mu - \Pi_{\mathcal{U}}(\mu - \alpha \nabla_{\mu}f_n^{\epsilon}(\theta, \mu))).\]\label{pgm}   
     Then there exists a $\kappa_{\alpha, \rho}(\theta):= 2\rho(\theta)/\alpha >0$ such that \[||G_{\alpha}^{\epsilon}(\mu;\theta)||_2^2 \geq \kappa_{\alpha, \rho} (\theta)\big[f_n^{\epsilon}(\theta, \mu) - f_n^{\epsilon}(\theta,\mu^{*}(\theta))\big].\]
\end{assumption}

 The objective functions $F_n$ and $f_n^{\epsilon}$ satisfy these assumptions.\footnote{I show the proof in \hyperref[verify]{Appendix A}.}

\subsection{Theorem}
The first theorem guarantees that the difference between two successive \hyperref[energy]{energy function} $\Phi(\theta_{t+1}, \mu_{t+1})$ and $\Phi(\theta_{t}, \mu_{t})$ decreases linearly in $\xi_t$ until the constraint $q^{\epsilon}(\theta_t, \mu_t)$ is greater than a constant as a function of theoretical parameters or certain iteration $t_0$. Moreover, after $t_0$, the difference between the two $\Phi$s will be $O(\xi_t^{1.5})$. This analysis extends the analysis in \cite{liu2022bome}, that studies only the overall optimization convergence rate of their algorithm. 

\begin{theorem}\label{thm1}\hspace{1mm}\\
Consider the \hyperref[algo:bi]{algorithm}, with $\xi_t, \alpha \in (0, 2/L_{n,\epsilon}]$ for $t\in[[T]]$. Define a Lyapunov-type energy function $\Phi: \Theta \times \mathcal{U} \rightarrow \mathbb{R}$. Furthermore, suppose that assumptions \ref{closedcvx} (closed and convex domains), \ref{smooth} (smoothness), and \ref{ppl} (projected PL inequality) hold. Then there exists a positive constant $C_K >0$, depending on $L_{n,\epsilon}, \rho, \alpha, \epsilon$, such that for the number of inner iterations $K \geq C_K$, 
\[\Phi_{t+1} - \Phi_t \leq -\frac{1}{2} \xi_t \mathcal{K}_t+ O(\xi_t^{1.5})\]
where $a_1$ is a positive constant depending on $L_{n,\epsilon}, \alpha, \kappa_{\alpha,\rho}.$
In other words, $\Phi$ strictly decreases at step $t$ for the first outer iteration up to $t_0$ and the remaining error after $t_0$ is bounded by $O(\xi_t^{1.5})$.\footnote{I prove the first theorem in \hyperref[proofthm1]{Appendix A}.}
\end{theorem}

 The second theorem is about the overall non-asymptotic optimization convergencerate of my algorithm. It proves that the average of a measure of stationarity $\mathcal{K}(\theta,\mu)$ over outer iteration $T$ is $O(T^{-1/4})$, the same rate \cite{liu2022bome} proved. 
\begin{theorem}\label{thm2}\hspace{1mm}\\
Consider the \hyperref[algo:bi]{algorithm}, with $\alpha \in (0, 2/L_{n,\epsilon}]$ for $t\in[[T]]$. Let $\xi_t = 1/\sqrt{T}$. Suppose that assumptions \ref{closedcvx} (closed and convex domains), \ref{smooth} (smoothness), and \ref{ppl} (projected PL inequality) hold. Then there exists a positive constant $C_K >0$, depending on $L_{n,\epsilon}, \rho, \alpha, \epsilon$, such that for the number of inner iterations $K \geq C_K$, 
\[\frac{1}{T}\sum_{t=0}^{T-1} \mathcal{K}_t \leq \frac{2}{\sqrt{T}}\big[\Phi_0 - \Phi_T \big] + O(T^{-1/4})=  O(T^{-1/4}).\footnote{I prove the second theorem in \hyperref[proofthm2]{Appendix A}.}\]
This theorem proves that the overall optimization convergence rate is achieved at a rate of $T^{-1/4}.$ This convergence rate is optimal in bilevel optimization with the nonconvex lower-level objective function (\cite{ghadimi2018approximation}, \cite{ji2021bilevel}). 
\end{theorem}
  
 \section{Numerical simulation}\label{sec7}
 \subsection{Performance}
 
The following section displays the comparison of summary statistics resulting from my estimator to MCMC-MLE, MPLE, and the one proposed by \cite{mele2023approximate}. All simulations are executed on UW's Hyak, a high-performance computing cluster service and accessed through a Slurm job scheduler.\footnote{ I containerize the computational environment using an Apptainer container image (collection\_trial:010925), which included Python 3.10.12, CUDA 12.5, PyTorch 2.4, and NumPy 1.24 for numerical computations. This setup ensures full consistency and reproducibility of my experiments. The container image is publicly available on Docker Hub: \href{https://hub.docker.com/r/lemineml/collection_trial/tags}{Docker Hub repository}.}  

 All the simulation results are based on 1000 Monte Carlo simulations. I use a simple model of potential as a function of the number of edges (direct utility) and of triangles (indirect utility) with homogeneous players (\cite{chatterjee2013estimating}, \cite{mele2017structural}): 
\begin{align*}
    &\pi_n(\theta\hspace{1mm}| g_n) = \frac{\text{exp}(Q_n(\theta\hspace{1mm}| g_n)}{\sum_{w \in \mathcal{G}_n} \text{exp}(Q_n(\theta\hspace{1mm}| w))} \propto \text{exp}(\theta_1 \sum_{i=1}^n\sum_{j= 1}^n g_{ij} +\frac{\theta_2}{6n}  \sum_{i=1}^n\sum_{j=1}^n\sum_{k=1}^ng_{ij}g_{jk}g_{ki})
\end{align*}

To generate 1000 simulated networks, I use the R package \texttt{ergm}, I sample 1,000 networks by initializing a network with the size of $n$ as an Erd\"os-R\'enyi graph with probability $p=\text{exp}(\theta_1)/(1+\text{exp}(\theta_1))$. The thinning number or the number of iterations for each sampled network is 10,000 after a burn-in of 10 million iterations.\footnote{I follow the network generation setting based on \cite{mele2023approximate}.} 
MCMC-MLE from the R package \texttt{ergm} estimates ERGM using the stochastic approximation approach by \cite{snijders2002markov}. The MPLE estimates the parameter of ERGM with the default setup.\footnote{I also use the setting in \cite{mele2023approximate}}.

The variational approximate algorithm in \cite{mele2023approximate} uses the following update rule to approximate the log-normalizing constant. First, choose a tolerance level $\varepsilon_{tol}$ and take any random $\mu_0\in[0,1]^{n \times n}$ as an initial point. At step $t$, compute \ref{lognormalizingmf} using $\mu_t$. Then update $\mu_{t+1}$ using the closed-form solution to the first-order condition of \ref{mf-ergm} and calculate $\psi_{n,t+1}^{MF}$ using $\mu_{t+1}$. Take difference between $\psi_{n,t+1}^{MF}$ and $\psi_{n,t}^{MF}$. If the difference is below $\varepsilon_{tol}$, the algorithm terminates, otherwise continue the algorithm until the condition is met, by setting $\psi_{n,t+1}^{MF}$ to $\psi_{n,t}^{MF}$.
    \begin{algorithm}
    \caption{Local optimization of mean-field approximation by \cite{mele2023approximate}} \label{algo:local-mf}
    \begin{algorithmic}[1]
    \Require Set the tolerance level $\varepsilon_{\text{tol}}$.
    \Require We provide a parameter $\theta = (\theta_1, \theta_2)$. 
    \State Set initial value of $\mu_0$ at $t = 0$. 
    \State Compute $\psi_{n,t}^{MF}$ via equation (\ref{lognormalizingmf}) and set $\text{diff} = 1$.
    \While {$\text{diff} > \epsilon$} 
        \State Given $\mu_{t}$, get $\mu_{t+1}$ via equation \begin{align*}
    \mu_{ij, t+1} =(1+ \exp(-(\theta_1 + \frac{\theta_2}{n} \sum_{k=1}^n \mu_{jk,t}\mu_{ki,t})))^{-1}
\end{align*}
        \State Compute $\psi_{n, t+1}^{MF}$ via equation (\ref{lognormalizingmf})
        \State $\text{diff} = \psi_{n, t+1}^{MF} - \psi_{n, t}^{MF}$
        \If{$\text{diff} < \varepsilon_{\text{tol}}$},
            \State \textbf{Break}
        \Else \State $\psi_{n, t}^{MF} = \psi_{n, t+1}^{MF}$
        \EndIf
    \EndWhile
    \end{algorithmic}
    \end{algorithm}
For the VRBEA, I select the inner step size $\alpha$ as 0.002 scaled by $n^2$, to cancel out the scaling $1/n^2$ of \hyperref[likelihood]{log-likelihood function} of ERGM. The outer step size, $\xi$, is 0.03. The number of outer iteration and inner iteration are $T =100,000$ and $K=10$, respectively. The regularization parameter $\epsilon$ is fixed at $10^{-2}$, and the constraint satisfaction control parameter $\eta$ is fixed at 0.8. I use the true parameter to initialize all the estimation algorithm.\footnote{\cite{mele2023approximate} takes this approach to decrease the computational time. For the simulation results with different initializations, see the \hyperref[appendix:c]{Appendix C}.}

The model with the number of edges and triangles has the true parameters $[-1,1].$ 
I display the results of the algorithms in \hyperref[table1]{Table 1}. I show estimation results for $n=50, 100, 200.$\footnote{Larger networks cannot be generated with these parameters due to the model degeneracy. When $n=500$, sampled networks are almost fully connected.} Performance is measured in terms of bias, mean, median, mean absolute deviation (MAD) and standard error. The VRBEA shows smaller bias and standard errors than other algorithms for both parameters. The MCMC-MLE and MPLE show small bias in the edge parameter, $\theta_1$. Their mean and median of estimates of $\theta_1$ are also close to the true parameter -1. Moreover, their estimates converge to the truth as $n$ increases. However, the bias and other performance indicators for the parameter of the number of triangles, $\theta_2$,  become significantly large. The standard errors are large compared to the VRBEA. Although the standard errors shrink as $n$ increases, they remain unstable. The variational approximate algorithm by \cite{mele2023approximate} shows substantial bias for  $\theta_2$, especially when $n=50.$ The algorithm exhibits good median estimates. This occurs for the two reasons mentioned in the introduction. First, the fixed-point iterate based on the sigmoid function fails to progress. The update is negligible when the upper-level variable, $\theta$, changes by a small amount. This leads to infinitesimal changes in the lower-level variable and the objective values. As a result, solvers such as \texttt{BFGS} or \texttt{L-BFGS-B} terminate the optimization because they cannot make progress in any direction within machine precision, approximately $10^{-16}$. The messages I recorded such as ``convergence due to precision error" or ``abnormal termination in line search" reflect this issue. Second, the inner-loop convergence criterion uses the $1/n^2-$scaled absolute difference between two successive mean-field approximation values. I observe that the difference again reaches near $10^{-9}$ at the first inner-loop iteration. The criterion $10^{-8}$ can be easily satisfied as $n$ increases. This indicates premature inner-loop convergence, causing the optimizers to stop the optimization progress. 

\begin{figure}[h]\label{figure1_edge_noperturb50}
      \centering
        \includegraphics[width=\linewidth]{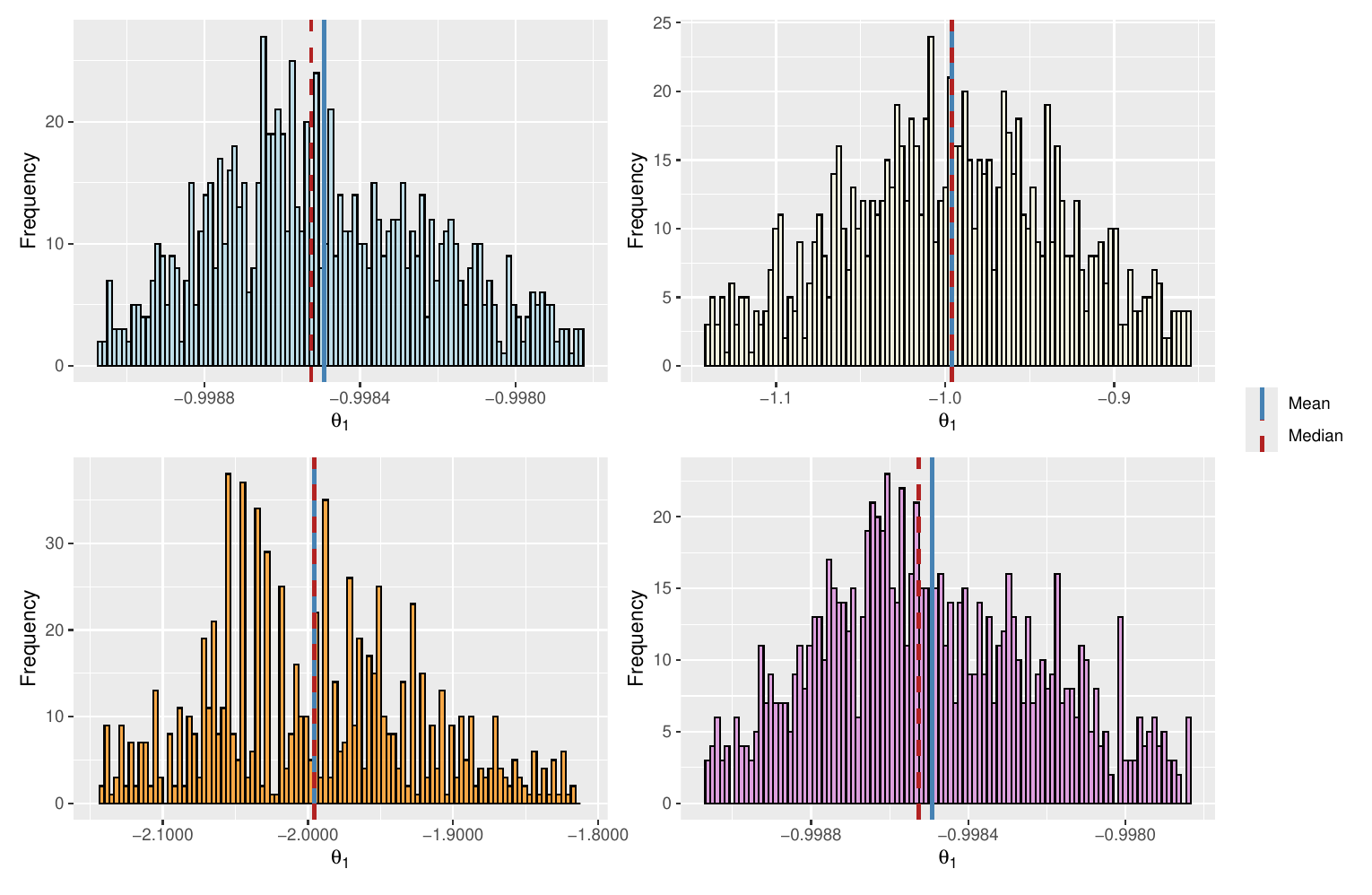}
        \caption{Histograms of 1,000 Monte Carlo simulations of four different algorithms to estimate edge parameter $\theta_1$ with $n=50$. Top left: MCMC-MLE, top right: MPLE, bottom left: Variational Approximate Estimation by \cite{mele2023approximate}, bottom right: VRBEA.}
\end{figure}

To illustrate these findings, I also visualize the Monte Carlo simulation results of four algorithms. \hyperref[figure1_edge_noperturb50]{Figure 1} shows that the MCMC-MLE, MPLE and VRBEA perform well when estimating the edge parameter $\theta_1$ when $n=$ 50. The MCMC-MLE and MPLE show larger variance than the VRBEA but the medians and means of estimates are close to the true parameter, indicating unbiasedness of the algorithms. On the other hand, the variational approximate algorithm by \cite{mele2023approximate} shows that the mean and median of its estimates are near -2, indicating a downward bias. This is due to the early stopping such that once the iteration halts, the updates fail to progress in a right direction to converge to the true parameters. 

\begin{figure}[h!]\label{figure2_triangle_noperturb50}
      \centering
        \includegraphics[width=\linewidth]{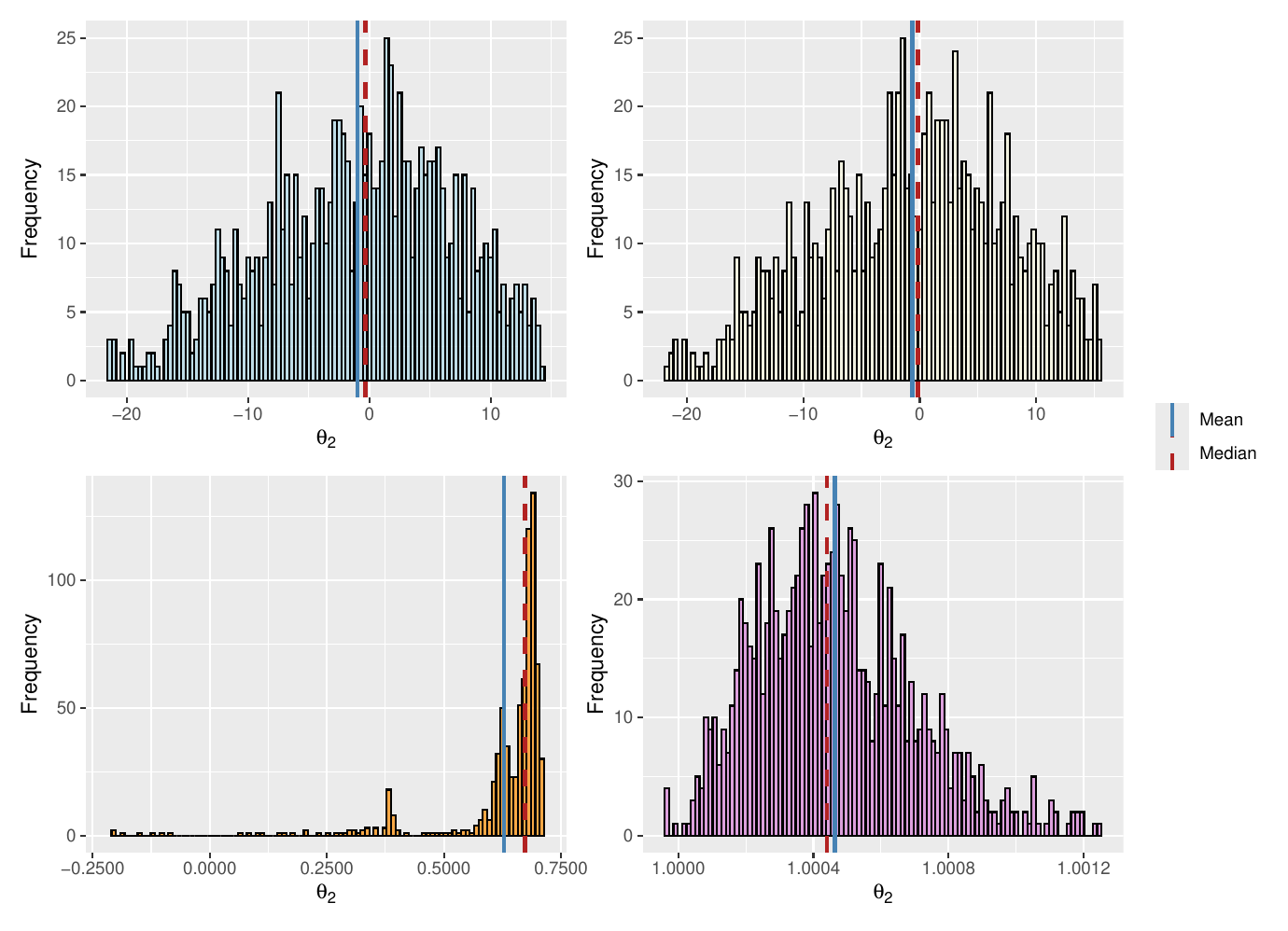}
        \caption{Histograms of 1,000 Monte Carlo simulations of four different algorithms to estimate triangle parameter $\theta_2$ with $n=50$. Top left: MCMC-MLE, top right: MPLE, bottom left: Variational Approximate Estimation by \cite{mele2023approximate}, bottom right: VRBEA.}
\end{figure}

\hyperref[figure2_triangle_noperturb50]{Figure 2} shows that both the MCMC-MLE and MPLE show large variance of estimates of the triangle parameter $\theta_2$ with $n=50$. This indicates that both algorithms suffer from unstable estimates. The algorithm by \cite{mele2023approximate} illustrates a left-skewed histogram, demonstrating that it produces biased and unstable estimates. On the other hand, the VRBEA shows small variance of estimates, confirming its stability. 

\begin{figure}[h!]\label{figure3_edge_noperturb200}
      \centering
        \includegraphics[width=\linewidth]{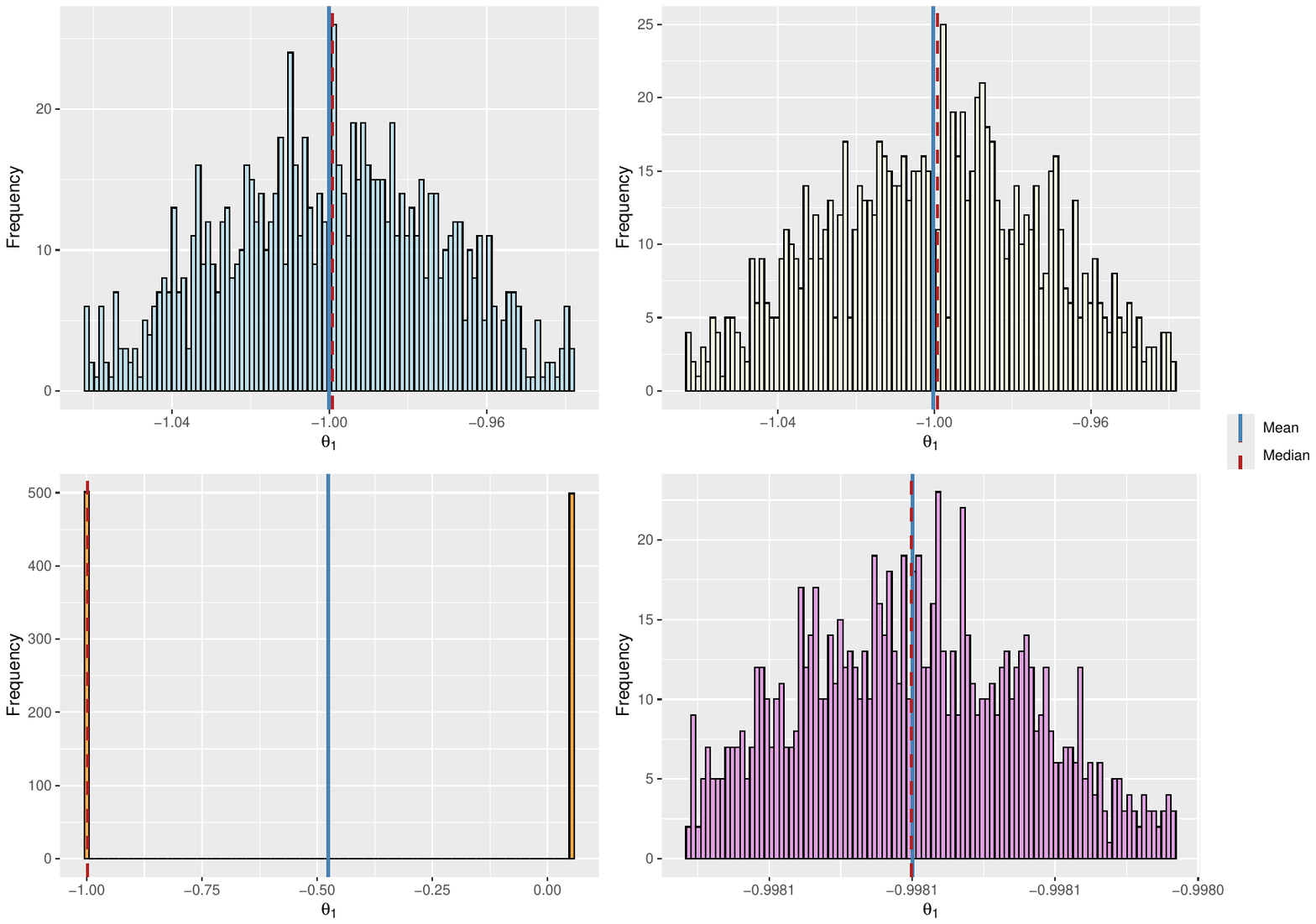}
        \caption{Histograms of 1,000 Monte Carlo simulations of four different algorithms to estimate edge parameter $\theta_1$ with $n=200$. top left: MCMC-MLE, top right: MPLE, bottom left: Variational Approximate Estimation by \cite{mele2023approximate}, bottom right: VRBEA.}
\end{figure}

\begin{figure}[h!]\label{figure4_triangle_noperturb200}
      \centering
        \includegraphics[width=\linewidth]{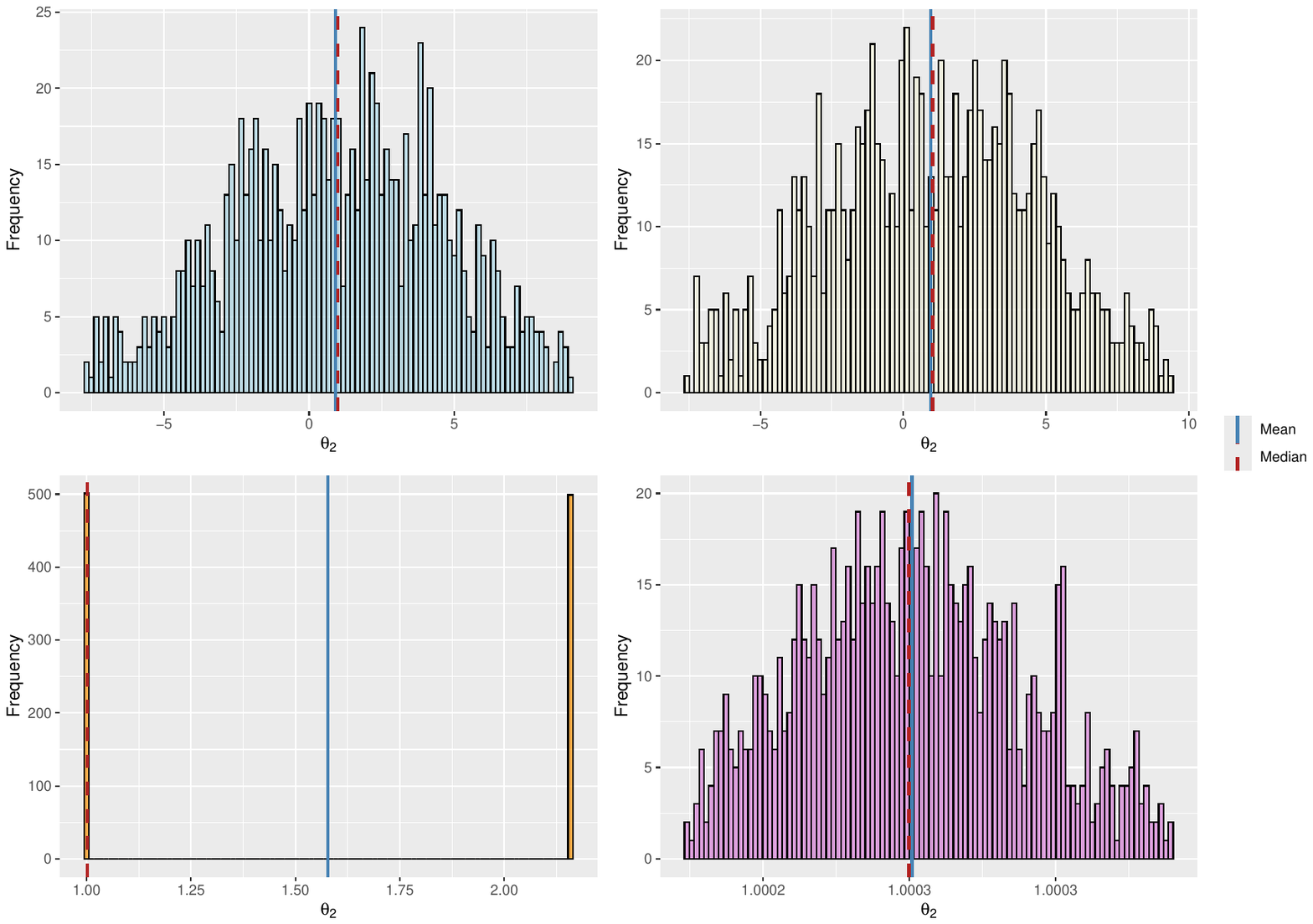}
        \caption{Histograms of 1,000 Monte Carlo simulations of four different algorithms to estimate triangle parameter $\theta_2$ with $n=200$. Top left: MCMC-MLE, top right: MPLE, bottom left: Variational Approximate Estimation by \cite{mele2023approximate}, bottom right: VRBEA.}
\end{figure}

\hyperref[figure3_edge_noperturb200]{Figure 3} and \hyperref[figure4_triangle_noperturb200]{Figure 4} present simulation results of four algorithms for $n=200$ using true parameter initialization. While the MCMC-MLE and MPLE show similar performance to $n=50$, the estimates by \cite{mele2023approximate} draw a bimodal histogram. It generates only two types of estimates. The first type is the initial value $[-1,1]$, meaning that the algorithm cannot identify any meaningful direction to estimate. The second type is the estimates around $[0.05, 2.15]$. These observations show that the algorithm prematurely halts. As evidence, the number of inner-loop iterations of this algorithm is either 0 or 1. This implies their algorithm cannot provide reliable estimates. In contrast, the VRBEA provides accurate and stable estimates across all network sizes $n=50, 100, $ and 200. \footnote{For detailed estimation time of each algorithm, see \hyperref[appendix:c]{Appendix C}.} 
\begin{table}[H]\label{table1}
\centering
\footnotesize

\caption{Monte Carlo Simulation Results: Comparison of algorithms, True parameter: [-1,1], No perturbation given}
\begin{threeparttable}
\begin{tabular}{c||cc||cc||cc||cc}
  \hline  
  $n=50$ & \multicolumn{2}{c||}{M \& Z Mean-Field}  & \multicolumn{2}{c||}{VRBEA} & \multicolumn{2}{c||}{MCMC-MLE}  & \multicolumn{2}{c}{MPLE}\\ 
  \hline
   No perturb& $\theta_1$ & $\theta_2$  &$\theta_1$ & $\theta_2$ &$\theta_1$ & $\theta_2$  &$\theta_1$ & $\theta_2$\\ 
  \hline
                
   bias &  0.4268 & 4.2455  &0.0015 & 0.0005 
                         
   & 0.0066 & 2.1821 &  0.0038 & 1.7953  \\ 
                 
  mean & -0.5732 & -3.2455  &-0.9985 & 1.0005
                         
  & -0.9934 & -1.1821 &  -0.9962 & -0.7953\\
                
  median & -2.0021 & 0.6624 & -0.9985 & 1.0004
                         
  & -0.9942 & -0.3290 &  -0.9960 & -0.1423\\  
                
  MAD & 2.8142 & 6.8624  &0.0003 & 0.0002
                         
  & 0.0571 & 7.1717 &  0.0594 & 7.4895  \\ 
                
  se & 17.7496 & 34.4923 & 0.0003 & 0.0002 
                         
   & 0.0723 & 9.0710 &  0.0750 & 9.4913 \\ 
\hline  
\hline  
  
   $n=100$  & $\theta_1$ & $\theta_2$  &$\theta_1$ & $\theta_2$ &$\theta_1$ & $\theta_2$  &$\theta_1$ & $\theta_2$\\ 
  \hline
                
   bias & 0.4059 & 1.2686  & 0.0019 & 0.0003 
                         
 & 0.0035 & 0.8574 &  0.0031 & 0.6830    \\ 
                 
  mean & -1.4059 & -0.2686  & -0.9981 & 1.0003
                         
  & -0.9965 & 0.1426 &  -0.9969 & 0.3170\\
                
  median & -1.9980 & 0.6591 & -0.9981 & 1.0003
                         
  & -0.9978 & 0.4701 & -0.9975 & 0.5269\\ 
                
  MAD & 1.1786 & 1.8638  & 0.0001 & 0.0000
                         
  & 0.0380 & 4.8223 &  0.0387 & 4.9584\\

  se & 10.8852 & 13.0867  & 0.0001 & 0.0001 
                         
  & 0.0477 & 6.0110 & 0.0485 & 6.1584  \\ 

\hline  
\hline
$n=200$  & $\theta_1$ & $\theta_2$  &$\theta_1$ & $\theta_2$ &$\theta_1$ & $\theta_2$  &$\theta_1$ & $\theta_2$\\ 
  \hline
                
   bias & 0.5244 & 0.5776  & 0.0019 & 0.0003
                         
   &  0.0002 & 0.0886 &  0.0003 & 0.0352   \\ 
                 
  mean & -0.4756 & 1.5776  & -0.9981 & 1.0003
                         
  & 1.0002 & 0.9114 &  -1.0003 & 0.9648\\
                
  median & -0.9980 & 1.0022 & -0.9981 & 1.0003
                         
  & -0.9993 & 0.9918 &  -0.9992 & 1.0320\\ 
                
  MAD & 0.5255 & 0.5787  & 0.0000 & 0.0000
                         
  & 0.0253 & 3.3256 &   0.0255 & 3.3716\\ 
                
  se & 0.5255 & 0.5787  & 0.0000 & 0.0000 
                         
  & 0.0316 & 4.1665 &   0.0318 & 4.1966 \\ 
\hline  
\end{tabular}
\begin{tablenotes}[para,flushleft]
 \footnotesize
  Note: Results of 1000 Monte Carlo estimates using the existing methods. The first column shows Approximate variational estimation algorithm of \cite{mele2023approximate}. The second column is my algorithm, VRBEA. The third column displays MCMC-MLE, the Markov Chain Monte Carlo Maximum Likelihood Estimation, with stochastic approximation by \cite{robbins1951stochastic}. The last column exhibits Maximum Pseudo Likelihood Estimation. 1000 networks are sampled by using the R package \texttt{ergm}.
  MAD is the mean absolute deviation, and se is the standard error. 
  \end{tablenotes}

\end{threeparttable}
\end{table}

\begin{table}[H]
\centering
\footnotesize

\caption{Monte Carlo Simulation Results: Comparison of algorithms, True parameter: [-1,-1]}
\begin{threeparttable}
\begin{tabular}{c||cc||cc||cc||cc}
  \hline  
  $n=50$ & \multicolumn{2}{c||}{M \& Z Mean-Field}  & \multicolumn{2}{c||}{VRBEA} & \multicolumn{2}{c||}{MCMC-MLE}  & \multicolumn{2}{c}{MPLE}\\ 
  \hline
   No perturb& $\theta_1$ & $\theta_2$  &$\theta_1$ & $\theta_2$ &$\theta_1$ & $\theta_2$  &$\theta_1$ & $\theta_2$\\ 
  \hline
                
   bias &   0.2607 & 1.8613 & 0.0014 & 0.0005
                         
   & 0.0049 & 1.6441 & 0.0022 & 1.2939  \\ 
                 
  mean & -1.2607 & -2.8613  &-0.9986 & -0.9995
                         
  & -0.9951 & -2.6441  & -0.9978 & -2.2939\\
                
  median & -1.9988 & -1.1906 & -0.9986 & -0.9996
                         
  & -0.9971 & -1.8372 & -0.9996 & -1.5659 \\  
                
  MAD & 1.5114 & 3.1491  & 0.0002 & 0.0002
                         
  & 0.0588 & 7.3441  & 0.0602 & 7.6100 \\ 
                
  se & 13.3173 & 20.7453 & 0.0003 & 0.0002
                         
   & 0.0737 & 9.2480  & 0.0755 & 9.6231 \\
\hline  
\hline  
  
   $n=100$  & $\theta_1$ & $\theta_2$  &$\theta_1$ & $\theta_2$ &$\theta_1$ & $\theta_2$  &$\theta_1$ & $\theta_2$\\ 
  \hline
                
   bias & 0.8126 & 0.6575 & 0.0018 & 0.0004
                         
 & 0.0026 & 0.7963  & 0.0022 & 0.6067   \\ 
                 
  mean & -0.1874 & -0.3425  & -0.9982 & -0.9996
                         
  & -0.9974 & -1.7963  & -0.9978 & -1.6067\\
                
  median & -0.1871 & -0.2661 & -0.9982 & -0.9996
                         
  & -0.9987 & -1.2951  & -0.9988 & -1.2063\\ 
                
  MAD & 0.1476 & 0.1653  & 0.0001 & 0.0000
                         
  & 0.0385 & 5.0953 & 0.0392 & 5.2549\\

  se & 0.7420 & 0.9466  & 0.0001 & 0.0001 
                         
  & 0.0484 & 6.4332 & 0.0494 & 6.6120 \\
\hline  
\hline
$n=200$  & $\theta_1$ & $\theta_2$  &$\theta_1$ & $\theta_2$ &$\theta_1$ & $\theta_2$  &$\theta_1$ & $\theta_2$\\ 
  \hline
                
   bias & 4.2656 & 3.0211  & 0.0019 & 0.0003
                         
   &  0.0019 & 0.4558  & 0.0019 & 0.4217   \\ 
                 
  mean & 3.2656 & 2.0211   & -0.9981 & -0.9997
                         
  &  -0.9981 & -1.4558  & -0.9981 & -1.4217\\
                
  median & -1.0000 & -1.0000 & -0.9981 & -0.9997 
                         
  & -0.9979 & -1.5357  & -0.9976 & -1.6201\\ 
                
  MAD &  4.3394 & 3.0869 & 0.0000 & 0.0000
                         
  & 0.0253 & 3.3256 &   0.0255 & 3.3716\\ 
                
  se & 4.6135 & 3.9613 & 0.0000 & 0.0000 
                         
  & 0.0317 & 4.3343  & 0.0320 & 4.3924  \\
\hline
\hline
\end{tabular}
\begin{tablenotes}[para,flushleft]
 \footnotesize
  Note: Results of 1000 Monte Carlo estimates using the existing methods. The first column shows Approximate variational estimation algorithm of \cite{mele2023approximate}. The second column is my algorithm, VRBEA. The third column displays MCMC-MLE, the Markov Chain Monte Carlo Maximum Likelihood Estimation, with stochastic approximation by \cite{robbins1951stochastic}. The last column exhibits Maximum Pseudo Likelihood Estimation. 1000 networks are sampled by using the R package \texttt{ergm}.
  MAD is the mean absolute deviation, and se is the standard error. 
  \end{tablenotes}

\end{threeparttable}
\end{table}

\subsection{Hyperparameter paths}
The following section describes changes in the estimates and function values with respect to the regularization parameter $\epsilon$ and constraint satisfaction parameter $\eta$ from Monte Carlo simulations. 
\subsubsection{Regularization path}
\begin{figure}[h!]\label{figure5_reg_path_noperturb}
      \centering
        \includegraphics[width=\linewidth]{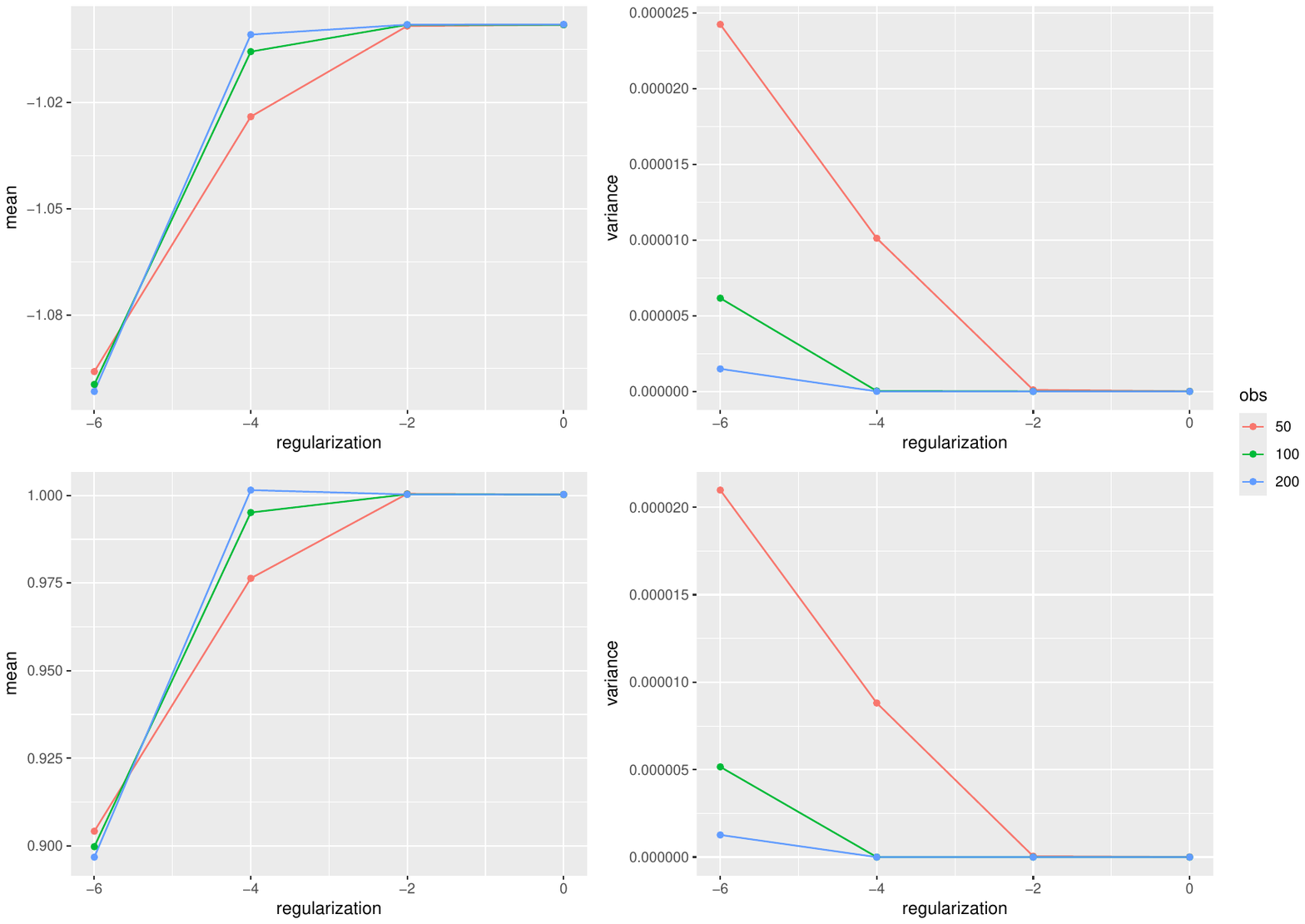}
      \caption{Regularization paths of edge estimates and triangle estimates of the VRBEA. The top panels display the mean and variance of edge estimates and the bottom panels the mean and variance of triangle estimates from 1,000 Monte Carlo simulations with no perturbation to the initialization of the algorithm and constraint satisfaction parameter $\eta$ fixed at 0.8. Regularization values are $0, 10^{-4}, 10^{-2}$, and 1. The x-axis are converted into $\log_10$ of regularization values}
\end{figure}

\begin{figure}[h!]\label{figure6_reg_path_perturb1}
      \centering
        \includegraphics[width=\linewidth]{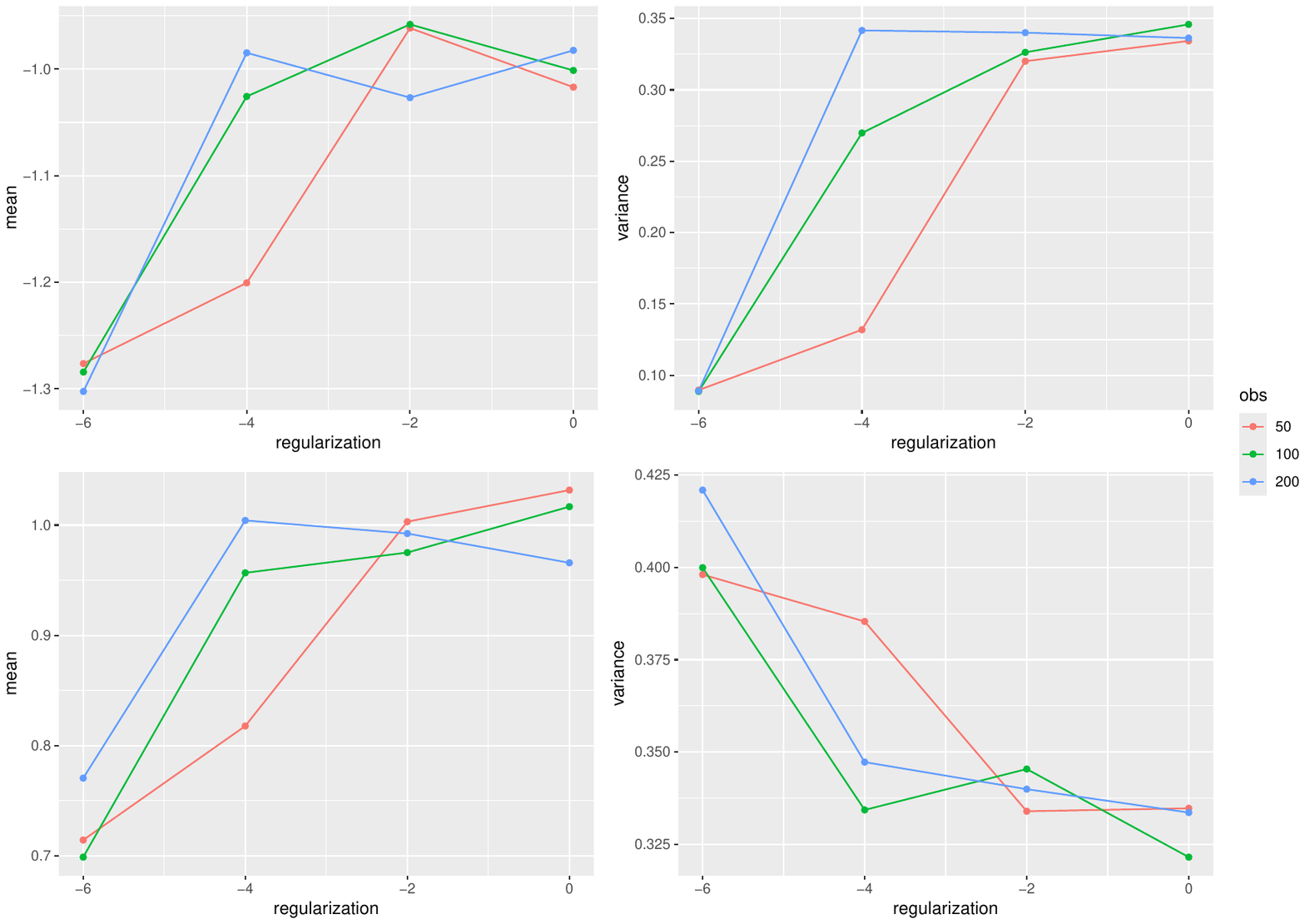}
        \caption{Regularization paths of edge estimates and triangle estimates of the VRBEA. The top panels display the mean and variance of edge estimates and the bottom panels the mean and variance of triangle estimates from 1,000 Monte Carlo simulations with perturbation of randomly drawn values from $U[-1,1]$ to the initialization of the algorithm and constraint satisfaction parameter $\eta$ fixed at 0.8. Regularization values are $0, 10^{-4}, 10^{-2}$, and 1. }
\end{figure}

\hyperref[figure5_reg_path_noperturb]{Figure 5} shows the regularization $\epsilon$ path of mean and variance of estimates from 1,000 Monte Carlo simulations with no perturbation to the initialization of the VRBEA. The paths illustrate that the mean of estimates of edge parameter $\theta_1$ converges to the true parameter -1 as the regularization increases from 0 to 1. The variance of estimates of $\theta_1$ decreases as the regularization rises. A similar trend is shown in the mean and variance of estimates of triangle parameter $\theta_2$. 
\hyperref[figure6_reg_path_perturb1]{Figure 6} displays interesting results. In contrast to the common knowledge that the variance of an estimator becomes smaller as the strength of regularization becomes larger. However, the top right corner of \hyperref[figure6_reg_path_perturb1]{Figure 6} shows a contradictory result to the bias-variance tradeoff of an estimator. This is because with a larger regularization the outer loop terminates in fewer iterations. This is because the regularization makes the lower-level variational problem easier to solve, so the feasibility term quickly reaches a small value under a high alignment parameter 
$\eta$. At the same time, the triangle component of the gradient becomes stabilized through its dependence on the stabilized mean-field variable 
$\mu$, while the edge component, which is largely independent of $\mu$, absorbs the remaining variability. This explains why, under perturbed initialization, the variance of edge estimates increases with the regularization, whereas the variance of triangle estimates decreases.

\subsubsection{Constraint satisfaction path}
\hyperref[{figure7_eta_path_n_100_param}]{Figure 7} illustrates that the constraint satisfaction parameter $\eta$ does not have influence on the means of estimates of edge and triangle parameters, $\theta_1$ and $\theta_2$, respectively. Their variances become larger as the amount of perturbation becomes larger. The size of variances remain unchanged as $\eta$ grows. 

On the other hand, \hyperref[figure8_eta_path_upper_q]{Figure 8} reveals an interesting result. The mean of values of upper-level function $F_n$ increases as $\eta$ rises. This is because as $\eta$ becomes larger, the algorithm requires the update direction of parameters $\delta$ to perfectly align with the gradient of $q^{\epsilon}$, $\nabla q^{\epsilon}$ when solving the \hyperref[direction]{optimization problem}
\[\delta_t := \argmin_{\delta} ||\nabla F_n(\theta_t\hspace{1mm}|g_n, \{X_i\}_{i=1}^n)-\delta||^2, \hspace{1mm}\text{subject to }\langle \nabla q^{\epsilon}(\theta_t, \mu_t), \delta \rangle \geq \eta \|\nabla q^{\epsilon}(\theta_t, \mu_t)\|^2.\]
Hence, this leads to high degree of discordance with the direction of purely updating $F_n$, $\nabla F_n$, leading to increasing the function value $F_n.$ It is concluded that a large $\eta$ does not necessarily mean minimizing the upper-level function $F_n$.
\begin{figure}[h!]\label{figure7_eta_path_n_100_param}
      \centering
        \includegraphics[width=\linewidth]{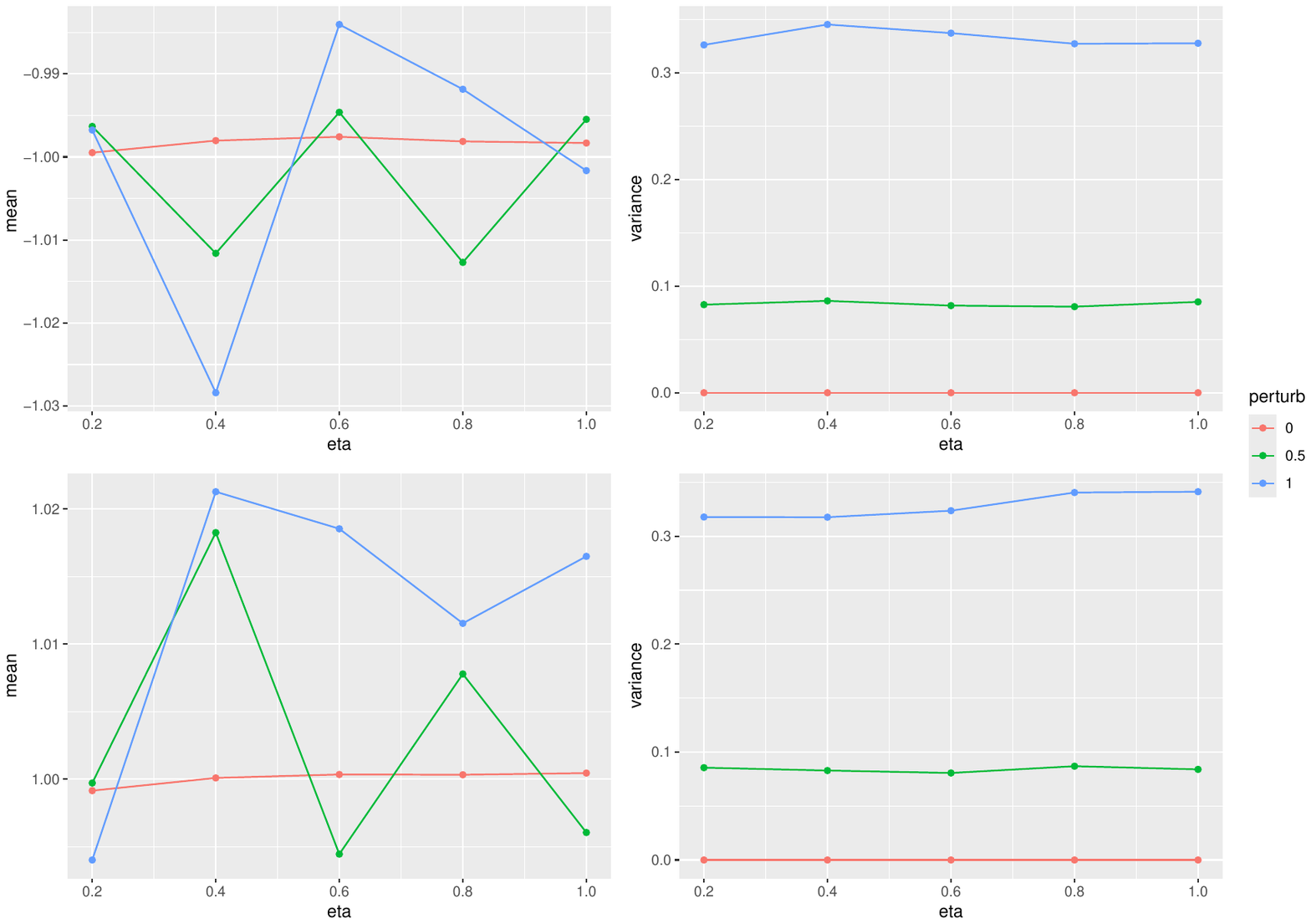}
        \caption{Constraint satisfaction paths of edge estimates and triangle estimates of the VRBEA. The top panels display the mean and variance of edge estimates and the bottom panels the mean and variance of triangle estimates from 1,000 Monte Carlo simulations with three different levels of perturbation to the initialization of the algorithm and regularization parameter $\epsilon$ fixed at 0.01. Constraint satisfaction parameter $\eta$ values on the x-axis of each plot varies from 0.2 to 1. }
\end{figure}

\begin{figure}[h!]\label{figure8_eta_path_upper_q}
      \centering
        \includegraphics[ width=\linewidth]{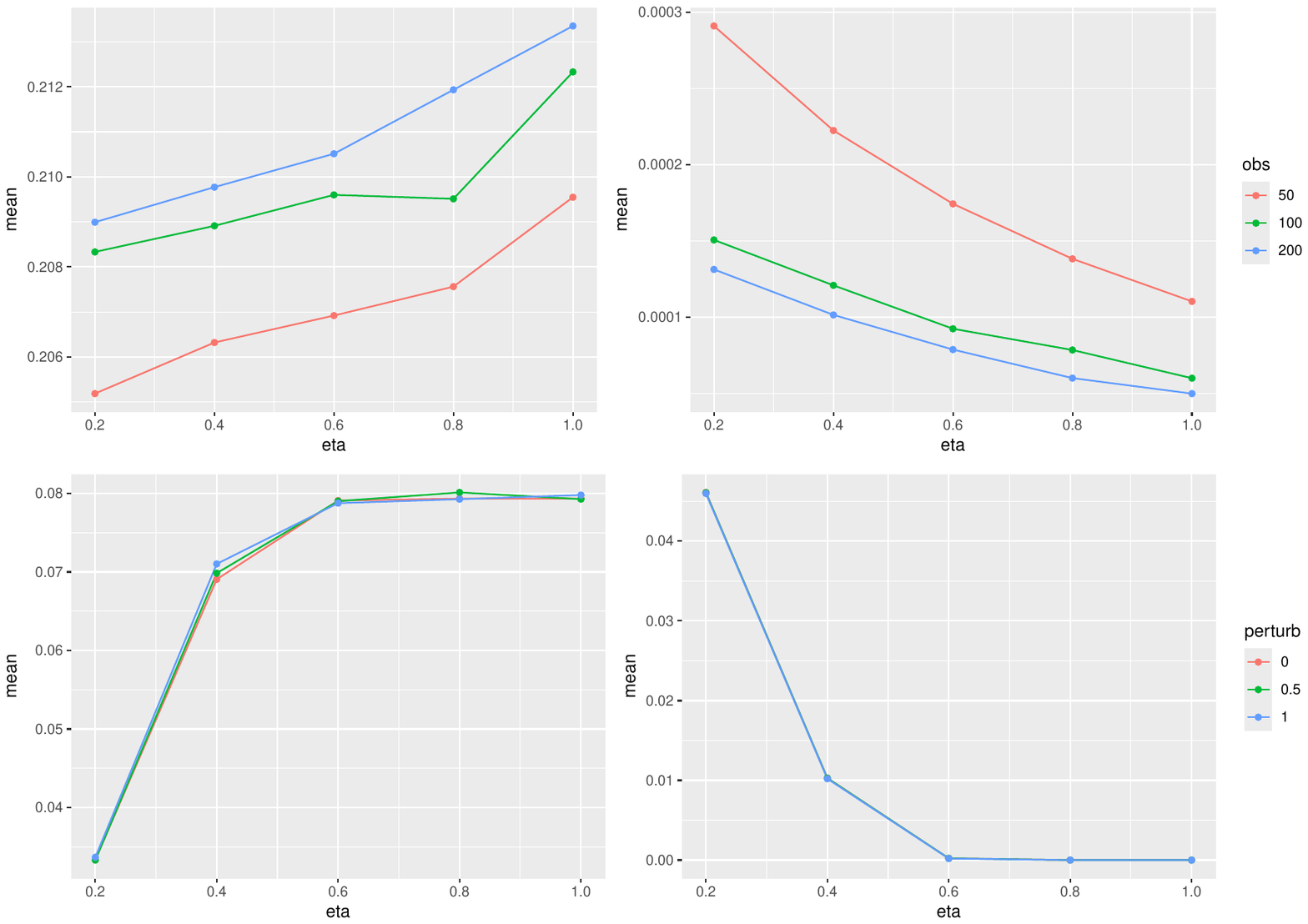}
        \caption{Constraint satisfaction paths of upper-level function $F_n$ and constraint $q^{\epsilon}$ of the VRBEA. The top left panel displays the mean of $F_n$ and the top right panel the mean of $q^{\epsilon}$ with three different network sizes $n=50, 100, 200$. The amount of perturbation to initialization is 0.5. The bottom left panel illustrates the mean of $F_n$ and the bottom right panel the mean of $q^{\epsilon}$ with three different perturbation to initialization,  $0, 0.5, 1$. The network size is fixed at 100. All the results are from 1,000 Monte Carlo simulations with the regularization parameter $\epsilon$ fixed at 0.01. Constraint satisfaction parameter $\eta$ value on the x-axis of each plot varies from 0.2 to 1. }
\end{figure}

\section{Conclusion}\label{sec8}

I develop a novel estimation algorithm for ERGMs by applying a value-function approach bilevel optimization technique proposed by \cite{liu2022bome}. By introducing $\ell_2$ regularization in the lower-level objective function, I address the nonconvexity of the optimization of the log-normalizing constant with respect to the lower-level variable $\mu$ and stabilize its optimization. In addition, I extend their non-asymptotic optimization convergenceanalysis with a unconstrained bilevel problem to the one with a constrained lower-level problem using the projected PL inequality and projected gradient descent. Finally, I demonstrate that the VRBEA enjoys more accurate and stable convergence to the true parameter of interest under appropriate initialization with judicious choices of hyper-parameters. 

There are several research questions to answer with regard to this research. First, the consistency of estimator using a mean-field approximated log-normalizing constant is not well-established (\cite{mele2023approximate}). The consistent structure estimation of ERGM using $M-$estimators such as MCMC-MLE has been recently developed under the assumption that the block structure is known (\cite{schweinberger2020concentration}). Moreover, the literature on a theoretical bound on the difference between the true log-normalizing constant and a regularized mean-field approximated log-normalizing constant in terms of network with complex structure has not been explored. A future research direction is to establish a theoretical bound on their difference. 
Second, the VRBEA relies heavily on the tuning and hyperparameters such as inner- and outer-step sizes, regularization parameter $\epsilon$ and the constraint satisfaction parameter $\eta$. Due to the property of network data, it is difficult to construct a partition of training, validation, and test data sets to enable us to obtain a data-driven set of hyper-parameters. Another future research direction is to develop a variational approach that includes a method to obtain a set of tuning parameters in a data-driven way.
Third, my algorithm constructs a variational approach due to the closed-form derivative of local dependence network topologies such as two-stars or triangles in a network. However, including these terms into the ERGM suffers from the model degeneracy. \cite{goodreau2009birds} suggests to include global dependence network topologies such as the geometrically weighted edgewise shared partner distribution (GWESP) to mitigate the degeneracy. This term does not have a closed-from derivative such that conventional variational approach may not work. A future research direction can be to develop a variational approximate algorithm that can estimate ERGM with globally dependent network topologies.  

\section*{Acknowledgment}
I acknowledge the use of Hyak at University of Washington for providing advanced computational resources essential to this research. I thank \href{https://hyak.uw.edu/team}{the Hyak Support Team} for their assistance in managing the computing infrastructure. 
 
\clearpage

\bibliography{reference.bib}

\clearpage


\onehalfspacing

\section*{Appendix A}\phantomsection\label{appendix:a}

\subsection*{Proof of Theorem 1}\label{proofthm1}\hspace{1mm}\\
\textbf{\textit{proof}.} 
Let $F_{n,t} = F_n(\theta_t)$, $\lambda_t = \lambda(\theta_t, \mu_t)$, $q_t^{\epsilon} = q^{\epsilon}(\theta_t, \mu_t)$ and $\delta_t = \nabla F_{n,t} + \lambda_t \hat{\nabla} q^{\epsilon}_t,$

where

\begin{align*}\lambda^{*}(\theta, \mu) =& \begin{cases}
    \max \bigg\{0, \eta - \frac{\langle \nabla F_n(\theta), \nabla q^{\epsilon}(\theta, \mu)\rangle}{||\nabla q^{\epsilon}(\theta, \mu)||^2}\bigg\}, & \text{for } ||\nabla q^{\epsilon}(\theta, \mu)|| >0 \\
    0 & \text{for } ||\nabla q^{\epsilon}(\theta, \mu)|| = 0.
  \end{cases}\vspace{3mm} \\
     \lambda(\theta, \mu) =& \begin{cases}
    \max \bigg\{0, \eta - \frac{\langle \nabla F_n(\theta), \hat{\nabla} q^{\epsilon}(\theta, \mu)\rangle}{||\hat{\nabla}q^{\epsilon}(\theta, \mu)||^2}\bigg\}, & \text{for } ||\hat{\nabla} q^{\epsilon}(\theta, \mu)|| >0 \\
    0 & \text{for } ||\hat{\nabla} q^{\epsilon}(\theta, \mu)|| = 0.\\
  \end{cases} 
  \vspace{10mm}\\
   \hat{\nabla}q^{\epsilon}(\theta, \mu) =& \nabla f_n^{\epsilon}(\theta, \mu) -\big[ \nabla_{\theta}^{\top} f_n^{\epsilon}(\theta, \mu^{(K)})) , \textbf{0}^{\top}\big]^{\top}
\end{align*}

Since $F_n$ is $L_{n, \epsilon}-$smooth in $\theta,$ $F_n$ is also $L_{n, \epsilon}-$ smooth in $(\theta, \mu)$, that is, for any $(\theta_i, \mu_i) \in \Theta \times \mathcal{U}$, $i = 1,2,$
\[\|\nabla F_n(\theta_1) - \nabla F_n(\theta_2) \| \leq L_{n, \epsilon} \| \theta_1 - \theta_2 \| \leq L_{n, \epsilon} \| (\theta_1, \mu_1) - (\theta_2, \mu_2)\|.\]
 Then, 
\begin{align*}
    F_{n, t+1} \leq& F_{n,t} + \langle \nabla F_{n,t}, (\theta_{t+1}, \mu_{t+1}) - (\theta_t, \mu_t)\rangle + \frac{L_{n,\epsilon}}{2}\|(\theta_{t+1}, \mu_{t+1}) - (\theta_t, \mu_t)\|^2\\
    =& F_{n,t} + \langle \nabla F_{n,t}, -\xi_t \delta_t \rangle + \frac{L_{n,\epsilon}}{2}\|-\xi_t \delta_t\|^2\tag{\hyperref[direction]{Update rule}}\\
    =& F_{n,t} -\xi_t\langle \nabla F_{n,t}, \delta_t \rangle + \frac{L_{n,\epsilon}}{2}\xi_t^2\| \delta_t\|^2\\
    \leq& F_{n,t} -\xi_t\langle \nabla F_{n,t} + \lambda_t \hat{\nabla}q^{\epsilon}_t- \lambda_t \hat{\nabla}q^{\epsilon}_t, \delta_t \rangle + \frac{L_{n,\epsilon}}{2}\xi_t^2\| \delta_t\|^2\\
    \leq& F_{n,t} -\xi_t\langle \delta_t - \lambda_t \hat{\nabla}q^{\epsilon}_t, \delta_t \rangle + \frac{L_{n,\epsilon}}{2}\xi_t^2\| \delta_t\|^2\\
    \leq& F_{n,t} -\xi_t\langle \delta_t, \delta_t \rangle + \xi_t | \langle \lambda_t \hat{\nabla}q^{\epsilon}_t, \delta_t \rangle| + \frac{L_{n,\epsilon}}{2}\xi_t^2\| \delta_t\|^2\\
    \leq& F_{n,t} -(\xi_t- \frac{L_{n,\epsilon}}{2}\xi_t^2) |\delta_t\|^2+ \xi_t \langle \lambda_t \hat{\nabla}q^{\epsilon}_t, \delta_t \rangle\\
    \leq& F_{n,t} -\frac{1}{2}\xi_t |\delta_t\|^2+ \xi_t \langle \lambda_t \hat{\nabla}q^{\epsilon}_t, \delta_t \rangle\tag{$\xi_t \leq \frac{1}{L_{n,\epsilon}}$}\\
    =&F_{n,t} -\frac{1}{2}\xi_t |\delta_t\|^2+  \eta\xi_t \lambda_t \|\hat{\nabla}q^{\epsilon}_t\|^2. 
\end{align*}
The last equality comes from the complementary slackness such that $\lambda_t(\langle \hat{\nabla}q^{\epsilon}_t, \delta_t\rangle - \eta \|\hat{\nabla}q^{\epsilon}_t\|^2) =0.$

Using \hyperref[lem9]{Lemma 8.9}, we have 
\begin{align*}
    F_{n, t+1} \leq& F_{n,t} -\frac{1}{2}\xi_t |\delta_t\|^2+  \eta\xi_t \lambda_t \|\hat{\nabla}q^{\epsilon}_t\|^2\\
    \leq& F_{n,t} -\frac{1}{2}\xi_t |\delta_t\|^2+  \eta\xi_t 
    \big[\eta \|\hat{\nabla}q^{\epsilon}_t\|^2 + M_{n, \epsilon}\|\hat{\nabla}q^{\epsilon}_t\|\big].
\end{align*}
Note that 
\begin{align*}
    \|\hat{\nabla}q^{\epsilon}_t\| =&\|\hat{\nabla}q^{\epsilon}_t - \nabla q^{\epsilon}_t + \nabla q^{\epsilon}_t\| \\
    \leq& \|\hat{\nabla}q^{\epsilon}_t - \nabla q^{\epsilon}_t\| + \|\nabla q^{\epsilon}_t\| \\
    \leq& \frac{L_{n, \epsilon, q^{\epsilon}}\sqrt{\alpha}}{\rho} \exp(-a_1 K/2)\|\nabla  q^{\epsilon}_t\| + \|\nabla q^{\epsilon}_t\| \tag{Lemma \ref{lem11}}\\
    \leq&\bigg[1+ \frac{L_{n, \epsilon, q^{\epsilon}}\sqrt{\alpha}}{\rho} \exp(-a_1 K/2)\bigg]\|\nabla  q^{\epsilon}_t\|\\
    \leq&\bigg[1+ \frac{L_{n, \epsilon, q^{\epsilon}}\sqrt{\alpha}}{\rho} \bigg]\|\nabla  q^{\epsilon}_t\|.
\end{align*}

Using Lemma \ref{lem10}, 
\begin{align*}
    F_{n, t+1} 
    \leq& F_{n,t} -\frac{1}{2}\xi_t |\delta_t\|^2+  \eta\xi_t 
    \big[\eta \|\hat{\nabla}q^{\epsilon}_t\|^2 + M_{n, \epsilon}\|\hat{\nabla}q^{\epsilon}_t\|\big]\\
    \leq& F_{n,t} -\frac{1}{2}\xi_t |\delta_t\|^2+ \eta\xi_t \Bigg[\eta \bigg[1+ \frac{L_{n, \epsilon, q^{\epsilon}}\sqrt{\alpha}}{\rho} \bigg]^2\frac{4L_{n, \epsilon, q^{\epsilon}}^2}{\kappa}q^{\epsilon}_t +\frac{2M_{n, \epsilon}L_{n, \epsilon, q^{\epsilon}}}{\sqrt{\kappa}}\sqrt{q^{\epsilon}_t} \Bigg]\tag{Lemma \ref{lem10}}
\end{align*}

Let $C_{\eta, \rho, \kappa} =\eta^2 \bigg[1+ \frac{L_{n, \epsilon, q^{\epsilon}}\sqrt{\alpha}}{\rho} \bigg]^2\frac{4L_{n, \epsilon, q^{\epsilon}}^2}{\kappa} $ and $C_{M, \eta} = 2\eta M_{n, \epsilon}\frac{L_{n, \epsilon, q^{\epsilon}}}{\sqrt{\kappa}}$. Then, 
\begin{align*}
    F_{n, t+1} 
    \leq& F_{n,t} -\frac{1}{2}\xi_t |\delta_t\|^2+ C_{\eta, \rho, \kappa} \xi_t  q^{\epsilon}_t +C_{M, \eta}\xi_t\sqrt{q^{\epsilon}_t}
\end{align*}

Using Lemma \ref{lem14}, we have 
\[\gamma(q_{t+1}^{\epsilon} - q_t^{\epsilon}) \leq -\frac{\gamma}{4}\eta  \kappa \xi_t q_t^{\epsilon}\textbf{1}\{t \leq t_0\} + \frac{\eta\kappa\gamma\xi_t}{4} b \textbf{1}\{t > t_0\}.\]

Adding the two inequality, we have 
\begin{align*}
    \Phi_{t+1} - \Phi_t \leq& -\frac{1}{2}\xi_t |\delta_t\|^2+ C_{\eta, \rho, \kappa} \xi_t  q^{\epsilon}_t +C_{M, \eta}\xi_t\sqrt{q^{\epsilon}_t} 
    -\frac{\gamma}{4}\eta  \kappa \xi_t q_t^{\epsilon}\textbf{1}\{t \leq t_0\} + \frac{\eta\kappa\gamma\xi_t}{4} b \textbf{1}\{t > t_0\}\\
    \leq& \xi_t\Bigg[-\frac{1}{2} |\delta_t\|^2+ C_{\eta, \rho, \kappa}  q^{\epsilon}_t  -\frac{\gamma}{4}\eta  \kappa q_t^{\epsilon}\textbf{1}\{t \leq t_0\} \Bigg] + \xi_t\Bigg[ C_{M, \eta}\sqrt{q^{\epsilon}_t} + \frac{\eta\kappa\gamma}{4} b \textbf{1}\{t > t_0\}\Bigg]\\
    \leq& \xi_t\Bigg[-\frac{1}{2} |\delta_t\|^2+ C_{\eta, \rho, \kappa}  q^{\epsilon}_t  -\frac{\gamma}{4}\eta  \kappa q_t^{\epsilon}\textbf{1}\{t \leq t_0\} \Bigg] + \xi_t\Bigg[ C_{M, \eta}\sqrt{b} + \frac{\eta\kappa\gamma}{4} b \textbf{1}\{t > t_0\}\Bigg].
\end{align*}
For $\gamma > \max \{\frac{2-4C_{\eta, \rho, \kappa}}{\eta \kappa}, 0\}$, we have 
\begin{align*}
    \Phi_{t+1} - \Phi_t \leq& \xi_t\Bigg[-\frac{1}{2} |\delta_t\|^2+ C_{\eta, \rho, \kappa}  q^{\epsilon}_t  -\frac{\gamma}{4}\eta  \kappa q_t^{\epsilon}\textbf{1}\{t \leq t_0\} \Bigg] + \xi_t\Bigg[ C_{M, \eta}\sqrt{b} + \frac{\eta\kappa\gamma}{4} b \textbf{1}\{t > t_0\}\Bigg]\\
    \leq& -\frac{1}{2} \xi_t \big[|\delta_t\|^2 +   q^{\epsilon}_t\big] + \xi_t \Bigg[ C_{M, \eta}\sqrt{b} + \frac{\eta\kappa\gamma}{4} b \textbf{1}\{t > t_0\}\Bigg]\\
    \leq& -\frac{1}{2} \xi_t \mathcal{K}_t+ O(\xi_t).
\end{align*}\hfill \qedsymbol

\subsection*{Proof of Theorem 2}\label{proofthm2}\hspace{1mm}\\
\textbf{\textit{proof}.} 
From Theorem \ref{thm1}, 
\begin{align*}
    \Phi_{T} - \Phi_{T-1} \leq&-\frac{1}{2} \xi_{T-1} \mathcal{K}_{T-1}+ O(\xi_{T-1})\\
    \hspace{10mm} &\vdots \\
    \Phi_{1} - \Phi_{0} \leq&-\frac{1}{2} \xi_{0} \mathcal{K}_{0}+ O(\xi_{0})
\end{align*}
We telescope $\Phi_t$ from $t=0$ through $t= T-1$ for some $T>0.$
\begin{align*}
    \Phi_T - \Phi_0 \leq& -\frac{1}{2}\sum_{t=0}^{T-1}\xi_t \mathcal{K}_t + \sum_{t=0}^{T-1} O(\xi_t)\\
\end{align*}
In fact, $O(\xi_t)$ includes $b = \max \{b_1, b_2\}$ and $\sqrt{b}$, where $b_1 = C_1  \exp(-a_1 K) = \frac{32^2(\eta+1)^2}{\eta^2 \kappa^5}L_{n,\epsilon, q^{\epsilon}}^2 M_{n,\epsilon}^2 \exp(-a_1 K)$ and $b_2 = C_2 \xi_t = \frac{8(\eta+1)L_{n,\epsilon, q^{\epsilon}}}{\eta \kappa}\xi_t$. We choose $K \geq \frac{1}{a_1}\log \frac{C_1}{C_2\xi_t}$ such that $b = b_2 \geq b_1$. Hence, 
rearranging the terms and using $\xi_t = \frac{1}{\sqrt{T}}$, 
\begin{align*}
    \frac{1}{2}\sum_{t=0}^{T-1} \frac{1}{\sqrt{T}}\mathcal{K}_t \leq& \Phi_0 - \Phi_T  + \sum_{t=0}^{T-1} O(T^{-3/4}) = \Phi_0 - \Phi_T  + O(T^{1/4}) 
    \end{align*}
 Multiplying both sides of inequality by $2/\sqrt{T},$ we have
\begin{align*}
    \frac{1}{T}\sum_{t=0}^{T-1} \mathcal{K}_t \leq \frac{2}{\sqrt{T}}\big[\Phi_0 - \Phi_T \big] + O(T^{-1/4})=O(T^{-1/4})
\end{align*}\hfill \qedsymbol

\subsection*{Lemmata}\label{lemmata}
This appendix provides self-contained proofs for the lemmata in \cite{liu2022bome}, in order to show theorem 1.

\begin{lemma}\label{lem1} (Quadratic growth)\hspace{1mm}\\
    Under \hyperref[smooth]{the smoothness}, suppose $f_n^{\epsilon}(\theta,\cdot)$ is $\rho(\theta)-$strongly convex for any $\theta \in \Theta.$ Then for all $\mu \in \mathcal{U}_{\zeta} \subseteq\mathcal{U}$, 
    \begin{align*}\label{QG}
        f_n^{\epsilon}(\theta,\mu) - f_n^{\epsilon}(\theta,\mu^{*}(\theta)) \geq \frac{\rho(\theta)}{2}||\mu-\mu^{*}(\theta)||^2 = \frac{\kappa(\theta)}{4}||\mu-\mu^{*}(\theta)||^2\tag{QG}
    \end{align*}
\end{lemma}

\textbf{\textit{proof}.} Since $f_n^{\epsilon}(\theta,\cdot)$ is $\rho(\theta)-$strongly convex, we apply the equivalence of the PL inequality to the quadratic growth (QG) under  \hyperref[smooth]{the smoothness} (\cite{karimi2016linear}). Since the update rule of $\mu$ is proceeded by the projection onto the compact subset $\mathcal{U}_{\zeta}$ of $\mathcal{U}$, \hyperref[ppl]{the projected PL inequality} and the projected gradient descent lemma can use the same constant. \hfill$\qedsymbol$

\begin{lemma}\label{lem2}(Projected gradient descent lemma)\hspace{1mm}\\
    Suppose $f_n^{\epsilon}(\theta,\cdot)$ is $L_{n,\epsilon}-$smooth. Then with a step size $\alpha \in (0, 1/L_{n, \epsilon}],$ 
    the \hyperref[pgd]{update} rule guarantees the following: 
    \[f_n^{\epsilon}(\theta, \mu^{(k+1)}) - f_n^{\epsilon}(\theta,\mu^{(k)}(\theta)) \leq -\alpha(1- \alpha \frac{L_{n,\epsilon}}{2})||G_{\alpha}^{\epsilon}(\mu^{(k)};\theta)||_2^2\]
\end{lemma}
\textbf{\textit{proof}.} Since $f_n^{\epsilon}(\theta,\cdot)$ is $L_{n,\epsilon}-$smooth, given $\theta \in \Theta$, for each step at $k\in [[K]] = \{0,1,2,....K-1\}$,
\begin{align*}
    f_n^{\epsilon}(\theta,\mu^{(k+1)}) \leq& f_n^{\epsilon}(\theta,\mu^{(k)}) + \langle \nabla_{\mu }f_n^{\epsilon}(\theta,\mu^{(k)}), \mu^{(k+1)} - \mu^{(k)}\rangle + \frac{L_{n,\epsilon}}{2}||\mu^{(k+1)} - \mu^{(k)}||_2^2.
\end{align*}
Due to the non-expansivity of projection $\Pi_{\mathcal{U}_{\zeta}}$, we have for any $x\in \mathcal{U}_{\zeta}$,
\[\langle y - \Pi_{\mathcal{U}_{\zeta}}(y) , x - \Pi_{\mathcal{U}_{\zeta}}(y) \rangle \leq 0.\]
Let $y= \mu^{(k)} - \alpha \nabla_{\mu }f_n^{\epsilon}(\theta,\mu^{(k)}) $ with step size $\alpha \in (0, 1/L_{n, \epsilon}]$, and $\mu^{(k+1)} = \Pi_{\mathcal{U}_{\zeta}}(y) $. Then, 
\[\langle \mu^{(k)} - \alpha \nabla_{\mu }f_n^{\epsilon}(\theta,\mu^{(k)}) - \mu^{(k+1)}, \mu^{(k)} - \mu^{(k+1)}\rangle \leq 0 .
\]
    
Rearranging the terms, we obtain
\[\langle \nabla_{\mu }f_n^{\epsilon}(\theta,\mu^{(k)}), \mu^{(k+1)} - \mu^{(k)}\rangle \leq -\frac{1}{\alpha}\langle \mu^{(k+1)} - \mu^{(k)}, \mu^{(k+1)} - \mu^{(k)}\rangle = -\frac{1}{\alpha}|| \mu^{(k+1)} - \mu^{(k)}||_2^2
\]
Hence, 
\begin{align*}
    f_n^{\epsilon}(\theta,\mu^{(k+1)}) \leq& f_n^{\epsilon}(\theta,\mu^{(k)}) + \langle \nabla_{\mu }f_n^{\epsilon}(\theta,\mu^{(k)}), \mu^{(k+1)} - \mu^{(k)}\rangle + \frac{L_{n,\epsilon}}{2}||\mu^{(k+1)} - \mu^{(k)}||_2^2\\
    \leq & f_n^{\epsilon}(\theta,\mu^{(k)}) -\frac{1}{\alpha}|| \mu^{(k+1)} - \mu^{(k)}||_2^2 + \frac{L_{n,\epsilon}}{2}||\mu^{(k+1)} - \mu^{(k)}||_2^2 
\end{align*}
Plugging in the \hyperref[pgd]{update} rule, we have 
\begin{align*}
    f_n^{\epsilon}(\theta,\mu^{(k+1)})
    \leq & f_n^{\epsilon}(\theta,\mu^{(k)}) -\alpha||G_{\alpha}^{\epsilon}(\mu^{(k)};\theta)||_2^2 + \alpha^2\frac{L_{n,\epsilon}}{2}||G_{\alpha}^{\epsilon}(\mu^{(k)};\theta)||_2^2\\
    \leq& f_n^{\epsilon}(\theta,\mu^{(k)}) -\alpha(1- \alpha \frac{L_{n,\epsilon}}{2})||G_{\alpha}^{\epsilon}(\mu^{(k)};\theta)||_2^2
\end{align*}\hfill \qedsymbol

\begin{lemma}\label{lem3}(Linear convergence rate)\hspace{1mm}\\
    Under the\hyperref[smooth]{smoothness}, suppose $f_n^{\epsilon}(\theta,\cdot)$ is $\rho(\theta)-$strongly convex for any $\theta \in \Theta.$ Also it has a non-empty solution set $M^{*}:=\{\mu^{*}(\theta)\}\subseteq \mathcal{U}_{\zeta} \subseteq\mathcal{U}$. Under the \hyperref[ppl]{projected PL inequality} with a step size $\alpha \in (0, 1/L_{n, \epsilon}],$ the \hyperref[pgd]{update} rule has a linear convergence rate, 
    \begin{align*}
        f_n^{\epsilon}(\theta, \mu^{(k)}) - f_n^{\epsilon}(\theta,\mu^{*}(\theta)) \leq [1-\alpha(1-\alpha \frac{L_{n,\epsilon}}{2})\kappa_{\alpha, \rho}]^{k}(f_n^{\epsilon}(\theta, \mu^{(0)}) - f_n^{\epsilon}(\theta,\mu^{*}(\theta))).
    \end{align*}
\end{lemma}
\textbf{\textit{proof}.} 
    We combine the \hyperref[lem2]{projected gradient descent lemma} and the \hyperref[ppl]{projected PL inequality.}
    \begin{align*}
        f_n^{\epsilon}(\theta,\mu^{(k+1)})-f_n^{\epsilon}(\theta,\mu^{(k)}) 
    \leq & -\alpha(1- \alpha \frac{L_{n,\epsilon}}{2})||G_{\alpha}^{\epsilon}(\mu^{(k)};\theta)||_2^2
    \leq -\alpha(1- \alpha \frac{L_{n,\epsilon}}{2}) \kappa_{\alpha, \rho}[f_n^{\epsilon}(\theta,\mu^{(k)}) - f_n^{\epsilon}(\theta,\mu^{*}(\theta))]
    \end{align*}
Subtracting $f_n^{\epsilon}(\theta,\mu^{*}(\theta))$ from both sides of the inequality and moving $f_n^{\epsilon}(\theta,\mu^{(k)}) $ to the LHS, we obtain
\begin{align*}
    f_n^{\epsilon}(\theta,\mu^{(k+1)})-f_n^{\epsilon}(\theta,\mu^{*}(\theta)) \leq&
f_n^{\epsilon}(\theta,\mu^{(k)}) -\alpha(1- \alpha \frac{L_{n,\epsilon}}{2}) \kappa_{\alpha, \rho}[f_n^{\epsilon}(\theta,\mu^{(k)}) - f_n^{\epsilon}(\theta,\mu^{*}(\theta))]-f_n^{\epsilon}(\theta,\mu^{*}(\theta))
\end{align*}
Rearranging the terms, 
\begin{align*}
    f_n^{\epsilon}(\theta,\mu^{(k+1)})-f_n^{\epsilon}(\theta,\mu^{*}(\theta)) \leq&
[1-\alpha(1- \alpha \frac{L_{n,\epsilon}}{2}) \kappa_{\alpha, \rho}][f_n^{\epsilon}(\theta,\mu^{(k)}) - f_n^{\epsilon}(\theta,\mu^{*}(\theta))]
\end{align*}
Iterating the process over $k$ results in
\begin{align*}
    f_n^{\epsilon}(\theta,\mu^{(k)})-f_n^{\epsilon}(\theta,\mu^{*}(\theta)) \leq&
[1-\alpha(1- \alpha \frac{L_{n,\epsilon}}{2}) \kappa_{\alpha, \rho}]^k [f_n^{\epsilon}(\theta,\mu^{(0)}) - f_n^{\epsilon}(\theta,\mu^{*}(\theta))].
\end{align*}

Moreover, notice that for any $k\in [[K]] = \{0,1,2,3,..., K-1\}$
\[q^{\epsilon}(\theta, \mu^{(k)}) = f_n^{\epsilon}(\theta,\mu^{(k)})-f_n^{\epsilon}(\theta,\mu^{*}(\theta)).\]
Therefore, we have 
\begin{align*}
    q^{\epsilon}(\theta, \mu^{(k)}) \leq \bigg[1-\alpha(1- \alpha \frac{L_{n,\epsilon}}{2}) \kappa_{\alpha, \rho}\bigg]^k q^{\epsilon}(\theta, \mu^{(0)})
\end{align*}
Using $1-x \leq \exp(-x),$ we have a positive constant $a_1 := a_1(\alpha, L_{n,\epsilon}, \kappa_{\alpha, \rho}) =\alpha \big\{1-\alpha \frac{L_{n,\epsilon}}{2}\big\} \kappa_{\alpha, \rho} >0$ as a function of parameters $\alpha, L_{n, \epsilon}, \kappa_{\alpha, \rho}$ such that the following inequality holds.
\begin{align*}
    q^{\epsilon}(\theta, \mu^{(k)}) \leq& \exp\{\bigg[-\alpha \big\{1-\alpha \frac{L_{n,\epsilon}}{2}\big\} \kappa_{\alpha, \rho}\bigg]\}^k q^{\epsilon}(\theta, \mu^{(0)})\\
    \leq& \exp\{k\bigg[-\alpha \big\{1-\alpha \frac{L_{n,\epsilon}}{2}\big\} \kappa_{\alpha, \rho}\bigg]\} q^{\epsilon}(\theta, \mu^{(0)})\\
    =& \exp\{-a_1(\alpha, L_{n,\epsilon}, \kappa_{\alpha, \rho})k\} q^{\epsilon}(\theta, \mu^{(0)}).\hspace{5mm}
\end{align*}\hfill \qedsymbol

\begin{lemma}\label{lem4}
    Under the \hyperref[smooth]{smoothness} 
    let $K$ be the maximum iteration for inner loop of VRBEA and $\mu^{*}(\theta) \in \text{int} (\mathcal{U}_{\zeta})$. Then for any $(\theta, \mu) \in \Theta \times \mathcal{U}_{\zeta}$
    \[||\nabla q^{\epsilon}(\theta, \mu) - \hat{\nabla}q^{\epsilon}(\theta, \mu)|| \leq L_{n, \epsilon}||\mu^{(K)}- \mu^{*}(\theta)||,\]
    
    where \[q^{\epsilon}(\theta, \mu) = f_n^{\epsilon}(\theta, \mu) - f_n^{\epsilon}(\theta, \mu^{*}(\theta)) \]\[\hat{\nabla}q^{\epsilon}(\theta, \mu) = \nabla f_n^{\epsilon}(\theta, \mu) -\big[ \nabla_{\theta}^{\top} f_n^{\epsilon}(\theta, \mu^{(K)})) , \textbf{0}^{\top}\big]^{\top}\]
\end{lemma}
\textbf{\textit{proof}.} 
In fact, 
\begin{align*}
    ||\nabla q^{\epsilon}(\theta, \mu) - \hat{\nabla}q^{\epsilon}(\theta, \mu)|| =& ||\begin{bmatrix}
        \nabla_{\theta} f_n^{\epsilon}(\theta, \mu) \\
        \nabla_{\mu} f_n^{\epsilon}(\theta, \mu)
    \end{bmatrix} - \begin{bmatrix}
        \nabla_{\theta} f_n^{\epsilon}(\theta, \mu^{*}(\theta)) \\
        \nabla_{\mu} f_n^{\epsilon}(\theta, \mu^{*}(\theta))
    \end{bmatrix}- \big(
    \begin{bmatrix}
    \nabla_{\theta} f_n^{\epsilon}(\theta, \mu) \\
        \nabla_{\mu} f_n^{\epsilon}(\theta, \mu)
    \end{bmatrix} - \begin{bmatrix}
        \nabla_{\theta} f_n^{\epsilon}(\theta, \mu^{(K)}) \\
        \nabla_{\mu} f_n^{\epsilon}(\theta, \mu^{(K)})\end{bmatrix}
    \big)||\\
    =& ||\begin{bmatrix}
        \nabla_{\theta} f_n^{\epsilon}(\theta, \mu) \\
        \nabla_{\mu} f_n^{\epsilon}(\theta, \mu)
    \end{bmatrix} - \begin{bmatrix}
        \nabla_{\theta} f_n^{\epsilon}(\theta, \mu^{*}(\theta)) \\
        \textbf{0}
    \end{bmatrix}- \big(
    \begin{bmatrix}
    \nabla_{\theta} f_n^{\epsilon}(\theta, \mu) \\
        \nabla_{\mu} f_n^{\epsilon}(\theta, \mu)
    \end{bmatrix} - \begin{bmatrix}
        \nabla_{\theta} f_n^{\epsilon}(\theta, \mu^{(K)}) \\
        \textbf{0}\end{bmatrix}
    \big)||\\
    =&||\nabla_{\theta} f_n^{\epsilon}(\theta, \mu^{(K)}) -  \nabla_{\theta} f_n^{\epsilon}(\theta, \mu^{*}(\theta))||.
\end{align*}
The second equality results from the definition of $\hat{\nabla}q^{\epsilon}(\theta, \mu)$ and the first order condition of 
$\nabla_{\mu} f_n^{\epsilon}(\theta, \mu^{*}(\theta))$. Hence, 
\begin{align*}
    ||\nabla q^{\epsilon}(\theta, \mu) - \hat{\nabla}q^{\epsilon}(\theta, \mu)|| =&||\nabla_{\theta} f_n^{\epsilon}(\theta, \mu^{(K)}) -  \nabla_{\theta} f_n^{\epsilon}(\theta, \mu^{*}(\theta))||\\
    \leq & L_{n, \epsilon}||(\theta, \mu^{(K)})- (\theta, \mu^{*}(\theta))|| \\
    =& L_{n, \epsilon}||\mu^{(K)}- \mu^{*}(\theta)||\hspace{5mm}
\end{align*}\hfill \qedsymbol

\begin{lemma}\label{lem5}
    Under the \hyperref[smooth]{smoothness} and the \hyperref[ppl]{projected PL inequality}, using \hyperref[lem1]{Quadratic growth} with a step size $\alpha \in (0, 1/L_{n, \epsilon}],$ for any $\theta_1, \theta_2 \in \Theta$, 
    \[||\mu^{*}(\theta_2) - \mu^{*}(\theta_1)|| \leq \frac{2 L_{n, \epsilon}}{\kappa}||\theta_1 - \theta_2||\]
\end{lemma}
\textbf{\textit{proof}.} Note that under the \hyperref[ppl]{projected PL inequality}, 
\[||G_{\alpha}^{\epsilon}(\mu; \theta)||^2 \geq \kappa_{\alpha, \rho}[f_n^{\epsilon}(\theta, \mu) - f_n^{\epsilon}(\theta, \mu^{*}(\theta))] \]
By the \hyperref[lem1]{quadratic growth}, we have for any $\theta$ and $\mu$, 
\[f_n^{\epsilon}(\theta, \mu) - f_n^{\epsilon}(\theta, \mu^{*}(\theta)) \geq \frac{\kappa}{4}||\mu - \mu^{*}(\theta)||^2.\]

Take $\theta = \theta_1$ and $\mu = \mu^{*}(\theta_2)$. Then,
\[f_n^{\epsilon}(\theta_1, \mu^{*}(\theta_2)) - f_n^{\epsilon}(\theta_1, \mu^{*}(\theta_1)) \geq \frac{\kappa}{4}||\mu^{*}(\theta_2) - \mu^{*}(\theta_1)||^2.\]
From the \hyperref[ppl]{projected PL inequality} and the definition of \hyperref[pgm]{projected gradient mapping},
\begin{align*}
    ||G_{\alpha}^{\epsilon}(\mu^{*}(\theta_2); \theta_1)||^2 =& ||G_{\alpha}^{\epsilon}(\mu^{*}(\theta_2); \theta_1) - G_{\alpha}^{\epsilon}(\mu^{*}(\theta_2); \theta_2)||^2\\
    =& \frac{1}{\alpha^2}||(\theta_2) - \Pi_{\mathcal{U}}(\mu^{*}(\theta_2)- \alpha \nabla_{\mu}f_n^{\epsilon}(\theta_1, \mu^{*}(\theta_2)))-  (\mu^{*}(\theta_2) - \Pi_{\mathcal{U}}(\mu^{*}(\theta_2)- \alpha \nabla_{\mu}f_n^{\epsilon}(\theta_2, \mu^{*}(\theta_2)))||^2\\
    =& \frac{1}{\alpha^2}|| - \Pi_{\mathcal{U}}(\mu^{*}(\theta_2)- \alpha \nabla_{\mu}f_n^{\epsilon}(\theta_1, \mu^{*}(\theta_2))) +  \Pi_{\mathcal{U}}(\mu^{*}(\theta_2)- \alpha \nabla_{\mu}f_n^{\epsilon}(\theta_2, \mu^{*}(\theta_2)))||^2
\end{align*}
Due to the non-expansivity of projection, in other words, $||\Pi_{\mathcal{U}}(x)-\Pi_{\mathcal{U}}(y)|| \leq ||x-y||,$
\begin{align*}
    &|| \Pi_{\mathcal{U}}(\mu^{*}(\theta_2)- \alpha \nabla_{\mu}f_n^{\epsilon}(\theta_2, \mu^{*}(\theta_2)))- \Pi_{\mathcal{U}}(\mu^{*}(\theta_2)- \alpha \nabla_{\mu}f_n^{\epsilon}(\theta_1, \mu^{*}(\theta_2)))||^2 \\
    =& ||\mu^{*}(\theta_2)- \alpha \nabla_{\mu}f_n^{\epsilon}(\theta_2, \mu^{*}(\theta_2)) - (\mu^{*}(\theta_2)- \alpha \nabla_{\mu}f_n^{\epsilon}(\theta_1, \mu^{*}(\theta_2)))||^2\\
    =& \alpha^2||\nabla_{\mu}f_n^{\epsilon}(\theta_1, \mu^{*}(\theta_2)) - \nabla_{\mu}f_n^{\epsilon}(\theta_2, \mu^{*}(\theta_2)) ||^2\\
    \leq& \alpha^2(L_{n, \epsilon})^2||(\theta_1, \mu^{*}(\theta_2)) - (\theta_2, \mu^{*}(\theta_2)) ||^2\\
    \leq&\alpha^2(L_{n, \epsilon})^2||\theta_1 - \theta_2||^2
\end{align*}
In sum, 
\[\frac{\kappa}{4}||\mu^{*}(\theta_2) - \mu^{*}(\theta_1)||^2\leq f_n^{\epsilon}(\theta_1, \mu^{*}(\theta_2)) - f_n^{\epsilon}(\theta_1, \mu^{*}(\theta_1)),\]
\[\kappa_{\alpha, \rho}[f_n^{\epsilon}(\theta_1, \mu^{*}(\theta_2)) - f_n^{\epsilon}(\theta_1, \mu^{*}(\theta_1))]\leq ||G_{\alpha}^{\epsilon}(\mu^{*}(\theta_2); \theta_1)||^2 \leq  L_{n, \epsilon}^2||\theta_1 - \theta_2||^2.\]
Hence, using $\kappa = 2\rho$, and $\kappa_{\alpha, \rho}= \frac{2\rho}{\alpha}$,
\begin{align*}
    \frac{2\rho}{\alpha}\frac{\rho}{2}||\mu^{*}(\theta_2) - \mu^{*}(\theta_1)||^2\leq&\frac{2\rho}{\alpha}[f_n^{\epsilon}(\theta_1, \mu^{*}(\theta_2)) - f_n^{\epsilon}(\theta_1, \mu^{*}(\theta_1))]\\ 
    \leq&||G_{\alpha}^{\epsilon}(\mu^{*}(\theta_2); \theta_1)||^2 \\
    \leq& L_{n, \epsilon}^2||\theta_1 - \theta_2||^2
\end{align*}
Multiplying both sides by $\alpha/\rho^2$ and take square root on them, 
\[||\mu^{*}(\theta_2) - \mu^{*}(\theta_1)||\leq \frac{\sqrt{\alpha}}{\rho}L_{n,\epsilon}\|\theta_1 - \theta_2||\leq \frac{L_{n,\epsilon}}{\rho}\|\theta_1 - \theta_2||\]

Hence, for any $\theta_1, \theta_2 \in \Theta$,\[||\mu^{*}(\theta_2) - \mu^{*}(\theta_1)||\leq \frac{2L_{n, \epsilon}}{\kappa}||\theta_1 - \theta_2||\hspace{5mm} \]\hfill \qedsymbol

\begin{lemma}\label{lem6}
    Under the \hyperref[smooth]{smoothness}, for any $\theta \in \Theta$. \[||\nabla_{\mu}q^{\epsilon}(\theta, \mu_1) - \nabla_{\mu}q^{\epsilon}(\theta, \mu_2)|| \leq L_{n,\epsilon}||\mu_1 - \mu_2||.\]
\end{lemma}
\textbf{\textit{proof}.} Notice that $\nabla_{\mu}q^{\epsilon}(\theta, \mu_1)  = \nabla_{\mu}f_n^{\epsilon}(\theta, \mu_1) -\nabla_{\mu}f_n^{\epsilon}(\theta, \mu^{*}(\theta_1))=  \nabla_{\mu}f_n^{\epsilon}(\theta, \mu_1).$ Therefore, \begin{align*}
    ||\nabla_{\mu}q^{\epsilon}(\theta, \mu_1) - \nabla_{\mu}q^{\epsilon}(\theta, \mu_2)|| \leq L_{n,\epsilon}||\mu_1 - \mu_2|| =& ||\nabla_{\mu}f_n^{\epsilon}(\theta, \mu_1) - \nabla_{\mu}f_n^{\epsilon}(\theta, \mu_2)|| \\
    \leq& L_{n,\epsilon}||(\theta, \mu_1) - (\theta, \mu_2)||\\
    =& L_{n,\epsilon}|| \mu_1-\mu_2||\hspace{5mm}
\end{align*}\hfill \qedsymbol

\begin{lemma}\label{lem7}
    Under the \hyperref[smooth]{smoothness} and the \hyperref[ppl]{projected PL inequality}, 
    \[||\nabla q^{\epsilon}(\theta_1, \mu_1) - \nabla q^{\epsilon}(\theta_2, \mu_2)|| \leq L_{n,\epsilon, q^{\epsilon}} ||(\theta_1, \mu_1) - (\theta_2, \mu_2)||, \tag{*}\]\label{lq}
    with $L_{n,\epsilon, q^{\epsilon}} = 2L_{n,\epsilon}(L_{n, \epsilon}/\kappa + 1).$
\end{lemma}\textbf{\textit{proof}.} 
\begin{align*}
    &||\nabla q^{\epsilon}(\theta_1, \mu_1) - \nabla q^{\epsilon}(\theta_2, \mu_2)|| \\
    =& ||\nabla f_n^{\epsilon}(\theta_1, \mu_1) - \nabla f_n^{\epsilon}(\theta_1, \mu^{*}(\theta_1)) - (\nabla f_n^{\epsilon}(\theta_2, \mu_2) - \nabla f_n^{\epsilon}(\theta_2, \mu^{*}(\theta_2)))||\\
    =&||\nabla f_n^{\epsilon}(\theta_1, \mu_1) -\nabla f_n^{\epsilon}(\theta_2, \mu_2) - (\nabla f_n^{\epsilon}(\theta_1, \mu^{*}(\theta_1)) - \nabla f_n^{\epsilon}(\theta_2, \mu^{*}(\theta_2)))||\\
    \leq& \underbrace{||\nabla f_n^{\epsilon}(\theta_1, \mu_1) -\nabla f_n^{\epsilon}(\theta_2, \mu_2)||}_{(1)} + \underbrace{||\nabla f_n^{\epsilon}(\theta_1, \mu^{*}(\theta_1)) - \nabla f_n^{\epsilon}(\theta_2, \mu^{*}(\theta_2))||}_{(2)}
\end{align*}
\begin{enumerate}[label = (\arabic*)]
    \item Under the \hyperref[smooth]{smoothness}, $||\nabla f_n^{\epsilon}(\theta_1, \mu_1) -\nabla f_n^{\epsilon}(\theta_2, \mu_2)|| \leq L_{n,\epsilon}||(\theta_1, \mu_1) - (\theta_2, \mu_2)||$.
    \item Under the \hyperref[smooth]{smoothness}, $||\nabla f_n^{\epsilon}(\theta_1, \mu^{*}(\theta_1)) -\nabla f_n^{\epsilon}(\theta_2, \mu^{*}(\theta_2)|| \leq L_{n,\epsilon}||(\theta_1, \mu^{*}(\theta_1)) - (\theta_2, \mu^{*}(\theta_2))||$.
    \begin{align*}
        |(\theta_1, \mu^{*}(\theta_1)) - (\theta_2, \mu^{*}(\theta_2))|| =& \big[||\theta_1 - \theta_2||^2 + ||\mu^{*}(\theta_1) - \mu^{*}(\theta_2)||^2\big]^{1/2}\\
        \leq& \big[||\theta_1 - \theta_2||^2 + 
        (\frac{2L_{n,\epsilon}}{\kappa})^2||\theta_1 - \theta_2|| ^2\big]^{1/2}\tag{ Lemma \ref{lem5}}\\
        =& \sqrt{1+ (\frac{2L_{n,\epsilon}}{\kappa})^2}\big[||\theta_1 - \theta_2|| ^2\big]^{1/2}\\
        \leq& \sqrt{(1+ \frac{2L_{n,\epsilon}}{\kappa})^2}\big[||\theta_1 - \theta_2||^2 + ||\mu_1 - \mu_2||^2\big]^{1/2}\\
        =& (1+ \frac{2L_{n,\epsilon}}{\kappa})||(\theta_1, \mu_1) - (\theta_2, \mu_2)||.
    \end{align*}
\end{enumerate}Putting altogether,
\begin{align*}
    ||\nabla q^{\epsilon}(\theta_1, \mu_1) - \nabla q^{\epsilon}(\theta_2, \mu_2)||  \leq& \underbrace{||\nabla f_n^{\epsilon}(\theta_1, \mu_1) -\nabla f_n^{\epsilon}(\theta_2, \mu_2)||}_{(1)} + \underbrace{||\nabla f_n^{\epsilon}(\theta_1, \mu^{*}(\theta_1)) - \nabla f_n^{\epsilon}(\theta_2, \mu^{*}(\theta_2))||}_{(2)} \\
    \leq& L_{n,\epsilon}||(\theta_1, \mu_1) - (\theta_2, \mu_2)|| + L_{n,\epsilon}(1+ \frac{2L_{n,\epsilon}}{\kappa})||(\theta_1, \mu_1) - (\theta_2, \mu_2)||\\
    =&  \big[L_{n,\epsilon} +  L_{n,\epsilon} + \frac{2L_{n,\epsilon}^2}{\kappa}\big]||(\theta_1, \mu_1) - (\theta_2, \mu_2)||\\
    =& \underbrace{2L_{n,\epsilon}(\frac{L_{n,\epsilon}}{\kappa}+1)}_{L_{n, \epsilon, q^{\epsilon}}}||(\theta_1, \mu_1) - (\theta_2, \mu_2)||\hspace{5mm}
\end{align*}\hfill \qedsymbol

\begin{lemma}\label{lem8}
    Under the \hyperref[bound]{boundedness}, for any $(\theta, \mu) \in \Theta \times \mathcal{U}_{\zeta}$,
    $||\delta^{*}(\theta, \mu)||, ||\nabla q^{\epsilon}(\theta, \mu)||, ||\hat{\nabla} q^{\epsilon}(\theta, \mu)|| \leq 2(\eta+1)M_{n,\epsilon}.$
\end{lemma}
\textbf{\textit{proof}.} 
First, we show the bound on $||\delta^{*}(\theta, \mu)||$. 
\begin{align*}
    ||\delta^{*}(\theta, \mu)|| =& ||\lambda^{*}(\theta, \mu) \hat{\nabla}q^{\epsilon}(\theta, \mu) + \nabla F_n(\theta, \mu)||\\
    \leq& \underbrace{|\lambda^{*}(\theta, \mu)|}_{(i)} ||\hat{\nabla}q^{\epsilon}(\theta, \mu)|| + ||\nabla F_n(\theta, \mu)||
\end{align*}
In fact, when $||\hat{\nabla}q^{\epsilon}(\theta, \mu)|| >0, \hspace{1mm}(i)$ is 
\begin{align*}
    |\lambda^{*}(\theta, \mu)| =& |\eta - \frac{\langle \hat{\nabla}q^{\epsilon}(\theta, \mu),\nabla F_n(\theta, \mu)\rangle }{||\hat{\nabla}q^{\epsilon}(\theta, \mu)||^2}|\\
    \leq& \eta + |\frac{\langle \hat{\nabla}q^{\epsilon}(\theta, \mu),\nabla F_n(\theta, \mu)\rangle }{||\hat{\nabla}q^{\epsilon}(\theta, \mu)||^2}\\
    =& \eta + \frac{1}{||\hat{\nabla}q^{\epsilon}(\theta, \mu)||^2}|\langle \hat{\nabla}q^{\epsilon}(\theta, \mu),\nabla F_n(\theta, \mu)\rangle|\\
    \leq& \eta + \frac{1}{||\hat{\nabla}q^{\epsilon}(\theta, \mu)||^2}|| \hat{\nabla}q^{\epsilon}(\theta, \mu)|| ||\nabla F_n(\theta, \mu)||
\end{align*}
Therefore, 
\begin{align*}
    ||\delta^{*}(\theta, \mu)||\leq& |\lambda^{*}(\theta, \mu)| ||\hat{\nabla}q^{\epsilon}(\theta, \mu)|| + ||\nabla F_n(\theta, \mu)||\\
    \leq&\big[\eta + \frac{||\nabla F_n(\theta, \mu)||}{||\hat{\nabla}q^{\epsilon}(\theta, \mu)||} \big]||\hat{\nabla}q^{\epsilon}(\theta, \mu)|| + ||\nabla F_n(\theta, \mu)||\\
    =&\eta ||\hat{\nabla}q^{\epsilon}(\theta, \mu)|| + 2||\nabla F_n(\theta, \mu)||
\end{align*}
Since $||\hat{\nabla}q^{\epsilon}(\theta, \mu)|| = ||\nabla f_n^{\epsilon}(\theta, \mu) -\big[ \nabla_{\theta}^{\top} f_n^{\epsilon}(\theta, \mu^{(K)})) , \textbf{0}^{\top}\big]^{\top}||, $
\[||\nabla f_n^{\epsilon}(\theta, \mu) -\big[ \nabla_{\theta}^{\top} f_n^{\epsilon}(\theta, \mu^{(K)})) , \textbf{0}^{\top}\big]^{\top}|| \leq ||\nabla f_n^{\epsilon}(\theta, \mu) ||+||\big[ \nabla_{\theta}^{\top} f_n^{\epsilon}(\theta, \mu^{(K)})) , \textbf{0}^{\top}\big]^{\top}||\leq 2M_{n,\epsilon}.\]
Therefore, 
\begin{align*}
    |\delta^{*}(\theta, \mu)| \leq \eta ||\hat{\nabla}q^{\epsilon}(\theta, \mu)|| + 2||\nabla F_n(\theta, \mu)|| \leq 2\eta M_{n,\epsilon} + 2M_{n,\epsilon} = 2(\eta+1)M_{n,\epsilon}\hspace{5mm}
\end{align*}\hfill \qedsymbol

\begin{lemma}\label{lem9}
Under the \hyperref[bound]{boundedness}, for any $(\theta, \mu) \in \Theta \times \mathcal{U}$, \[\lambda^{*}(\theta, \mu)||\hat{\nabla}q^{\epsilon}(\theta, \mu)||^2 \leq \eta ||\hat{\nabla}q^{\epsilon}(\theta, \mu)||^2 + M_{n, \epsilon}||\hat{\nabla}q^{\epsilon}(\theta, \mu)||\]
\end{lemma}
\textbf{\textit{proof}.} 
    \begin{align*}
        \lambda^{*}(\theta, \mu)||\hat{\nabla}q^{\epsilon}(\theta, \mu)||^2 =&\bigg[\eta - \frac{\langle \hat{\nabla}q^{\epsilon}(\theta, \mu),\nabla F_n(\theta, \mu)\rangle }{||\hat{\nabla}q^{\epsilon}(\theta, \mu)||^2}\bigg]||\hat{\nabla}q^{\epsilon}(\theta, \mu)||^2\\
        \leq&|\eta - \frac{\langle \hat{\nabla}q^{\epsilon}(\theta, \mu),\nabla F_n(\theta, \mu)\rangle }{||\hat{\nabla}q^{\epsilon}(\theta, \mu)||^2}| ||\hat{\nabla}q^{\epsilon}(\theta, \mu)||^2\\
        \leq& \eta ||\hat{\nabla}q^{\epsilon}(\theta, \mu)||^2 + |\frac{\langle \hat{\nabla}q^{\epsilon}(\theta, \mu),\nabla F_n(\theta, \mu)\rangle }{||\hat{\nabla}q^{\epsilon}(\theta, \mu)||^2}|||\hat{\nabla}q^{\epsilon}(\theta, \mu)||^2\\
        =&\eta ||\hat{\nabla}q^{\epsilon}(\theta, \mu)||^2 +|\langle \hat{\nabla}q^{\epsilon}(\theta, \mu),\nabla F_n(\theta, \mu)\rangle|\\
        \leq& \eta ||\hat{\nabla}q^{\epsilon}(\theta, \mu)||^2 + ||\hat{\nabla}q^{\epsilon}(\theta, \mu)||||\nabla F_n(\theta, \mu)||\tag{Cauchy-schwarz}\\
        \leq&\eta ||\hat{\nabla}q^{\epsilon}(\theta, \mu)||^2 + M_{n, \epsilon}||\hat{\nabla}q^{\epsilon}(\theta, \mu)||\hspace{5mm}
    \end{align*}\hfill \qedsymbol

\begin{lemma}\label{lem10}
Under the \hyperref[lem1]{quadratic growth} and \hyperref[smooth]{smoothness}, 
\[||\nabla q^{\epsilon}(\theta,\mu)|| \leq \frac{2L_{n, \epsilon, q^{\epsilon}}}{\sqrt{\kappa}}\sqrt{q^{\epsilon}(\theta,\mu)}\]
\end{lemma}
\textbf{\textit{proof}.} Notice that for any $\theta \in \Theta$, 
\[\nabla q (\theta, \mu^{*}(\theta)) = \nabla f_n^{\epsilon} (\theta, \mu^{*}(\theta)) - \nabla f_n^{\epsilon} (\theta, \mu^{*}(\theta)) = 0.\]
Then by Lemma \ref{lem7}, 
\begin{align*}
    ||\nabla q (\theta, \mu)||=& ||\nabla q (\theta, \mu) - \nabla q (\theta, \mu^{*}(\theta))|| \\
    \leq& L_{n, \epsilon, q^{\epsilon}}||(\theta, \mu) - (\theta, \mu^{*}(\theta))|| \\
    =& L_{n, \epsilon, q^{\epsilon}}||\mu -  \mu^{*}(\theta)||\\
    \leq& L_{n, \epsilon, q^{\epsilon}} \frac{2}{\sqrt{\kappa}}\sqrt{f_n^{\epsilon}(\theta, \mu) - f_n^{\epsilon}(\theta, \mu^{*}(\theta))} \tag{Quadratic Growth}\\
    =& \frac{2L_{n, \epsilon, q^{\epsilon}}}{\sqrt{\kappa}}\sqrt{q^{\epsilon}(\theta,\mu)} \hspace{5mm}
\end{align*}\hfill \qedsymbol
\clearpage
\begin{lemma}\label{lem11} (Approximation Error Control)\hspace{1mm}\\
For any $(\theta, \mu) \in \Theta \times \mathcal{U}$ and inner iteration $K$, there exists a positive constant $C_0$, which depends on $L_{n, \epsilon, q^{\epsilon}}, \rho, \alpha, M_{n,\epsilon}$ and $\eta$, such that 
\[||\lambda(\theta, \mu)(\nabla q^{\epsilon}(\theta, \mu) - \hat{\nabla} q^{\epsilon}(\theta,\mu))||\leq C_0 \exp(-a_1 K/2)\]
\end{lemma}
\textbf{\textit{proof}.} We want to show
\begin{align*}
    ||\lambda(\theta, \mu)(\nabla q^{\epsilon}(\theta, \mu) - \hat{\nabla} q^{\epsilon}(\theta,\mu))||  = \underbrace{|\lambda(\theta, \mu)|}_{(i)} \underbrace{|| \nabla q^{\epsilon}(\theta, \mu) - \hat{\nabla} q^{\epsilon}(\theta,\mu)||}_{(ii)}
\end{align*}
\begin{enumerate}[label = (\roman*)]
    \item From  Lemma \ref{lem8}
    \[|\lambda(\theta, \mu)|
    \leq \eta + \frac{||\nabla F_n(\theta, \mu)||}{||\hat{\nabla}q^{\epsilon}(\theta, \mu)||}.\]
    \item From  Lemma \ref{lem7}, 
    \begin{align*}
        || \nabla q^{\epsilon}(\theta, \mu) - \hat{\nabla} q^{\epsilon}(\theta,\mu)|| \leq& 
        L_{n, \epsilon, q^{\epsilon}}||\mu^{(K)} - \mu^{*}(\theta)||\\
        \leq& \frac{2L_{n, \epsilon, q^{\epsilon}}}{\sqrt{\kappa}}\sqrt{q^{\epsilon}(\theta, \mu^{(K)})}\tag{Lemma \ref{lem10}}\\
        \leq& \frac{2L_{n, \epsilon, q^{\epsilon}}}{\sqrt{\kappa}} \exp(-a_1 K/2)\sqrt{q^{\epsilon}(\theta, \mu^{(0)})}\tag{Lemma \ref{lem3}}
    \end{align*}
    Hence we need to bound $\sqrt{q^{\epsilon}(\theta, \mu^{(0)})}$. In fact,
    \begin{align*}
        q^{\epsilon}(\theta, \mu^{(0)}) = f_n^{\epsilon}(\theta, \mu^{(0)}) - (\theta, \mu^{*}(\theta)) \leq \frac{1}{\kappa_{\alpha, \rho}}||G_{\alpha}^{\epsilon}(\mu^{(0)};\theta)||^2
    \end{align*}
    Using the non-expansivity of projection $\Pi_{\mathcal{U}},$ for any $z in \mathcal{U}$ and $y$,
    \begin{align*}
        \langle y - \Pi_{\mathcal{U}}(y), z - \Pi_{\mathcal{U}}(y)\rangle \leq 0
    \end{align*}
    Let $y = \mu^{(0)} - \alpha \nabla_{\mu} q^{\epsilon}(\theta, \mu^{(0)})$, and $z = \mu^{(0)}$ where $\mu^{(0)}\in \mathcal{U}$. Then, 
    \begin{align*}
        \langle  \mu^{(0)} - \alpha \nabla_{\mu} q^{\epsilon}(\theta, \mu^{(0)}) - \Pi_{\mathcal{U}}( \mu^{(0)} - \alpha \nabla_{\mu} q^{\epsilon}(\theta, \mu^{(0)})), \mu^{(0)} - \Pi_{\mathcal{U}}( \mu^{(0)} - \alpha \nabla_{\mu} q^{\epsilon}(\theta, \mu^{(0)}))\rangle \leq 0
    \end{align*}
    Rearranging the terms and using the definition of \hyperref[pgd]{projected gradient mapping}, we have 
    \begin{align*}
        \langle \alpha G_{\alpha}^{\epsilon}(\mu^{(0)};\theta), \alpha G_{\alpha}^{\epsilon}(\mu^{(0)};\theta) \rangle \leq \langle \alpha \nabla_{\mu} q^{\epsilon}(\theta, \mu^{(0)}, \alpha G_{\alpha}^{\epsilon}(\mu^{(0)};\theta)\rangle 
    \end{align*}
    \clearpage
    Then, 
    \begin{align*}
        \langle \alpha G_{\alpha}^{\epsilon}(\mu^{(0)};\theta), \alpha G_{\alpha}^{\epsilon}(\mu^{(0)};\theta) \rangle =&  \alpha^2 ||G_{\alpha}^{\epsilon}(\mu^{(0)};\theta)||^2 \\
        \leq& \alpha^2 \langle \nabla_{\mu} q^{\epsilon}(\theta, \mu^{(0)}), G_{\alpha}^{\epsilon}(\mu^{(0)};\theta)\rangle 
        \\
        \leq & \alpha^2|\langle \nabla_{\mu} q^{\epsilon}(\theta, \mu^{(0)}), G_{\alpha}^{\epsilon}(\mu^{(0)};\theta)\rangle|\\
        \leq& \alpha^2||\nabla_{\mu} q^{\epsilon}(\theta, \mu^{(0)}|| || G_{\alpha}^{\epsilon}(\mu^{(0)};\theta)||
    \end{align*}
    Diving both sides by $\alpha^2 ||G_{\alpha}^{\epsilon}(\mu^{(0)};\theta)||$ (assuming $||G_{\alpha}^{\epsilon}(\mu^{(0)};\theta)||>0$), we have
    \[||G_{\alpha}^{\epsilon}(\mu^{(0)};\theta)|| \leq ||\nabla_{\mu} q^{\epsilon}(\theta, \mu^{(0)}||.\]
    
    Note that $||\nabla_{\mu} q^{\epsilon}(\theta, \mu^{(0)}||\leq ||\nabla q^{\epsilon}(\theta, \mu^{(0)}||$. Therefore, 
    \begin{align*}
        q^{\epsilon}(\theta, \mu^{(0)}) = f_n^{\epsilon}(\theta, \mu^{(0)}) - f_n^{\epsilon}(\theta, \mu^{*}(\theta)) \leq \frac{1}{\kappa_{\alpha, \rho}}||G_{\alpha}^{\epsilon}(\mu^{(0)};\theta)||^2 \leq \frac{1}{\kappa_{\alpha, \rho}}||\nabla_{\mu} q^{\epsilon}(\theta, \mu^{(0)})||^2 \leq\frac{1}{\kappa_{\alpha, \rho}}||\nabla  q^{\epsilon}(\theta, \mu^{(0)})||^2 
    \end{align*}
    Taking square root on both sides, 
    \[\sqrt{q^{\epsilon}(\theta, \mu^{(0)}) }\leq \frac{1}{\sqrt{\kappa_{\alpha, \rho}}}||\nabla  q^{\epsilon}(\theta, \mu^{(0)})||,\]
    leading to 
    \begin{align*}
        || \nabla q^{\epsilon}(\theta, \mu) - \hat{\nabla} q^{\epsilon}(\theta,\mu)|| \leq& \frac{2L_{n, \epsilon, q^{\epsilon}}}{\sqrt{\kappa}} \exp(-a_1 K/2)\sqrt{q^{\epsilon}(\theta, \mu^{(0)})}\\
        \leq& \frac{2L_{n, \epsilon, q^{\epsilon}}}{\sqrt{\kappa}} \exp(-a_1 K/2)\frac{1}{\sqrt{\kappa_{\alpha, \rho}}}||\nabla  q^{\epsilon}(\theta, \mu^{(0)})|| \\
        =&\frac{L_{n, \epsilon, q^{\epsilon}}\sqrt{\alpha}}{\rho} \exp(-a_1 K/2)||\nabla  q^{\epsilon}(\theta, \mu^{(0)})||. 
    \end{align*}
    Moreover, let $c = \frac{L_{n, \epsilon, q^{\epsilon}}\sqrt{\alpha}}{\rho}.$ Then for $K \geq \frac{2}{a_1}\log(2c),$ we have $(1-\frac{L_{n, \epsilon, q^{\epsilon}}\sqrt{\alpha}}{\rho} \exp(-a_1 K/2))\geq 1/2.$\\ 
    \begin{align*}
        ||\hat{\nabla}q^{\epsilon}(\theta, \mu)|| =& ||\hat{\nabla}q^{\epsilon}(\theta, \mu) - \nabla q^{\epsilon}(\theta, \mu) + \nabla q^{\epsilon}(\theta, \mu) ||\\
        \geq & ||\nabla q^{\epsilon}(\theta, \mu)|| - ||\nabla q^{\epsilon}(\theta, \mu) - \hat{\nabla}q^{\epsilon}(\theta, \mu)||\\
        \geq& (1- \frac{L_{n, \epsilon, q^{\epsilon}}\sqrt{\alpha}}{\rho} \exp(-a_1 K/2))||\nabla  q^{\epsilon}(\theta, \mu^{(0)})||\\
        \geq& \frac{1}{2}||\nabla  q^{\epsilon}(\theta, \mu^{(0)})||
    \end{align*}
  
\end{enumerate}
Putting altogether, 
\begin{align*}
    &||\lambda(\theta, \mu)(\nabla q^{\epsilon}(\theta, \mu) - \hat{\nabla} q^{\epsilon}(\theta,\mu))||\\   =& |\lambda(\theta, \mu)||| \nabla q^{\epsilon}(\theta, \mu) - \hat{\nabla} q^{\epsilon}(\theta,\mu)||\\
    \leq& (\eta + \frac{||\nabla F_n(\theta, \mu)||}{||\hat{\nabla}q^{\epsilon}(\theta, \mu)||})|| \nabla q^{\epsilon}(\theta, \mu) - \hat{\nabla} q^{\epsilon}(\theta,\mu)||\\
    =&\eta || \nabla q^{\epsilon}(\theta, \mu) - \hat{\nabla} q^{\epsilon}(\theta,\mu)|| +\frac{||\nabla F_n(\theta, \mu)||}{||\hat{\nabla}q^{\epsilon}(\theta, \mu)||} || \nabla q^{\epsilon}(\theta, \mu) - \hat{\nabla} q^{\epsilon}(\theta,\mu)||\\
    \leq&\eta \frac{L_{n, \epsilon, q^{\epsilon}}\sqrt{\alpha}}{\rho} \exp(-a_1 K/2)||\nabla  q^{\epsilon}(\theta, \mu^{(0)})|| + \frac{||\nabla F_n(\theta, \mu)||}{||\hat{\nabla}q^{\epsilon}(\theta, \mu)||} \frac{L_{n, \epsilon, q^{\epsilon}}\sqrt{\alpha}}{\rho} \exp(-a_1 K/2)||\nabla  q^{\epsilon}(\theta, \mu^{(0)}||\\
    \leq& \frac{L_{n, \epsilon, q^{\epsilon}}\sqrt{\alpha}}{\rho} \exp(-a_1 K/2)\bigg[\eta ||\nabla  q^{\epsilon}(\theta, \mu^{(0)})||+ \frac{M_{n,\epsilon}}{||\hat{\nabla}q^{\epsilon}(\theta, \mu)||} ||\nabla  q^{\epsilon}(\theta, \mu^{(0)}||\bigg] \\
    \leq& \frac{L_{n, \epsilon, q^{\epsilon}}\sqrt{\alpha}}{\rho} \exp(-a_1 K/2)\bigg[\eta ||\nabla  q^{\epsilon}(\theta, \mu^{(0)})||+ M_{n,\epsilon} (1-\frac{L_{n, \epsilon, q^{\epsilon}}\sqrt{\alpha}}{\rho} \exp(-a_1 K/2))^{-1}||\nabla q^{\epsilon}(\theta, \mu)||^{-1} ||\nabla  q^{\epsilon}(\theta, \mu^{(0)}||\bigg] \\
    =&\frac{L_{n, \epsilon, q^{\epsilon}}\sqrt{\alpha}}{\rho} \exp(-a_1 K/2)\bigg[\eta ||\nabla  q^{\epsilon}(\theta, \mu^{(0)})||+ 2M_{n,\epsilon} \bigg] \\
    \leq& \frac{L_{n, \epsilon, q^{\epsilon}}\sqrt{\alpha}}{\rho} \exp(-a_1 K/2)\bigg[2\eta(\eta +1) M_{n,\epsilon}  +  2M_{n,\epsilon}\bigg] \tag{Lemma \ref{lem8}}\\
    \leq& \frac{L_{n, \epsilon, q^{\epsilon}}\sqrt{\alpha}}{\rho} \exp(-a_1 K/2)[2\eta(\eta +1)   +  2]M_{n,\epsilon}\\
    =& C_0 \exp(-a_1 K/2)
\end{align*}
where $C_0 = (2\eta(\eta +1)   +  2)\frac{L_{n, \epsilon, q^{\epsilon}M_{n,\epsilon}}\sqrt{\alpha}}{\rho}.$ \hfill \qedsymbol

\begin{lemma}\label{lem12}
    For any $(\theta, \mu) \in \Theta \times \mathcal{U}$ and inner iteration $K$, 
    \[| \lambda(\theta, \mu) - \lambda^{*}(\theta, \mu) | \|\nabla q^{\epsilon}(\theta, \mu) \| \leq 5\|\nabla F_{n}(\theta) \|\]
\end{lemma}
\textbf{\textit{proof}.} We want to show 
\[| \lambda(\theta, \mu) - \lambda^{*}(\theta, \mu) | \|\nabla q^{\epsilon}(\theta, \mu) \| \leq 5\|\nabla F_{n}(\theta) \|, \] where
\begin{align*}\lambda^{*}(\theta, \mu) =& \begin{cases}
    \max \bigg\{0, \eta - \frac{\langle \nabla F_n(\theta), \nabla q^{\epsilon}(\theta, \mu)\rangle}{||\nabla q^{\epsilon}(\theta, \mu)||^2}\bigg\}, & \text{for } ||\nabla q^{\epsilon}(\theta, \mu)|| >0 \\
    0 & \text{for } ||\nabla q^{\epsilon}(\theta, \mu)|| = 0.
  \end{cases}\vspace{3mm} \\
     \lambda(\theta, \mu) =& \begin{cases}
    \max \bigg\{0, \eta - \frac{\langle \nabla F_n(\theta), \hat{\nabla} q^{\epsilon}(\theta, \mu)\rangle}{||\hat{\nabla}q^{\epsilon}(\theta, \mu)||^2}\bigg\}, & \text{for } ||\hat{\nabla} q^{\epsilon}(\theta, \mu)|| >0 \\
    0 & \text{for } ||\hat{\nabla} q^{\epsilon}(\theta, \mu)|| = 0.
  \end{cases} \vspace{3mm}\\
   \hat{\nabla}q^{\epsilon}(\theta, \mu) =& \nabla f_n^{\epsilon}(\theta, \mu) -\big[ \nabla_{\theta}^{\top} f_n^{\epsilon}(\theta, \mu^{(K)}) , \textbf{0}^{\top}\big]^{\top}\\
    \nabla q^{\epsilon}(\theta, \mu) =& \nabla f_n^{\epsilon}(\theta, \mu) -\big[ \nabla_{\theta}^{\top} f_n^{\epsilon}(\theta, \mu^{*}(\theta)) , \textbf{0}^{\top}\big]^{\top}.
\end{align*}

For simplicity, let 
\[\delta(\theta, \mu) = \nabla F_{n}(\theta) + \lambda(\theta, \mu)\hat{\nabla} q^{\epsilon}(\theta, \mu), \qquad  g(\theta, \mu) = \nabla F_{n}(\theta) + \lambda^{*}(\theta, \mu)\nabla q^{\epsilon}(\theta, \mu),\qquad \overrightarrow{\bigtriangleup}(\theta, \mu) = g(\theta, \mu) - \delta(\theta, \mu). \]
Then, 
\begin{align*}
    \|\overrightarrow{\bigtriangleup}(\theta, \mu)\| =& \|g(\theta, \mu) - \delta(\theta, \mu)\| \\
    =& \|\nabla F_{n}(\theta) + \lambda^{*}(\theta, \mu)\nabla q^{\epsilon}(\theta, \mu) - (\nabla F_{n}(\theta) + \lambda(\theta, \mu)\hat{\nabla} q^{\epsilon}(\theta, \mu)) )\| \\
    =& \|\lambda^{*}(\theta, \mu)\nabla q^{\epsilon}(\theta, \mu) - \lambda(\theta, \mu)\hat{\nabla} q^{\epsilon}(\theta, \mu)\|\\
    =& \|\lambda^{*}(\theta, \mu)\nabla q^{\epsilon}(\theta, \mu) - \lambda(\theta, \mu)\nabla q^{\epsilon}(\theta, \mu) + \lambda(\theta, \mu)\nabla q^{\epsilon}(\theta, \mu) - \lambda(\theta, \mu)\hat{\nabla} q^{\epsilon}(\theta, \mu)\|\\
    =& \|(\lambda^{*}(\theta, \mu) - \lambda(\theta, \mu)) \nabla q^{\epsilon}(\theta, \mu) + \lambda(\theta, \mu) (\nabla q^{\epsilon}(\theta, \mu) - \hat{\nabla}q^{\epsilon}(\theta, \mu)) \|\\
    \leq& \underbrace{\|(\lambda^{*}(\theta, \mu) - \lambda(\theta, \mu)) \nabla q^{\epsilon}(\theta, \mu)\|}_{(i)} + \underbrace{\|\lambda(\theta, \mu) (\nabla q^{\epsilon}(\theta, \mu) - \hat{\nabla}q^{\epsilon}(\theta, \mu)) \|}_{(ii)}.
\end{align*}
Since we proved $(ii)$ in  Lemma \ref{lem11}, we only need to show $(i). $ 
\begin{align*}
    \|(\lambda^{*}(\theta, \mu) - \lambda(\theta, \mu)) \nabla q^{\epsilon}(\theta, \mu)\| =\underbrace{| \lambda(\theta, \mu) - \lambda^{*}(\theta, \mu)|}_{(*)} \| \nabla q^{\epsilon}(\theta, \mu)\|. 
\end{align*}       
\begin{enumerate}[listparindent=0pt, label = (\alph*)]
    \item When \[\lambda(\theta, \mu) =\eta - \frac{\langle \nabla F_n(\theta), \hat{\nabla}q^{\epsilon}(\theta,\mu) \rangle}{\|\hat{\nabla}q^{\epsilon}(\theta,\mu)\|^2} ,\qquad  \lambda^{*}(\theta, \mu) =  \eta - \frac{\langle \nabla F_n(\theta), \nabla q^{\epsilon}(\theta,\mu) \rangle}{\|\nabla q^{\epsilon}(\theta,\mu)\|^2},\]
    \begin{align*}
    &| \lambda(\theta, \mu) - \lambda^{*}(\theta, \mu)|\\
    =&|\eta - \frac{\langle \nabla F_n(\theta), \hat{\nabla}q^{\epsilon}(\theta,\mu) \rangle}{\|\hat{\nabla}q^{\epsilon}(\theta,\mu)\|^2} - (\eta - \frac{\langle \nabla F_n(\theta), \nabla q^{\epsilon}(\theta,\mu) \rangle}{\|\nabla q^{\epsilon}(\theta,\mu)\|^2})|\\
    =& |\frac{\langle \nabla F_n(\theta), \nabla q^{\epsilon}(\theta,\mu) \rangle}{\|\nabla q^{\epsilon}(\theta,\mu)\|^2} - \frac{\langle \nabla F_n(\theta), \hat{\nabla}q^{\epsilon}(\theta,\mu) \rangle}{\|\hat{\nabla}q^{\epsilon}(\theta,\mu)\|^2}|\\
    =& |\frac{\langle \nabla F_n(\theta), \hat{\nabla}q^{\epsilon}(\theta,\mu) \rangle}{\|\hat{\nabla}q^{\epsilon}(\theta,\mu)\|^2} - \frac{\langle \nabla F_n(\theta), \nabla q^{\epsilon}(\theta,\mu) \rangle}{\|\nabla q^{\epsilon}(\theta,\mu)\|^2}|\\
    =& |\frac{\langle \nabla F_n(\theta), \hat{\nabla}q^{\epsilon}(\theta,\mu) \rangle}{\|\hat{\nabla}q^{\epsilon}(\theta,\mu)\|^2}  - \frac{\langle \nabla F_n(\theta), \nabla q^{\epsilon}(\theta,\mu) \rangle}{\|\hat{\nabla} q^{\epsilon}(\theta,\mu)\|^2} + \frac{\langle \nabla F_n(\theta), \nabla q^{\epsilon}(\theta,\mu) \rangle}{\|\hat{\nabla} q^{\epsilon}(\theta,\mu)\|^2} - \frac{\langle \nabla F_n(\theta), \nabla q^{\epsilon}(\theta,\mu) \rangle}{\|\nabla q^{\epsilon}(\theta,\mu)\|^2}|\\
    \leq& \underbrace{|\frac{\langle \nabla F_n(\theta), \hat{\nabla}q^{\epsilon}(\theta,\mu) \rangle}{\|\hat{\nabla}q^{\epsilon}(\theta,\mu)\|^2}  - \frac{\langle \nabla F_n(\theta), \nabla q^{\epsilon}(\theta,\mu) \rangle}{\|\hat{\nabla} q^{\epsilon}(\theta,\mu)\|^2}|}_{(1)} + \underbrace{|\frac{\langle \nabla F_n(\theta), \nabla q^{\epsilon}(\theta,\mu) \rangle}{\|\hat{\nabla} q^{\epsilon}(\theta,\mu)\|^2} - \frac{\langle \nabla F_n(\theta), \nabla q^{\epsilon}(\theta,\mu) \rangle}{\|\nabla q^{\epsilon}(\theta,\mu)\|^2}|}_{(2)}
\end{align*}
\begin{enumerate}[listparindent=0pt, label = (\arabic*)]
    \item \begin{align*}
        |\frac{\langle \nabla F_n(\theta), \hat{\nabla}q^{\epsilon}(\theta,\mu) \rangle}{\|\hat{\nabla}q^{\epsilon}(\theta,\mu)\|^2}  - \frac{\langle \nabla F_n(\theta), \nabla q^{\epsilon}(\theta,\mu) \rangle}{\|\hat{\nabla} q^{\epsilon}(\theta,\mu)\|^2}|=& \frac{1}{\|\hat{\nabla}q^{\epsilon}(\theta,\mu)\|^2}|\langle \nabla F_n(\theta), \nabla q^{\epsilon}(\theta,\mu) - \hat{\nabla} q^{\epsilon}(\theta,\mu) \rangle| \\
        \leq& \frac{1}{\|\hat{\nabla} q^{\epsilon}(\theta,\mu)\|^2} \|\nabla F_n(\theta)\| \|\nabla q^{\epsilon}(\theta,\mu) - \hat{\nabla} q^{\epsilon}(\theta,\mu)\|.
    \end{align*}
    As a byproduct in Lemma \ref{lem11}, when $K \geq \frac{2}{a_1}\log(2c),$ where $c = \frac{L_{n, \epsilon, q^{\epsilon}}\sqrt{\alpha}}{\rho} $, 
    \[|| \nabla q^{\epsilon}(\theta, \mu) - \hat{\nabla} q^{\epsilon}(\theta,\mu)|| \leq \frac{L_{n, \epsilon, q^{\epsilon}}\sqrt{\alpha}}{\rho} \exp(-a_1 K/2)||\nabla  q^{\epsilon}(\theta, \mu)||, \qquad \|\hat{\nabla} q^{\epsilon}(\theta,\mu)\| \geq \frac{1}{2} \|\nabla q^{\epsilon}(\theta, \mu) \|.\]
    
    Thus, 
    \begin{align*}
        &|\frac{\langle \nabla F_n(\theta), \hat{\nabla}q^{\epsilon}(\theta,\mu) \rangle}{\|\hat{\nabla}q^{\epsilon}(\theta,\mu)\|^2}  - \frac{\langle \nabla F_n(\theta), \nabla q^{\epsilon}(\theta,\mu) \rangle}{\|\hat{\nabla} q^{\epsilon}(\theta,\mu)\|^2}|\\
        \leq&\frac{1}{\|\hat{\nabla} q^{\epsilon}(\theta,\mu)\|^2} \|\nabla F_n(\theta)\| \|\nabla q^{\epsilon}(\theta,\mu) - \hat{\nabla} q^{\epsilon}(\theta,\mu)\|\\
        \leq& \frac{1}{\|\hat{\nabla} q^{\epsilon}(\theta,\mu)\|^2} \|\nabla F_n(\theta)\|\frac{L_{n, \epsilon, q^{\epsilon}}\sqrt{\alpha}}{\rho} \exp(-a_1 K/2)||\nabla  q^{\epsilon}(\theta, \mu)||\\        \leq& \frac{2}{\|\nabla q^{\epsilon}(\theta,\mu)\|^2}\|\nabla F_n(\theta)\|\|\nabla q^{\epsilon}(\theta,\mu)\|\\
        \leq& \frac{2}{\|\nabla q^{\epsilon}(\theta,\mu)\|}\|\nabla F_n(\theta)\|.
    \end{align*}
    \item 
    \begin{align*}
        &|\frac{\langle \nabla F_n(\theta), \nabla q^{\epsilon}(\theta,\mu) \rangle}{\|\hat{\nabla} q^{\epsilon}(\theta,\mu)\|^2} - \frac{\langle \nabla F_n(\theta), \nabla q^{\epsilon}(\theta,\mu) \rangle}{\|\nabla q^{\epsilon}(\theta,\mu)\|^2}|\\
        \leq& |(\frac{1}{\|\hat{\nabla} q^{\epsilon}(\theta,\mu)\|^2} - \frac{1}{\|\nabla q^{\epsilon}(\theta,\mu)\|^2} ) \langle\nabla F_n(\theta), \nabla q^{\epsilon}(\theta,\mu) \rangle|\\
        =&|(\frac{1}{\|\hat{\nabla} q^{\epsilon}(\theta,\mu)\|^2} - \frac{1}{\|\nabla q^{\epsilon}(\theta,\mu)\|^2} )| |\langle\nabla F_n(\theta), \nabla q^{\epsilon}(\theta,\mu) \rangle|\\
        \leq& |(\frac{1}{\|\hat{\nabla} q^{\epsilon}(\theta,\mu)\|^2} - \frac{1}{\|\nabla q^{\epsilon}(\theta,\mu)\|^2} )| \|\nabla F_n(\theta)\| \|\nabla q^{\epsilon}(\theta,\mu)\|\\
        \leq&|(\frac{4}{\|\nabla q^{\epsilon}(\theta,\mu)\|^2} - \frac{1}{\|\nabla q^{\epsilon}(\theta,\mu)\|^2} )| \|\nabla F_n(\theta)\| \|\nabla q^{\epsilon}(\theta,\mu)\|\\
        \leq&\frac{3}{\|\nabla q^{\epsilon}(\theta,\mu)\|}  \|\nabla F_n(\theta)\|
    \end{align*}
    
\end{enumerate}
    Hence, 
\begin{align*}
    \|(\lambda^{*}(\theta, \mu) - \lambda(\theta, \mu)) \nabla q^{\epsilon}(\theta, \mu)\| = &\underbrace{| \lambda(\theta, \mu) - \lambda^{*}(\theta, \mu)|}_{(*)} \| \nabla q^{\epsilon}(\theta, \mu)\|\\
    \leq& \big[\frac{2}{\|\nabla q^{\epsilon}(\theta,\mu)\|}\|\nabla F_n(\theta)\| + \frac{3}{\|\nabla q^{\epsilon}(\theta,\mu)\|}  \|\nabla F_n(\theta)\|\big] \| \nabla q^{\epsilon}(\theta, \mu)\|\\
    =& 5 \|\nabla F_n(\theta)\|
\end{align*}

\end{enumerate}\hfill \qedsymbol

\begin{lemma}\label{lem13} 
Under the \hyperref[ppl]{projected PL inequality}, \hyperref[bound]{boundedness} and \hyperref[smooth]{smoothness}, let $q_t^{\epsilon}= q^{\epsilon}(\theta_t, \mu_t)$. Then when $\|\hat{\nabla}q^{\epsilon}_t\| > 0$, we have 
\begin{align*}
    q_{t+1}^{\epsilon}- q_t^{\epsilon} \leq& - \eta \xi_t \|\nabla q_t^{\epsilon}\|^2 + \eta \xi_t L_{n,\epsilon}\|\mu_t^{(K)} - \mu^{*}(\theta_t)\|\big[L_{n,\epsilon}\|\mu_t^{(K)} - \mu^{*}(\theta_t)\| + 2L_{n,\epsilon, q^{\epsilon}}\|\mu_t - \mu^{*}(\theta_t)\|\big]\\ +& 2(\eta+1)\xi_t L_{n,\epsilon}\|\mu_t^{(K)} - \mu^{*}(\theta_t)\| M_{n,\epsilon}+ (\eta+1)L_{n,\epsilon, q^{\epsilon}}\xi_t^2 M_{n, \epsilon}
\end{align*}
When $\|\hat{\nabla}q^{\epsilon}_t\| = 0$, 
\[ q_{t+1}^{\epsilon} - q_t^{\epsilon}
    \leq (\eta+1)L_{n,\epsilon, q^{\epsilon}} \xi_t^2 M_{n,\epsilon}\]
\end{lemma}
\textbf{\textit{proof}.} We know that $q^{\epsilon}$ is $L_{n,\epsilon, q^{\epsilon}}-$smooth. Hence, 
\[q_{t+1}^{\epsilon} = q^{\epsilon}(\theta_{t+1}, \mu_{t+1})\leq q_t^{\epsilon} + \langle \nabla q_t^{\epsilon},(\theta_{t+1}, \mu_{t+1})- (\theta_t, \mu_t)\rangle + \frac{L_{n,\epsilon, q^{\epsilon}}}{2}\|(\theta_{t+1}, \mu_{t+1})- (\theta_t, \mu_t)\|^2.\]
Then, 
\begin{align*}
    q_{t+1}^{\epsilon} - q_t^{\epsilon} \leq&\langle \nabla q_t^{\epsilon},(\theta_{t+1}, \mu_{t+1})- (\theta_t, \mu_t)\rangle + \frac{L_{n,\epsilon, q^{\epsilon}}}{2}\|(\theta_{t+1}, \mu_{t+1})- (\theta_t, \mu_t)\|^2 \\
    \leq& -\xi_t\langle \nabla q_t^{\epsilon},\delta_t \rangle + \frac{L_{n,\epsilon, q^{\epsilon}}}{2}\xi_t^2\|\delta_t\|^2
\end{align*}
where $(\theta_{t+1}, \mu_{t+1})- (\theta_t, \mu_t) = -\xi_t \delta_t$ and $\delta_t = \nabla F_{n,t} + \lambda_t \hat{\nabla}q^{\epsilon}_t$. 
\begin{align*}
    q_{t+1}^{\epsilon} - q_t^{\epsilon} \leq&-\xi_t\langle \nabla q_t^{\epsilon} + \hat{\nabla}q^{\epsilon}_t - \hat{\nabla}q^{\epsilon}_t,\delta_t \rangle + \frac{L_{n,\epsilon, q^{\epsilon}}}{2}\xi_t^2\|\delta_t\|^2\\
    \leq&-\xi_t\langle  \hat{\nabla}q^{\epsilon}_t ,\delta_t \rangle -\xi_t\langle \nabla q_t^{\epsilon} - \hat{\nabla}q^{\epsilon}_t,\delta_t \rangle+ \frac{L_{n,\epsilon, q^{\epsilon}}}{2}\xi_t^2\|\delta_t\|^2
\end{align*}
Note that $\langle  \hat{\nabla}q^{\epsilon}_t ,\delta_t \rangle \geq \eta \|\hat{\nabla}q^{\epsilon}_t\|^2$ by the constraint of the problem to find \hyperref[direction]{update direction}. Moreover, by the Cauchy-Schwarz inequality on $\langle \nabla q_t^{\epsilon} - \hat{\nabla}q^{\epsilon}_t,\delta_t \rangle \geq - \|\nabla q_t^{\epsilon} - \hat{\nabla}q^{\epsilon}_t\| \|\delta_t\|$, we have, 
\begin{align*}
    q_{t+1}^{\epsilon} - q_t^{\epsilon} \leq& -\eta \xi_t \|\hat{\nabla}q^{\epsilon}_t\|^2 + \xi_t \|\nabla q_t^{\epsilon} - \hat{\nabla}q^{\epsilon}_t\| \|\delta_t\| + \frac{L_{n,\epsilon, q^{\epsilon}}}{2}\xi_t^2\|\delta_t\|^2\\
    \leq& -\eta \xi_t \|\hat{\nabla}q^{\epsilon}_t\|^2 + \xi_t \|\nabla q_t^{\epsilon} - \hat{\nabla}q^{\epsilon}_t\| \|\delta_t\| + 
    L_{n,\epsilon, q^{\epsilon}} \xi_t^2 (\eta+1) M_{n,\epsilon} \tag{Lemma \ref{lem8}}\\
    \leq& -\eta \xi_t \|\hat{\nabla}q^{\epsilon}_t\|^2 + \xi_t \|\nabla q_t^{\epsilon} - \hat{\nabla}q^{\epsilon}_t\| 2(\eta+1)M_{n,\epsilon} + 
    L_{n,\epsilon, q^{\epsilon}} \xi_t^2 (\eta+1) M_{n,\epsilon} \tag{Lemma \ref{lem8}}\\
    \leq& -\eta \xi_t \|\hat{\nabla}q^{\epsilon}_t\|^2 + 2(\eta+1)\xi_t L_{n,\epsilon}\|\mu_t^{(K)}-\mu^{*}(\theta_t)\| M_{n,\epsilon} + 
    L_{n,\epsilon, q^{\epsilon}} \xi_t^2 (\eta+1) M_{n,\epsilon} \tag{Lemma \ref{lem4}}
\end{align*}

Note that
\begin{align*}
    |\|\hat{\nabla}q^{\epsilon}_t\|^2 - \|\nabla q^{\epsilon}_t\|^2| \leq& \|\nabla q^{\epsilon}_t - \hat{\nabla}q^{\epsilon}_t\|\|\nabla q^{\epsilon}_t+ \hat{\nabla}q^{\epsilon}_t\|\\
    \leq& L_{n,\epsilon}\|\mu_{\theta_t}^{(K)} - \mu^{*}(\theta_t)\|\|\nabla q^{\epsilon}_t+ \hat{\nabla}q^{\epsilon}_t\|\tag{Lemma \ref{lem4}}\\
    =& L_{n,\epsilon}\|\mu_{\theta_t}^{(K)} - \mu^{*}(\theta_t)\|\big[\|\nabla q^{\epsilon}_t+\nabla q^{\epsilon}_t - \nabla q^{\epsilon}_t + \hat{\nabla}q^{\epsilon}_t\|\big]\\
    \leq& L_{n,\epsilon}\|\mu_{\theta_t}^{(K)} - \mu^{*}(\theta_t)\|\big[\| \hat{\nabla}q^{\epsilon}_t - \nabla q^{\epsilon}_t\| + 2\|\nabla q^{\epsilon}_t\|\big]\\
    \leq& L_{n,\epsilon}\|\mu_{\theta_t}^{(K)} - \mu^{*}(\theta_t)\|\big[L_{n,\epsilon}\|\mu_{\theta_t}^{(K)} - \mu^{*}(\theta_t)\| + 2\|\nabla q^{\epsilon}_t\|\big]\tag{Lemma \ref{lem4}}\\
    \leq& L_{n,\epsilon}\|\mu_{\theta_t}^{(K)} - \mu^{*}(\theta_t)\|\big[L_{n,\epsilon}\|\mu_{\theta_t}^{(K)} - \mu^{*}(\theta_t)\| + 2\|\nabla q^{\epsilon}_t - \nabla q^{\epsilon} (\theta_t, \mu^{*}(\theta_t))\|\big]\tag{$\nabla q^{\epsilon}(\theta_t, \mu^{*}(\theta_t)) = 0$}\\
    \leq&L_{n,\epsilon}\|\mu_{\theta_t}^{(K)} - \mu^{*}(\theta_t)\|\big[L_{n,\epsilon}\|\mu_{\theta_t}^{(K)} - \mu^{*}(\theta_t)\| + 2L_{n,\epsilon, q^{\epsilon]}}\|\mu_t - \mu^{*}(\theta_t)\|\big]\tag{Lemma \ref{lem7}}
\end{align*}
Hence, 
\[\|\hat{\nabla}q^{\epsilon}_t\|^2 - \|\nabla q^{\epsilon}_t\|^2 \geq -L_{n,\epsilon}\|\mu_{\theta_t}^{(K)} - \mu^{*}(\theta_t)\|\bigg[L_{n,\epsilon}\|\mu_{\theta_t}^{(K)} - \mu^{*}(\theta_t)\| + 2L_{n,\epsilon, q^{\epsilon]}}\|\mu_t - \mu^{*}(\theta_t)\|\bigg].\]
Putting altogether, 
\begin{align*}
    q_{t+1}^{\epsilon} - q_t^{\epsilon}\leq& -\eta \xi_t \|\hat{\nabla}q^{\epsilon}_t\|^2 + 2(\eta+1)\xi_t L_{n,\epsilon}\|\mu_t^{(K)}-\mu^{*}(\theta_t)\| M_{n,\epsilon} + 
    L_{n,\epsilon, q^{\epsilon}} \xi_t^2 (\eta+1) M_{n,\epsilon}\\
    \leq& -\eta \xi_t \|\nabla q^{\epsilon}_t\|^2 +\eta \xi_tL_{n,\epsilon}\|\mu_{\theta_t}^{(K)} - \mu^{*}(\theta_t)\|\bigg[L_{n,\epsilon}\|\mu_{\theta_t}^{(K)} - \mu^{*}(\theta_t)\| + 2L_{n,\epsilon, q^{\epsilon]}}\|\mu_t - \mu^{*}(\theta_t)\|\bigg] \\
    + &2(\eta+1)\xi_t L_{n,\epsilon}\|\mu_t^{(K)}-\mu^{*}(\theta_t)\| M_{n,\epsilon} + 
    (\eta+1)L_{n,\epsilon, q^{\epsilon}} \xi_t^2 M_{n,\epsilon}
\end{align*}
When $\|\hat{\nabla}q^{\epsilon}_t\| = 0$, then this implies that given $\theta_t,$ $\mu_t = \mu^{*}(\theta_t)$, leading $q^{\epsilon}(\theta_t, \mu_t) = q^{\epsilon}(\theta_t, \mu^{*}(\theta_t)) = f_n^{\epsilon}(\theta_t, \mu_t) - f_n^{\epsilon}(\theta_t, \mu^{*}(\theta_t)) = 0$ and $\nabla q^{\epsilon}_t = 0.$
Therefore, 
\begin{align*}
    q_{t+1}^{\epsilon} - q_t^{\epsilon}\leq&-\xi_t\langle \nabla q_t^{\epsilon} + \hat{\nabla}q^{\epsilon}_t - \hat{\nabla}q^{\epsilon}_t,\delta_t \rangle + \frac{L_{n,\epsilon, q^{\epsilon}}}{2}\xi_t^2\|\delta_t\|^2\\
    \leq& (\eta+1)^2L_{n,\epsilon, q^{\epsilon}} ^2\xi_t^2 M_{n,\epsilon}^2
\end{align*}\hfill\qedsymbol

\begin{lemma}\label{lem14}
Under the \hyperref[ppl]{projected PL inequality}, \hyperref[bound]{boundedness} and \hyperref[smooth]{smoothness}, let $q_t^{\epsilon}= q^{\epsilon}(\theta_t, \mu_t)$. Then for some $t_0 \in [[T]]$ and $K\geq \frac{1}{a_1}\log(\frac{72L_{n,\epsilon, q^{\epsilon}}^2}{\kappa^2}),$ we have 
\begin{align*}
    q_{t+1}^{\epsilon}- q_t^{\epsilon} \leq& -\frac{1}{4}\eta  \kappa \xi_t q_t^{\epsilon}\textbf{1}\{t \leq t_0\} + \frac{\eta\kappa\xi_t}{4} b \textbf{1}\{t >t_0\}.
\end{align*}
Moreover, 
\begin{align*}
     q_{t}^{\epsilon} \leq (1- \frac{1}{4}\eta  \kappa \xi_t)^{t} q_0^{\epsilon}\textbf{1}\{t \leq t_0\} + (1+\frac{1}{4}\eta \kappa\xi_t) b \textbf{1}\{t >t_0\},
\end{align*}
where $q_0^{\epsilon} = q^{\epsilon}(\theta_0, \mu_0).$

\end{lemma}
\textbf{\textit{proof}.} We know that from Lemma \ref{lem1} and Lemma \ref{lem3}, 
\begin{align*}
    \|\mu_t^{(K)} - \mu^{*}(\theta_t)\| \leq& \frac{2}{\sqrt{\kappa}}\sqrt{q^{\epsilon}(\theta_t, \mu_t^{(K)})}\tag{Lemma \ref{lem1}}\\
    \leq& \frac{2}{\sqrt{\kappa}} \exp(-a_1 K/2) \sqrt{q^{\epsilon}(\theta_t, \mu_t)}\tag{Lemma \ref{lem3}}\\
    \|\mu_t - \mu^{*}(\theta_t)\| \leq&\frac{2}{\sqrt{\kappa}} \sqrt{q^{\epsilon}(\theta_t, \mu_t)}.
\end{align*}
Note that $L_{n,\epsilon, q^{\epsilon}}- L_{n, \epsilon} = L_{n,\epsilon}\big[\frac{2L_{n,\epsilon}}{\kappa}+2- 1\big]>0.$ Hence, 
\begin{align*}
    &L_{n,\epsilon}\|\mu_t^{(K)} - \mu^{*}(\theta_t)\|\bigg[L_{n,\epsilon}\|\mu_t^{(K)} - \mu^{*}(\theta_t)\| + 2L_{n,\epsilon, q^{\epsilon}}\|\mu_t - \mu^{*}(\theta_t)\|\bigg] \\
    \leq&L_{n,\epsilon, q^{\epsilon}}\|\mu_t^{(K)} - \mu^{*}(\theta_t)\|\bigg[L_{n,\epsilon, q^{\epsilon}}\|\mu_t^{(K)} - \mu^{*}(\theta_t)\| + 2L_{n,\epsilon, q^{\epsilon}}\|\mu_t - \mu^{*}(\theta_t)\|\bigg]\\
    \leq& L_{n,\epsilon, q^{\epsilon}}\frac{2}{\sqrt{\kappa}} \exp(-a_1 K/2) \sqrt{q^{\epsilon}_t}\bigg[ L_{n,\epsilon, q^{\epsilon}}\frac{2}{\sqrt{\kappa}} \exp(-a_1 K/2) \sqrt{q^{\epsilon}_t} + 2L_{n,\epsilon, q^{\epsilon}}\frac{2}{\sqrt{\kappa}} \sqrt{q^{\epsilon}_t}\bigg]\\
    =& L_{n,\epsilon, q^{\epsilon}}^2\frac{4}{\kappa} \exp(-a_1 K) q^{\epsilon}_t\bigg[ 1 + 2\exp(a_1 K/2)\bigg]
\end{align*}
We have 
\begin{align*}
    L_{n,\epsilon, q^{\epsilon}}\|\mu_t^{(K)} - \mu^{*}(\theta_t)\|\bigg[L_{n,\epsilon, q^{\epsilon}}\|\mu_t^{(K)} - \mu^{*}(\theta_t)\| + 2L_{n,\epsilon, q^{\epsilon}}\|\mu_t - \mu^{*}(\theta_t)\|\bigg]\
    \leq& \frac{36}{\kappa}L_{n,\epsilon, q^{\epsilon}}^2 \exp(-a_1 K) q^{\epsilon}_t.
\end{align*}
Then,
\begin{align*}
    q_{t+1}^{\epsilon}- q_t^{\epsilon} \leq& - \eta \xi_t \|\nabla q_t^{\epsilon}\|^2 + \frac{36}{\kappa}\eta \xi_t L_{n,\epsilon, q^{\epsilon}}^2 \exp(-a_1 K) q^{\epsilon}_t \\
    +& 2(\eta+1)\xi_t L_{n,\epsilon, q^{\epsilon}}M_{n,\epsilon} \frac{2}{\sqrt{\kappa}} \exp(-a_1 K/2) \sqrt{q^{\epsilon}_t} + (\eta+1)^2L_{n,\epsilon, q^{\epsilon}} ^2\xi_t^2 M_{n,\epsilon}^2\\
    \leq& - \eta \xi_t \kappa q_t^{\epsilon} + \frac{36}{\kappa}\eta \xi_t L_{n,\epsilon, q^{\epsilon}}^2 \exp(-a_1 K) q^{\epsilon}_t \\
    +& 2(\eta+1)\xi_t L_{n,\epsilon, q^{\epsilon}}M_{n,\epsilon} \frac{2}{\sqrt{\kappa}} \exp(-a_1 K/2) \sqrt{q^{\epsilon}_t} + (\eta+1)^2L_{n,\epsilon, q^{\epsilon}} ^2\xi_t^2 M_{n,\epsilon}^2,
\end{align*}
where the last inequality comes from 
\[\kappa q_t^{\epsilon} \leq \|G_{\alpha}(\mu_t;\theta_t)\|^2 \leq \|\nabla_{\mu}q^{\epsilon}_t\|^2 \leq \|\nabla q^{\epsilon}_t\|^2.\]

To have $-1+ \frac{36}{\kappa}L_{n,\epsilon, q^{\epsilon}}^2 \exp(-a_1 K) \leq -1/2$, we choose $K\geq \frac{1}{a_1}\log(\frac{72L_{n,\epsilon, q^{\epsilon}}^2}{\kappa^2}).$
Then, 
\begin{align*}
    q_{t+1}^{\epsilon}- q_t^{\epsilon} \leq&- \frac{1}{2}\eta \xi_t \kappa q_t^{\epsilon} 
    + \frac{4(\eta+1)\xi_t}{\sqrt{\kappa}} L_{n,\epsilon, q^{\epsilon}}M_{n,\epsilon}  \exp(-a_1 K/2) \sqrt{q^{\epsilon}_t} + (\eta+1)^2L_{n,\epsilon, q^{\epsilon}} ^2\xi_t^2 M_{n,\epsilon}^2,
\end{align*}
Let $b_1 = \frac{32^2(\eta+1)^2}{\eta^2 \kappa^5}L_{n,\epsilon, q^{\epsilon}}^2 M_{n,\epsilon}^2 \exp(-a_1 K)$ and $b_2 = \frac{8(\eta+1)\xi_tL_{n,\epsilon, q^{\epsilon}}}{\eta \kappa}$.
If $ b_1 \leq q_t^{\epsilon}$ and $b_2 \leq q_t^{\epsilon}$, then 

\[q_{t+1}^{\epsilon}- q_t^{\epsilon} \leq - \frac{1}{4}\eta \xi_t \kappa q_t^{\epsilon}. \]
Let $b = \max \{b_1, b_2\}.$ If $q_t^{\epsilon}<b,$
\begin{align*}
    q_{t+1}^{\epsilon} - q_t^{\epsilon} \leq& - \frac{1}{2}\eta \xi_t \kappa q_t^{\epsilon} 
    + \frac{4(\eta+1)\xi_t}{\sqrt{\kappa}} L_{n,\epsilon, q^{\epsilon}}M_{n,\epsilon}  \exp(-a_1 K/2) \sqrt{q^{\epsilon}_t} + (\eta+1)^2L_{n,\epsilon, q^{\epsilon}} ^2\xi_t^2 M_{n,\epsilon}^2\\
    \leq&  \frac{4(\eta+1)\xi_t}{\sqrt{\kappa}} L_{n,\epsilon, q^{\epsilon}}M_{n,\epsilon}  \exp(-a_1 K/2) \sqrt{b} + (\eta+1)^2L_{n,\epsilon, q^{\epsilon}} ^2\xi_t^2 M_{n,\epsilon}^2\\
    \leq& 
    \frac{1}{4}\eta \kappa \xi_t b 
\end{align*}

This implies that for the first step $t_0$ that satisfies $q_t^{\epsilon} <b$, the difference between the values from two successive value functions evaluated at $t$ and $t+1$ decreases proportional to the value $q_t^{\epsilon}$. 
In sum, when $q_t^{\epsilon} \geq b$ or $t \leq t_0,$
\begin{align*}
    q_{t}^{\epsilon} \leq    (1- \frac{1}{4}\eta  \kappa \xi_t)^{t} q_0^{\epsilon}.
\end{align*}
Moreover, when $q_t^{\epsilon} < b$ or for any $t > t_0,$  
\begin{align*}
    q_{t}^{\epsilon} \leq (1+\frac{1}{4}\eta \kappa)\xi_t b. 
\end{align*}

Therefore, for $K\geq \frac{1}{a_1}\log(\frac{72L_{n,\epsilon, q^{\epsilon}}^2}{\kappa^2}),$ 
\begin{align*}
    q_{t+1}^{\epsilon} - q_t^{\epsilon} \leq&  -\frac{1}{4}\eta  \kappa \xi_t q_t^{\epsilon}\textbf{1}\{q_t^{\epsilon} > b\} + \frac{\eta\kappa\xi_t}{4} b \textbf{1}\{q_t^{\epsilon} < b\}\\
    \leq& -\frac{1}{4}\eta  \kappa \xi_t q_t^{\epsilon}\textbf{1}\{t \leq t_0\} + \frac{\eta\kappa\xi_t}{4} b \textbf{1}\{t >t_0\}.
\end{align*}

Moreover, 
\begin{align*}
     q_{t}^{\epsilon} \leq (1- \frac{1}{4}\eta  \kappa \xi_t)^{t} q_0^{\epsilon}\textbf{1}\{t \leq t_0\} + (1+\frac{1}{4}\eta \kappa\xi_t) b \textbf{1}\{t >t_0\}. 
\end{align*}\hfill\qedsymbol

\clearpage

\subsection*{Verification of \hyperref[smooth]{smoothness}, \hyperref[pl]{the PL inequality} and \hyperref[bound]{boundedness}.\label{verify}
}
We need to verify that my model satisfies the assumptions \ref{smooth}, \ref{ppl} and \ref{bound}. We define the following: 
 \begin{align*}
        &T:\mathcal{U} \rightarrow \mathbb{R}^k \text{ is a vector-valued function on a space of square matrices with size n and constraints.}\\
        &T_n := T_n(\theta \hspace{1mm}|g_n, \{X_i\}_{i=1}^n) = \langle \theta, \frac{1}{n^2}T(g_n)\rangle\\
        &\psi_n^{MF} :=\Gamma_n (\theta, \mu^{*}) = \sup_{\mu} \Gamma_n (\theta, \mu) = \sup_{\mu} \frac{1}{n^2}\bigg\{\langle \theta, T(\mu)\rangle - H(\mu)\bigg\}\\
        &\psi_n^{\epsilon} :=\Gamma_n (\theta, \mu^{*}) - \frac{\epsilon}{2n^2}||\mu^{*}||_F^2= \sup_{\mu} \Gamma_n (\theta, \mu) = \sup_{\mu} \bigg\{\frac{1}{n^2} \big[\langle \theta, T(\mu)\rangle - H(\mu)\big] - \frac{\epsilon}{2n^2}||\mu||_F^2\bigg\}\\
        &H(\mu) := \frac{1}{2}\sum_{i=1}^n\sum_{j=1}^n \bigg\{\mu_{ij}\text{log}\mu_{ij}+ (1-\mu_{ij})\text{log}(1-\mu_{ij})\bigg\}= \frac{1}{2}\textbf{1}_n^{\top}\big\{\mu\odot \text{log}\mu + (\textbf{1}_n\textbf{1}_n^{\top}-\mu) \odot\text{log}(\textbf{1}_n\textbf{1}_n^{\top}-\mu)\big\}\textbf{1}_n
    \end{align*}
    where $\odot$ is the Hadamard product (element-wise matrix multiplication), and $\textbf{1}_n = [1,1,...,1]^{\top} \in \mathbb{R}^n$.
\begin{enumerate}
    \item Assumption \hyperref[smooth]{smoothness}.
    \begin{enumerate}[label=(\alph*)]
    \item  $F_n$: We want to show for any $\theta_1, \theta_2 \in \Theta, $ there exists a positive constant $L_{n}^{(1)} >0$, which may depend on some fixed $n \in \mathbb{N}$ such that
    \[|F_n(\theta_1) - F_n(\theta_2)| \leq L_{n}^{(1)}||\theta_1 - \theta_2||,\]
    where $F_n(\theta) = -\ell_n^{MF}(\theta) = -T_n(\theta) + \psi_n^{\epsilon}(\theta)$.
    For simplicity, let $F_n^{(i)} = F_n(\theta_i)$ and $\psi_n^{(i)} = \psi_n^{\epsilon}(\theta_i)$, and $\mu_i^{*} = \mu^{*}(\theta_i)$ for $i=1,2$. Then, 
    \begin{align*}
        |F_n^{(1)} - F_n^{(2)}| =& |T_n^{(2)} - T_n^{(1)} + (\psi_n^{(1)} - \psi_n^{(2)})|\\
        =& |T_n^{(2)} - T_n^{(1)} + \Gamma_n (\theta_1, \mu_1^{*}) - \frac{\epsilon}{2n^2}||\mu_1^{*}||_2^2 - \Gamma_n (\theta_2, \mu_2^{*}) + \frac{\epsilon}{2n^2}||\mu_2^{*}||_2^2|\\
        =& |\{\langle \theta_2, \frac{1}{n^2}T(g_n)\rangle - \langle \theta_1,\frac{1}{n^2} T(g_n)\rangle\} + \Gamma_n (\theta_1, \mu_1^{*}) - \frac{\epsilon}{2n^2}||\mu_1^{*}||_2^2 - \Gamma_n (\theta_2, \mu_2^{*}) + \frac{\epsilon}{2n^2}||\mu_2^{*}||_2^2|\\
        \leq&  \big\{\underbrace{|\langle \theta_1 - \theta_2, \frac{1}{n^2}T(g_n)\rangle|}_{(i)} + \underbrace{\Gamma_n (\theta_1, \mu_1^{*}) - \frac{\epsilon}{2n^2}||\mu_1^{*}||_2^2 - \Gamma_n (\theta_2, \mu_2^{*}) + \frac{\epsilon}{2n^2}||\mu_2^{*}||_2^2}_{(ii)}\big\}
    \end{align*}
    \begin{enumerate}[label=(\roman*)]
        \item 
            \[|\langle \theta_1 - \theta_2, \frac{1}{n^2}T(g_n)\rangle| \leq ||\theta_1 - \theta_2|| ||\frac{1}{n^2}T(g_n)|| \tag{\text{Cauchy-schwarz}}\]
        \item For the upper bound, \begin{align*}
            &\Gamma_n (\theta_1, \mu_1^{*}) - \frac{\epsilon}{2n^2}||\mu_1^{*}||_2^2 - \Gamma_n (\theta_2, \mu_2^{*}) - \frac{\epsilon}{2n^2}||\mu_2^{*}||_2^2\\\leq& \Gamma_n (\theta_1, \mu_1^{*}) - \frac{\epsilon}{2n^2}||\mu_1^{*}||_2^2 - \Gamma_n (\theta_2, \mu_1^{*}) + \frac{\epsilon}{2n^2}||\mu_1^{*}||_2^2\\
            = & \langle \theta_1, \frac{1}{n^2}T(\mu_1^{*})\rangle - H(\mu_1^{*}) -\langle \theta_2, \frac{1}{n^2}T(\mu_1^{*})\rangle + H(\mu_1^{*})\\ 
            =& \langle \theta_1 - \theta_2, \frac{1}{n^2}T(\mu_1^{*})\rangle 
        \end{align*}
        For the lower bound, 
        \begin{align*}
           & \Gamma_n (\theta_1, \mu_1^{*}) - \frac{\epsilon}{2n^2}||\mu_1^{*}||_2^2 - \Gamma_n (\theta_2, \mu_2^{*}) - \frac{\epsilon}{2n^2}||\mu_2^{*}||_2^2\\\geq& \Gamma_n (\theta_1, \mu_2^{*}) - \frac{\epsilon}{2n^2}||\mu_2^{*}||_2^2 - \Gamma_n (\theta_2, \mu_2^{*}) - \frac{\epsilon}{2n^2}||\mu_2^{*}||_2^2\\
            = & \langle \theta_1, \frac{1}{n^2}T(\mu_2^{*})-\rangle - H(\mu_2^{*}) \langle \theta_2, \frac{1}{n^2}T(\mu_2^{*})\rangle+ H(\mu_2^{*})\\ 
            =& \langle \theta_1 - \theta_2, \frac{1}{n^2}T(\mu_2^{*})\rangle 
        \end{align*}
        Therefore, for any $\theta_1, \theta_2 \in \Theta$
        \begin{align*}
            |\Gamma_n (\theta_1, \mu_1^{*}) - \Gamma_n (\theta_2, \mu_2^{*})| \leq & \max \{|\langle \theta_1 - \theta_2, \frac{1}{n^2}T(\mu_1^{*})\rangle|, |\langle \theta_1 - \theta_2, \frac{1}{n^2}T(\mu_2^{*})\rangle|\}\\
            \leq & \max\{||\theta_1- \theta_2|| ||\frac{1}{n^2}T(\mu_1^{*})||, ||\theta_1- \theta_2|| ||\frac{1}{n^2}T(\mu_2^{*})||\}\tag{\text{Cauchy-schwarz}}\\
            =& ||\theta_1- \theta_2||\frac{1}{n^2} \max \{||T(\mu_1^{*})||, ||T(\mu_2^{*})||\}\\
            \leq & ||\theta_1- \theta_2|| \frac{1}{n^2}||T^{*}||, 
        \end{align*}
        where 
        \[||T^{*}|| = \max_{\mu \in \mathcal{U}} \{||T(\mu)||\}. \]
        (In fact, T is a vector-valued function whose components are network statistics, polynomials of components of element $M\in \mathcal{U}$. Let $g(z):= ||z||, z \in \mathbb{R}^k$. Then $g$ is a norm in the finite-dimensional Euclidean space, which is continuous. $h: = ||T||= g\circ T: \mathcal{U} \rightarrow \mathbb{R}$ is a composite function of two continuous functions on the compact set $\mathcal{U}$. Hence, by the extreme value theorem, it attains minima and maxima on its domain. Therefore $||T||$ is bounded above by some $||T^{*}||$. )
        Then, 
        \begin{align*}
            |\Gamma_n (\theta_1, \mu_1^{*}) - \Gamma_n (\theta_2, \mu_2^{*})| 
            \leq & ||\theta_1 - \theta_2|| \frac{1}{n^2}||T^{*}||.
        \end{align*}

    \end{enumerate}Hence, 
        \begin{align*}
            |F_n^{(1)} - F_n^{(2)}| \leq&  \big\{|\langle \theta_1 - \theta_2, \frac{1}{n^2}T(g_n)\rangle| + |\Gamma_n (\theta_1, \mu_1^{*}) - \frac{\epsilon}{2n^2}||\mu_1^{*}||_2^2 - \Gamma_n (\theta_2, \mu_2^{*}) + \frac{\epsilon}{2n^2}||\mu_2^{*}||_2^2|\big\}\\
            \leq&   \frac{1}{n^2}\big[||T(g_n)|| + ||T^{*}||\big]||\theta_1 -\theta_2||
            \end{align*}
    Let $L_{n}^{(1)} = \frac{1}{n^2}\big[||T(g_n)|| + ||T^{*}||\big]$. Hence, we have a positive constant $L_{n}^{(1)}>0$ such that
    \[ |F_n^{(1)} - F_n^{(2)}| \leq L_n^{(1)}||\theta_1 -\theta_2|| \]

    \item $\nabla F_n$: We want to show for any $\theta_1, \theta_2 \in \Theta, $ there exists a positive constant $L_{n}^{(2)} >0$, which may depend on some fixed $n \in \mathbb{N}$ such that \[||\nabla F_n( \theta_1)-\nabla F_n(\theta_2)||\leq L_{n}^{(2)} ||\theta_1- \theta_2||.\]
    \begin{align*}
        \nabla F_n(\theta_1) =& -\frac{\partial}{\partial \theta_1} T_n(\theta_1) + \frac{\partial}{\partial \theta_1} \psi_n^{\epsilon}(\theta_1)\\ 
        =& -\frac{\partial}{\partial \theta_1} T_n(\theta_1) + \frac{\partial}{\partial \theta_1}\big[ \Gamma_n (\theta_1, \mu_1^{*}) -\frac{\epsilon}{2n^2}||\mu_1^{*}||_2^2\big]\\
        =& -\frac{\partial}{\partial \theta_1} \langle \theta_1, \frac{1}{n^2}T(g_n) \rangle + \frac{\partial}{\partial \theta_1} \bigg(\langle \theta_1, \frac{1}{n^2} T(\mu_1^{*}) \rangle - H(\mu_1^{*})\bigg)\\
        =& -\frac{1}{n^2} T(g_n)+\frac{1}{n^2} T(\mu_1^{*})
    \end{align*}
    Hence, 
    \begin{align*}
        ||\nabla F_n( \theta_1)-\nabla F_n(\theta_2)|| = ||\frac{1}{n^2}\big[ T(\mu_1^{*}) - T(\mu_2^{*})\big]||
    \end{align*}
    We need to show that for a given $\theta_1 \in \Theta,$
    \[\mu_1^{*}:= \mu^{*}(\theta_1) \in \argsup_{\mu}\Gamma_n (\theta_1, \mu) - \frac{\epsilon}{2n^2}||\mu||_2^2 = \argsup_{\mu}\frac{1}{n^2}\bigg\{\langle \theta_1, T(\mu)\rangle - H(\mu) \bigg\}- \frac{\epsilon}{2n^2}||\mu||_2^2 \] is continuous. In other words, the distance between any pair of two solutions sets $\argsup_{\mu}\Gamma_n (\theta_1, \mu)- \frac{\epsilon}{2n^2}||\mu||_2^2 $ and $\argsup_{\mu}\Gamma_n (\theta_2, \mu)- \frac{\epsilon}{2n^2}||\mu||_2^2 $ defined by $\theta_1$ and $\theta_2$, is close enough whenever the two $\theta$s are close enough. 

    Define 
    \[S(\theta_i) = \argsup_{\mu'} \Gamma_n (\theta_i, \mu')- \frac{\epsilon}{2n^2}||\mu'||_2^2 , \hspace{3mm}i = 1,2\]
    as the solution mapping for $\theta_1, \theta_2 \in \Theta$. 
    We want to show that there exists a positive constant $R>0$ such that the solution mapping is $R-$ Lipschitz, that is, 
    \[ d_H (S(\theta_1), S(\theta_2)) \leq R||\theta_1 - \theta_2||, \]
    where $d_H (A, B) = \max \{d_h(A,B), d_h(B,A)\}$ is the Hausdorff distance between two closed sets $A$ and $B$, and $d_h(A,B) = \sup_{a\in A} \bigg\{\inf_{b\in B}||a-b||\bigg\}$. $S (\theta)$ is a subset of the compact set $\mathcal{U_{\zeta}}$ and is nonempty, because the domain $\mathcal{U_{\zeta}}$ is compact, since $\mathcal{U}_{\zeta} := \{M \in [\zeta, 1-\zeta]^{n^2} \hspace{1mm}|\hspace{1mm}\mu_{ij} \in [\zeta, 1- \zeta], \hspace{1mm} \mu_{ii} = 0 \hspace{2mm}\text{for } i, j \in [n] \}$, for some small enough $\zeta >0$ and the function $T(\cdot)$ is continuous (a vector of polynomials of elements of $\mu$). 

    In fact, under a suitable choice of the regularization parameter $\epsilon$, that is, if $\epsilon$ is large enough to dominate the minimum eigenvalue of the Hessian matrix $\nabla^2_{
    \mu \mu}f_n(\theta, \mu)$ in order to make the entire Hessian matrix $\nabla^2_{\mu\mu}f_n^{\epsilon}$ positive definite, the regularized lower-level objective function $f_n^{\epsilon}$ becomes strongly convex in $\mu$ for any given $\theta$. This means that there exists a unique solution $\mu^{*}$ to $f_n^{\epsilon}$ and instead of establishing the Lipschitz continuity of the solution mapping $S(\cdot)$, we only need to show for any $\theta_1, \theta_2 \in \Theta$, the distance between the two corresponding unique solutions $\mu_1^{*} := \mu^{*}(\theta_1)$ and $\mu_2^{*} := \mu^{*}(\theta_2)$ are close enough whenever $\theta_1 $ and $\theta_2$ are, i.e., there exists a positive constant $P_{n,\epsilon}>0$, such that 
    \[||\mu_1 - \mu_2||_F \leq P_{n,\epsilon} ||\theta_1 - \theta_2||_2\tag{L}\label{L}\]

    By definition, a function $f:\mathcal{C} \subseteq \mathbb{R}^d \rightarrow \mathbb{R}$ is strongly convex with parameter $\rho >0$ such that for any $x,y \in \mathcal{C}$
    \[f(y) \geq f(x) + \nabla f(x)^{\top}(y-x) + \frac{\rho}{2}||y-x||_2^2.\]
    We are going to show \hyperref[L]{(L)}, using one of the equivalent statements to the above definition of strong convexity. For any given $\theta$, 
    \[(\nabla_{\mu}f_n^{\epsilon}(\theta, \mu_1) - \nabla_{\mu}f_n^{\epsilon}(\theta, \mu_2))^{\top}(\mu_1 - \mu_2) \geq \rho ||\mu_1 - \mu_2||_F^2.\tag{E}\label{E}\]
    Since the Frobenius norm $||\cdot||_F$ is equivalent to the 2-norm $||\cdot||_2$ after vectorization of arguments, we vectorize all the terms of gradients or Jacobian matrix $\nabla_{\mu}f_n^{\epsilon}\in \mathbb{R}^{n \times n}$ to $\nabla_{\mu}f_n^{\epsilon} :=vec(\nabla_{\mu}f_n^{\epsilon})\in \mathbb{R}^{n^2}$, and $\mu_i:=vec(\mu_i) \in \mathbb{R}^{n^2}$ for $i=1,2.$
    
    Note that 
    \begin{align*}
        \nabla_{\mu}f_n^{\epsilon}(\theta_1, \mu_1) - \nabla_{\mu}f_n^{\epsilon}(\theta_2, \mu_2) =& \underbrace{\nabla_{\mu}f_n^{\epsilon}(\theta_1, \mu_1) - \nabla_{\mu}f_n^{\epsilon}(\theta_1, \mu_2)}_{(*)}+ \nabla_{\mu}f_n^{\epsilon}(\theta_1, \mu_2) -\nabla_{\mu}f_n^{\epsilon}(\theta_2, \mu_2)=0
    \end{align*}
    We know that $(*)$ is the first term of inner product in the left-hand side of \hyperref[E]{(E)}. Rearraging the terms yields 
    \[\nabla_{\mu}f_n^{\epsilon}(\theta_1, \mu_1) - \nabla_{\mu}f_n^{\epsilon}(\theta_1, \mu_2) = \nabla_{\mu}f_n^{\epsilon}(\theta_2, \mu_2) -\nabla_{\mu}f_n^{\epsilon}(\theta_1, \mu_2),\]
    where
    \begin{align*}
        \nabla_{\mu}f_n^{\epsilon}(\theta_2, \mu_2) =& \frac{-1}{n^2}\bigg[\langle \theta_2, \nabla_{\mu}T(\mu_2)\rangle - \nabla_{\mu}H(\mu_2)\bigg] + \frac{\epsilon}{n^2}\mu_2\\
        \nabla_{\mu}f_n^{\epsilon}(\theta_1, \mu_2) =& \frac{-1}{n^2}\bigg[\langle \theta_1, \nabla_{\mu}T(\mu_2)\rangle - \nabla_{\mu}H(\mu_2)\bigg] + \frac{\epsilon}{n^2}\mu_2,
    \end{align*}
    and $\langle \theta, \nabla_{\mu}T(\mu)\rangle = \nabla_{\mu}T(\mu) \theta \in \mathbb{R}^{(n^2 \times d) \times d}$. Their difference is 
    \[\nabla_{\mu}f_n^{\epsilon}(\theta_2, \mu_2) - \nabla_{\mu}f_n^{\epsilon}(\theta_1, \mu_2) = \frac{1}{n^2}\langle \theta_1 - \theta_2, \nabla_{\mu}T(\mu_2)\rangle\]
    Hence, plugging the difference into \hyperref[E]{(E)} yields
    \begin{align*}
        (\nabla_{\mu}f_n^{\epsilon}(\theta, \mu_1) - \nabla_{\mu}f_n^{\epsilon}(\theta, \mu_2))^{\top}(\mu_1 - \mu_2) =& (\nabla_{\mu}f_n^{\epsilon}(\theta_2, \mu_2) - \nabla_{\mu}f_n^{\epsilon}(\theta_1, \mu_2))^{\top}(\mu_1 - \mu_2) \\
        =& \frac{1}{n^2}\bigg[\langle \theta_1 - \theta_2, \nabla_{\mu}T(\mu_2)\rangle\bigg]^{\top}(\mu_1 - \mu_2)\\
        \geq& \rho ||\mu_1 - \mu_2||_2^2\\
         =& (\lambda_m + \epsilon) ||\mu_1 - \mu_2||_2^2
    \end{align*}
    Rearranging both sides, we obtain
    \begin{align*}
        \rho n^2 ||\mu_1 - \mu_2||_2^2 \leq& \bigg[\langle \theta_1 - \theta_2, \nabla_{\mu}T(\mu_2)\rangle\bigg]^{\top}(\mu_1 - \mu_2)\\
        \leq& |\bigg[\langle \theta_1 - \theta_2, \nabla_{\mu}T(\mu_2)\rangle\bigg]^{\top}(\mu_1 - \mu_2)| \\
        =& |( \theta_1 - \theta_2)^{\top} \nabla_{\mu}T(\mu_2)^{\top}(\mu_1 - \mu_2)|\\
        \leq&||( \theta_1 - \theta_2)||_2  ||\nabla_{\mu}T(\mu_2)^{\top}(\mu_1 - \mu_2)||_2 \tag{Cauchy-schwarz} \\
        \leq&||( \theta_1 - \theta_2)||_2  ||\nabla_{\mu}T(\mu_2)||_2||(\mu_1 - \mu_2)||_2 \tag{Cauchy-schwarz}
    \end{align*}
    Dividing both sides by $\rho n^2$ and $||\mu_1 - \mu_2||_2$ leads to
    \begin{align*}
        ||\mu_1 - \mu_2||_2 \leq \frac{1}{\rho n^2}||\nabla_{\mu}T(\mu_2)||_2  ||\theta_1 - \theta_2||_2.
    \end{align*}
    Since $\mu \in \mathcal{U}_{\zeta}$, which is compact in the Euclidean space, and $T$ is a vector of arbitrary polynomials of $\mu$ or a vector of smooth functions, its Jacobian is also continuous in $\mu$. By the extreme value theorem, it attains the minimum and maximum on its domain $\mathcal{U}_{\zeta}$.
    Let $\nabla_{\mu}T(\mu^{*}):= \max_{\mu'\in \mathcal{U}_{\zeta}}||\nabla_{\mu}T(\mu')||$ and $P_{n,\epsilon}:= \frac{1}{\rho n^2}||\nabla_{\mu}T(\mu^{*})||_2 = \frac{1}{(\lambda_m + \epsilon) n^2}||\nabla_{\mu}T(\mu^{*})||_2>0.$ Then there exists a Lipschitz constant $P_{n,\epsilon}>0$ such that 
    \[||\mu_1 - \mu_2||_2 \leq P_{n,\epsilon}  ||\theta_1 - \theta_2||_2.\]

    We know that $T$ is a vector of polynomials and every polynomial on a closed and bounded set is Lipschitz continuous. Therefore, there exists a positive constant $P_n>0$ such that for any $\mu_1, \mu_2 \in \mathcal{U}_{\zeta}$,
    \[||T(\mu_1) - T(\mu_2)||_2\leq P_n ||\mu_1 - \mu_2||_2.\]

    Putting altogether, 
    \begin{align*}
        ||\nabla F_n( \theta_1)-\nabla F_n(\theta_2)|| =& ||\frac{1}{n^2}\big[ T(\mu_1^{*}) - T(\mu_2^{*})\big]|| \leq \frac{P_n}{n^2} ||\mu_1^{*} - \mu_2^{*}||_2 \\
        \leq&  \frac{P_n}{(\lambda_m + \epsilon) n^4}||\nabla_{\mu}T(\mu^{*})||_2||\theta_1 - \theta_2||_2      
        \end{align*}

    Let $L_{n,\epsilon}^{(2)} := \frac{P_n}{(\lambda_m + \epsilon) n^4}||\nabla_{\mu}T(\mu^{*})||_2 $. Then there exists a positive constant $L_{n,\epsilon}^{(2)}>0$, such that
    \[||\nabla F_n( \theta_1)-\nabla F_n(\theta_2)|| \leq  L_{n,\epsilon}^{(2)}  ||\theta_1 - \theta_2||_2  \]

    \item $f_n^{\epsilon}$:  We want to show that $f_n^{\epsilon}:= f_n^{\epsilon}(\theta, \mu) = -\Gamma_n(\theta, \mu) + \frac{\epsilon}{2n^2}||\mu||_2^2$ is Lipschitz continuous over $\Theta \times \mathcal{U}_{\zeta}$, for some positive constant $L_{n,\epsilon}^{(3)} > 0$, where $\mathcal{U}_{\zeta} := \{M \in [\zeta, 1-\zeta]^{n^2} \hspace{1mm}|\mu_{ij} \in [\zeta, 1- \zeta], \hspace{1mm} \mu_{ii} = 0 \hspace{2mm}\text{for } i, j \in [n] \}$, for some small enough $\zeta >0.$ In other words, for any $(\theta_1, \mu_1), (\theta_2, \mu_2) \in \Theta \times\mathcal{U}_{\zeta},$ we want to show
    \[|f_n^{\epsilon}(\theta_1, \mu_1) - f_n^{\epsilon}(\theta_2, \mu_2)| \leq L_{n,\epsilon}^{(3)} ||(\theta_1, \mu_1) - (\theta_2, \mu_2)||.\]
    The restriction on $\mathcal{U}$ by $\zeta$ is required to control for the behavior of derivative of $H(\mu)$. Otherwise, the derivative is undefined at the boundary of the original set $\mathcal{U}$.  
    In fact, we only need to check that the function is differentiable and its derivative with respect to the argument is bounded, thanks to the Mean Value theorem.
    We already checked that $f_n^{\epsilon}$ is differentiable with respect to $\theta$ and $\mu$. We only need to show $\nabla f_n^{\epsilon}$ is bounded. 

    For some positive $L_{n,\epsilon}^{(3)}>0,$ we want to show that the gradient of $f_n^{\epsilon}$ is bounded, i.e.,
    \begin{align*}
        ||\nabla f_n^{\epsilon}(\theta, \mu)||_F \leq L_{n,\epsilon}^{(3)}
    \end{align*}
    In fact,
    \begin{align*}
        ||\nabla f_n^{\epsilon}(\theta, \mu)|| = || \underbrace{\big[\nabla_{\theta_1} f_n^{\epsilon}(\theta, \mu), \cdots, \nabla_{\theta_d} f_n^{\epsilon}(\theta, \mu)]^{\top}}_{(i)}, \hspace{1mm} \underbrace{vec(\nabla_{\mu} f_n^{\epsilon} (\theta, \mu))}_{(ii)}||.
    \end{align*}
    
    \begin{enumerate}[label=(\roman*)]
        \item For each $l \in [d]$,
        \[\nabla_{\theta_l} f_n^{\epsilon}(\theta, \mu) = -\frac{1}{n^2} T_l(\mu).\]
        We know that $T_l(\cdot)$ is a polynomial whose components come from the compact set $\mathcal{U}$. Hence, by the extreme value theorem, $|\nabla_{\theta_l}f_n^{\epsilon}(\theta, \mu)|$ is bounded, leading to $(i)$ is bounded. 
        \item We want to show that $\nabla_{\mu} f_n^{\epsilon} (\theta, \mu)$ is bounded. Using the fact that the two-norm $||\cdot||_2$ is compatible with the Frobenius norm $||\cdot||_F$, 
        \[||\nabla_{\mu} f_n^{\epsilon} (\theta, \mu)||_F = ||vec(\nabla_{\mu} f_n^{\epsilon} (\theta, \mu))||_2.\]
        Then, 
        \begin{align*}
            ||vec(\nabla_{\mu} f_n^{\epsilon} (\theta, \mu))||_2 = ||\bigg[\frac{\partial}{\partial \mu_{11}}f_n^{\epsilon}(\theta, \mu), \frac{\partial}{\partial \mu_{12}}f_n^{\epsilon}(\theta, \mu), \cdots, \frac{\partial}{\partial \mu_{n,n-1}}f_n^{\epsilon}(\theta, \mu), \frac{\partial}{\partial \mu_{nn}}f_n^{\epsilon}(\theta, \mu)                     \bigg]^{\top}||_2
        \end{align*}
        Investigating each element $\frac{\partial}{\partial \mu_{ij}} f_n^{\epsilon}(\theta, \mu)$ will give us the proof. \\

        For each $i,j \in [n], i\neq j$, 
        \begin{align*}
            \frac{\partial}{\partial \mu_{ij}}f_n^{\epsilon}(\theta, \mu) = -\frac{1}{n^2} \bigg[\langle \theta, \frac{\partial}{\partial \mu_{ij}} T(\mu)\rangle - \frac{\partial}{\partial \mu_{ij}} H(\mu)\bigg] + 2\epsilon \mu_{ij}
        \end{align*}
        Using the triangle inequality, we have 
        \begin{align*}
            |\langle \theta, \frac{\partial}{\partial \mu_{ij}} T(\mu)\rangle - \frac{\partial}{\partial \mu_{ij}} H(\mu) | \leq& \underbrace{|\langle \theta, \frac{\partial}{\partial \mu_{ij}} T(\mu)\rangle|}_{(a)} + \underbrace{| \frac{\partial}{\partial \mu_{ij}} H(\mu)|}_{(b)}
        \end{align*}
        \begin{enumerate}[label = (\alph*)]
            \item \[|\langle \theta, \frac{\partial}{\partial \mu_{ij}} T(\mu)\rangle| \leq||\theta||\hspace{1mm} ||\frac{\partial}{\partial \mu_{ij}} T(\mu)||\tag{Cauchy-schwarz}\]
            \begin{align*}
                ||\frac{\partial}{\partial \mu_{ij}} T(\mu)|| = \bigg(|\frac{\partial}{\partial \mu_{ij}} T_1(\mu)|^2 + |\frac{\partial}{\partial \mu_{ij}} T_2(\mu)|^2+ \cdots + |\frac{\partial}{\partial \mu_{ij}} T_d(\mu)|^2\bigg)^{1/2}
            \end{align*}
            We know that for each $l\in [d]$, $T_l$ is a continuously differential polynomial with respect to the elements from the compact interval $[\zeta, 1-\zeta]$, $\frac{\partial}{\partial \mu_{ij}} T_l$ is bounded by some positive $M_l>0$. Hence, 
            \begin{align*}
                ||\frac{\partial}{\partial \mu_{ij}} T(\mu)|| =& \bigg(|\frac{\partial}{\partial \mu_{ij}} T_1(\mu)|^2 + |\frac{\partial}{\partial \mu_{ij}} T_2(\mu)|^2+ \cdots + |\frac{\partial}{\partial \mu_{ij}} T_d(\mu)|^2\bigg)^{1/2}\\
                =&\bigg(M_1^2 + M_2^2+ \cdots + M_d^2\bigg)^{1/2}\\
                \leq& \sqrt{d}M^{(1)}
            \end{align*}
            where $M^{(1)} = \max_{l\in[d]} \{M_l\}$.
            
            \item \[| \frac{\partial}{\partial \mu_{ij}} H(\mu)| = |\text{log}\frac{\mu_{ij}}{(1-\mu_{ij})}|\]
            We know that $|\text{log}\frac{\mu_{ij}}{(1-\mu_{ij})}|$ is continuous on $[\zeta, 1-\zeta]$, we apply the extreme value theorem such that it attains the maximum on the compact set. Let $M^{(2)} = \max |\text{log}\frac{\zeta}{1-\zeta}| $.
            Then, \[| \frac{\partial}{\partial \mu_{ij}} H(\mu)| = |\text{log}\frac{\mu_{ij}}{(1-\mu_{ij})}|\leq M^{(2)} \]
        \end{enumerate}
       For each $i\neq j \in [n],$
        \[|\langle \theta, \frac{\partial}{\partial \mu_{ij}} T(\mu)\rangle - \frac{\partial}{\partial \mu_{ij}} H(\mu)|+ 2\epsilon \mu_{ij} \leq \sqrt{d}M^{(1)} + M^{(2)} +2\epsilon \]
        Therefore, for fixed $n\in \mathbb{N}$
        \begin{align*}
            ||\nabla_{\mu}f_n(\theta, \mu)||_F =& ||vec(\nabla_{\mu}f_n(\theta, \mu))||_2\\ 
            =& \bigg[\sum_{i,j}^n |\frac{\partial}{\partial \mu_{ij}}f_n(\theta, \mu)|^2\bigg]^{1/2}\\ 
            =& \bigg[\sum_{i,j}^n | -\frac{1}{n^2} \bigg[\langle \theta, \frac{\partial}{\partial \mu_{ij}} T(\mu)\rangle - \frac{\partial}{\partial \mu_{ij}} H(\mu)\bigg] +2\epsilon \mu_{ij}|^2\bigg]^{1/2}\\ 
            \leq& \frac{1}{n^2}\bigg[\sum_{i,j}^n  \bigg[ ||\theta|| \sqrt{d}M^{(1)}+M^{(2)} +2\epsilon n^2  \bigg]^2\bigg]^{1/2}\\
            =& \frac{1}{n^2}\bigg[n(n-1) \bigg[ ||\theta|| \sqrt{d}M^{(1)}+M^{(2)}+2\epsilon n^2\bigg]^2\bigg]^{1/2}\\
            \leq& \frac{\sqrt{n(n-1)}}{n^2}\big(\sqrt{d}M_{\theta}M^{(1)} + M^{(2)} +2\epsilon n^2\big):= L_{n,\epsilon}^{(3)}
        \end{align*}
        where $M_{\theta} = \max_{\theta \in \Theta}||\theta||$.
        Hence $f_n^{\epsilon}(\theta, \mu)$ is Lipschitz continuous on $\Theta \times \mathcal{U}_{\zeta}$, i.e., 
       \[|f_n^{\epsilon}(\theta_1, \mu_1) - f_n^{\epsilon}(\theta_2, \mu_2)| \leq L_{n,\epsilon}^{(3)} ||(\theta_1, \mu_1) - (\theta_2, \mu_2)||.\]
    \end{enumerate}
   
    \item $\nabla f_n^{\epsilon}$: We want to show that $\nabla f_n^{\epsilon}$ is Lipschitz continuous with for some Lipschitz constant $L_{n,\epsilon}^{(4)} >0$, with fixed $n\in \mathbb{N}$. To show this, we only need to show that $\nabla f_n^{\epsilon}$ is (continuously) differentiable and $\nabla^2 f_n^{\epsilon}$ is bounded. 
    \begin{enumerate}[label = (\arabic*)]
        \item We want to show that for any $(\theta_1, \mu_1), (\theta_2, \mu_2) \in \Theta \times \mathcal{U}_{\zeta}$ for some small enough $\zeta>0$, 
        \[||\nabla f_{n,1}^{\epsilon} - \nabla f_{n,2}^{\epsilon}|| \leq L_{n,\epsilon}^{(4)}||(\theta_1, vec(\mu_1)) - (\theta_2, vec(\mu_2)) || ,\]
        where $\nabla f_{n,l}^{\epsilon} = [\nabla_{\theta} f_n^{\epsilon}(\theta_l, \mu_l), vec(\nabla_{\mu} f_n^{\epsilon}(\theta_l, \mu_l))]^{\top} $ for $l \in \{1,2\}.$
        In fact, we showed that $\nabla_{\theta} f_n^{\epsilon}(\theta, \mu) = \frac{-1}{n^2}T(\mu)$ is a vector-valued function of $\mu$, whose components are polynomials. They are $C^{\infty}$ functions, so continuously differentiable on $\mathcal{U}_{\zeta}.$ Also,  $\nabla_{\mu} f_n^{\epsilon}(\theta,\mu)$ is (continuously) differentiable, because for each $i,j \in [n], i\neq j$, each element of $\nabla_{\mu} f_n^{\epsilon} (\theta, \mu)$ is
        \begin{align*}
            \frac{\partial}{\partial \mu_{ij}}f_n^{\epsilon}(\theta, \mu) =& -\frac{1}{n^2} \bigg[\langle \theta, \frac{\partial}{\partial \mu_{ij}} T(\mu)\rangle - \frac{\partial}{\partial \mu_{ij}} H(\mu)\bigg] + 2 \epsilon \mu_{ij}\\
            =& -\frac{1}{n^2} \bigg[\langle \theta, \frac{\partial}{\partial \mu_{ij}} T(\mu)\rangle - \text{log}\big(\frac{\mu_{ij}}{1-\mu_{ij}}\big)\bigg]+ 2 \epsilon \mu_{ij}
        \end{align*}
            
        where the first term is a linear combination of polynomials and the second term a (continuously) differentiable function $\text{log}(\cdot)$. Therefore, $\nabla f_n^{\epsilon}$ is (continuously) differentiable on $\Theta \times \mathcal{U}_{\zeta}.$
        \item We need to show that there exists a positive constant $L_{n,\epsilon}^{(3)}  >0$ such that 
        \[||\nabla^2 f_n^{\epsilon}(\theta, \mu)||_F \leq L_{n,\epsilon}^{(3)} ,\]
        where 
        \begin{align*}
            \nabla^2 f_n^{\epsilon}(\theta, \mu) =& 
            \begin{bmatrix}
                \overbrace{\nabla_{\theta}^2 f_n^{\epsilon}(\theta, \mu)}^{(i)} & \overbrace{\nabla_{\mu \theta} f_n^{\epsilon}(\theta, \mu)}^{(ii)} \\
                \nabla_{\theta \mu} f_n^{\epsilon}(\theta, \mu) & \underbrace{\nabla_{\mu}^2 f_n^{\epsilon}(\theta, \mu)}_{(iii)}
            \end{bmatrix} \in \mathbb{R}^{(d+n(n-1))^2}
        \end{align*}
        \begin{enumerate}[label = (\roman*)]
            \item $\nabla_{\theta}^2 f_n^{\epsilon}(\theta, \mu)\in \mathbb{R}^{d\times d}$
            \[\nabla_{\theta}^2 f_n^{\epsilon}(\theta, \mu) = \nabla_{\theta}\nabla_{\theta}f_n^{\epsilon}(\theta, \mu) =  \frac{-1}{n^2}\nabla_{\theta}T(\mu)=\textbf{0} \]
            
            \item $\nabla_{\mu \theta} f_n^{\epsilon}(\theta, \mu) = \nabla_{\theta \mu} f_n^{\epsilon}(\theta, \mu)^{\top} \in \mathbb{R}^{d\times n(n-1)}$

            \begin{align*}
                \nabla_{\mu \theta} f_n^{\epsilon}(\theta, \mu) = \frac{-1}{n^2}
                \begin{bmatrix}
                    \frac{\partial}{\partial \mu_{11}}T_1(\mu) & \frac{\partial}{\partial \mu_{12}}T_1(\mu) & \cdots & \frac{\partial}{\partial \mu_{nn}}T_1(\mu)\\
                    \frac{\partial}{\partial \mu_{11}}T_2(\mu) & \frac{\partial}{\partial \mu_{12}}T_2(\mu) & \cdots & \frac{\partial}{\partial \mu_{nn}}T_2(\mu)\\
                    \vdots\\
                    \frac{\partial}{\partial \mu_{11}}T_k(\mu) & \frac{\partial}{\partial \mu_{12}}T_k(\mu) & \cdots & \frac{\partial}{\partial \mu_{nn}}T_k(\mu)\\
                \end{bmatrix}
            \end{align*}
            For all $i,j \in [n]$ and for each $l\in[d]$, let 
            \[M_l = \max_{[\zeta, 1-\zeta]}\frac{\partial}{\partial \mu_{ij}}T_l(\mu), \] and 
            \[M^{(3)} = \max_l \{M_l\}.\]
            Then 
            \begin{align*}
                ||\nabla_{\mu \theta} f_n^{\epsilon}(\theta, \mu)||_F =& \bigg[\sum_{l=1}^d ||\frac{1}{n^2}vec(\nabla_{\mu} T_l (\mu)||_2^2\bigg] ^{1/2}\\
                \leq& \bigg[\sum_{l=1}^d \frac{n(n-1)}{n^4}M_l^2\bigg]^{1/2}\\
                 =& \big[\frac{n(n-1)}{n^4}\big(M_1^2+ M_2^2 + \cdots M_d^2\big) \big] ^{1/2}\\
                 \leq& \big[\frac{n(n-1)}{n^4}d(M^{(3)})^2 \big] ^{1/2}\\
                 =&\frac{\sqrt{n(n-1)}}{n^2}\sqrt{d}M^{(3)}
            \end{align*}
            \item $\nabla_{\mu}^2 f_n^{\epsilon}(\theta, \mu) \in \mathbb{R}^{n(n-1) \times n(n-1)}$\\
    
            We want to show that $||\nabla_{\mu}^2 f_n^{\epsilon}(\theta, \mu)||_F \leq L_{n,\epsilon}^{(4)}$ for some positive constant $L_{n,\epsilon}^{(4)}>0$ for some fixed $n\in \mathbb{N}.$ 
            \begin{align*}
                \nabla_{\mu}^2 f_n^{\epsilon}(\theta, \mu) =& \begin{bmatrix}
                    \frac{\partial}{\partial \mu_{11}} vec(\nabla_{\mu} f_n^{\epsilon}(\theta, \mu))^{\top}\\
                    \frac{\partial}{\partial \mu_{12}} vec(\nabla_{\mu} f_n^{\epsilon}(\theta, \mu))^{\top}\\
                    \vdots \\
                    \frac{\partial}{\partial \mu_{n,n-1}} vec(\nabla_{\mu} f_n^{\epsilon}(\theta, \mu))^{\top}\\
                    \frac{\partial}{\partial \mu_{nn}} vec(\nabla_{\mu} f_n^{\epsilon}(\theta, \mu))^{\top}\\
                \end{bmatrix}\\
                =& \begin{bmatrix}
                    \frac{\partial}{\partial \mu_{11}} \frac{\partial}{\partial \mu_{11}}f_n^{\epsilon}(\theta, \mu) & \frac{\partial}{\partial \mu_{11}} \frac{\partial}{\partial \mu_{12}}f_n^{\epsilon}(\theta, \mu) & \cdots  & \frac{\partial}{\partial \mu_{11}} \frac{\partial}{\partial \mu_{nn}}f_n^{\epsilon}(\theta, \mu)\\
                    \frac{\partial}{\partial \mu_{12}} \frac{\partial}{\partial \mu_{11}}f_n^{\epsilon}(\theta, \mu) & \frac{\partial}{\partial \mu_{12}} \frac{\partial}{\partial \mu_{12}}f_n^{\epsilon}(\theta, \mu) & \cdots  & \frac{\partial}{\partial \mu_{12}} \frac{\partial}{\partial \mu_{nn}}f_n^{\epsilon}(\theta, \mu)\\
                    \cdots & & &  \\
                    \frac{\partial}{\partial \mu_{nn}} \frac{\partial}{\partial \mu_{11}}f_n^{\epsilon}(\theta, \mu) & \frac{\partial}{\partial \mu_{nn}} \frac{\partial}{\partial \mu_{12}}f_n^{\epsilon}(\theta, \mu) & \cdots  & \frac{\partial}{\partial \mu_{nn}} \frac{\partial}{\partial \mu_{nn}}f_n^{\epsilon}(\theta, \mu)\\
                \end{bmatrix}
            \end{align*}
            Hence, 
            \begin{align*}
                ||\nabla_{\mu}^2 f_n^{\epsilon}(\theta, \mu)||_F = &\Bigg[\sum_{k,l,i,j}|\frac{\partial}{\partial \mu_{kl}}\frac{\partial}{\partial \mu_{ij}} f_n^{\epsilon}(\theta, \mu)|^2\Bigg]^{1/2}\\
                =& \Bigg[\underbrace{\sum_{\substack{(k,l) = (i,j)\\ (k,l) = (j,i)} }|\frac{\partial}{\partial \mu_{kl}}\frac{\partial}{\partial \mu_{ij}} f_n^{\epsilon}(\theta, \mu)|^2}_{(a)} + \underbrace{\sum_{\substack{(k,l) \neq (i,j)\\ (k,l) \neq (j,i)} }|\frac{\partial}{\partial \mu_{kl}}\frac{\partial}{\partial \mu_{ij}} f_n^{\epsilon}(\theta, \mu)|^2}_{(b)}\Bigg]^{1/2}
            \end{align*}
            \begin{itemize}
                \item (a): When $(k,l) = (i,j)$ or $(k,l) = (j,i)$,
                \begin{align*}
                    |\frac{\partial}{\partial \mu_{kl}}\frac{\partial}{\partial \mu_{ij}} f_n^{\epsilon}(\theta, \mu)| =& |\frac{-1}{n^2}\big[\langle \theta, \frac{\partial}{\partial \mu_{kl}}\frac{\partial}{\partial \mu_{ij}} T(\mu) \rangle - \frac{\partial}{\partial \mu_{kl}}\text{log}\frac{\mu_{ij}}{1-\mu_{ij}} \big]+ 2\epsilon \frac{\partial}{\partial \mu_{kl}}\mu_{ij}|\\
                    =& \frac{1}{n^2}|\theta_1 \frac{\partial^2}{\partial \mu_{ij}^2} T_1(\mu) + \cdots + \theta_k \frac{\partial^2}{\partial \mu_{ij}^2} T_k(\mu) - \frac{1}{\mu_{ij}(1-\mu_{ij})}|+ 2\epsilon 
                \end{align*}
                Let 
                \begin{align*}
                    M_l^{(4)} = \max \frac{\partial^2}{\partial \mu_{ij}^2} T_l(\mu) \hspace{5mm} M^{(5)} = \max_{\mu_{ij}\in [\zeta, 1-\zeta]}\frac{1}{\mu_{ij}(1-\mu_{ij})} 
                \end{align*}
                Both maxima exist because $T_l$ and $\frac{1}{x(1-x)}$ are continuous functions on the compact interval on the real line. 
                Then, 
                \begin{align*}
                    |\frac{\partial}{\partial \mu_{kl}}\frac{\partial}{\partial \mu_{ij}} f_n^{\epsilon}(\theta, \mu)| =& \frac{1}{n^2}|\theta_1 \frac{\partial^2}{\partial \mu_{ij}^2} T_1(\mu) + \cdots + \theta_k \frac{\partial^2}{\partial \mu_{ij}^2} T_k(\mu) - \frac{1}{\mu_{ij}(1-\mu_{ij})}|+2\epsilon \\
                     \leq&\frac{1}{n^2}|\theta_1 \frac{\partial^2}{\partial \mu_{ij}^2} T_1(\mu) + \cdots + \theta_k \frac{\partial^2}{\partial \mu_{ij}^2} T_k(\mu)| +\frac{1}{n^2}| \frac{1}{\mu_{ij}(1-\mu_{ij})}|+2\epsilon \\
                    \leq& \frac{1}{n^2}\bigg[|\theta_1 M_1^{(4)} + \cdots + \theta_k M_k^{(4)} | + M^{(5)}\bigg]+2\epsilon \\
                    =& \frac{1}{n^2}\bigg[|\langle \theta, \bar{M}^{(4)}\rangle | + M^{(5)}\bigg]+2\epsilon\tag{$\bar{M}^{(4)} = [M_1^{(4)}, \cdots, M_k^{(4)}]^{\top}$}\\
                     \leq& \frac{1}{n^2}\bigg[||\theta|| ||\bar{M}^{(4)}|| + M^{(5)}\bigg]+2\epsilon\tag{\text{Cauchy-schwarz}}\\
                     \leq& \frac{1}{n^2}\bigg[M_{\theta}\sqrt{k}M^{(4)} + M^{(5)}+2\epsilon n^2\bigg]\tag{$M^{(4)} = \max_{l \in [k]}\{M_l^{(4)}\}$}
                \end{align*}
                \item (b): When $(k,l) \neq (i,j)$ and $(k,l)\neq (j,i)$,
                \begin{align*}
                    |\frac{\partial}{\partial \mu_{kl}}\frac{\partial}{\partial \mu_{ij}} f_n^{\epsilon}(\theta, \mu)| =& |\frac{-1}{n^2}\big[\langle \theta, \frac{\partial}{\partial \mu_{kl}}\frac{\partial}{\partial \mu_{ij}} T(\mu)\big] \rangle|\\
                    \leq&\frac{1}{n^2}||\theta|| ||\frac{\partial}{\partial \mu_{kl}}\frac{\partial}{\partial \mu_{ij}} T(\mu)\big] ||\tag{\text{Cauchy-schwarz}}\\
                    \leq& \frac{1}{n^2} M_{\theta}M^{(6)}
                \end{align*}
                where $M_l^{(6)} = \max_{\substack{(k,l) \neq (i,j) \\ (k,l)\neq (i,j)}}|\frac{\partial}{\partial \mu_{kl}}\frac{\partial}{\partial \mu_{ij}} T_l(\mu)| $ and $M^{(6)} = \max_{l\in[k]} \{M_l^{(6)}\}$.
            \end{itemize}
        Hence, \begin{align*}
            ||\nabla_{\mu}^2 f_n^{\epsilon}(\theta, \mu)||_F =&\Bigg[\underbrace{\sum_{\substack{(k,l) = (i,j)\\ (k,l) = (j,i)} }|\frac{\partial}{\partial \mu_{kl}}\frac{\partial}{\partial \mu_{ij}} f_n^{\epsilon}(\theta, \mu)|^2}_{(a)} + \underbrace{\sum_{\substack{(k,l) \neq (i,j)\\ (k,l) \neq (j,i)} }|\frac{\partial}{\partial \mu_{kl}}\frac{\partial}{\partial \mu_{ij}} f_n^{\epsilon}(\theta, \mu)|^2}_{(b)}\Bigg]^{1/2}\\
            \leq& \Bigg[ \sum_{\substack{(k,l) = (i,j) \\ (k,l)= (i,j)}} \big\{\frac{1}{n^2}\big(M_{\theta}\sqrt{k}M^{(4)} + M^{(5)}+2\epsilon n^2\big)\big\}^2 + \sum_{\substack{(k,l) \neq (i,j) \\ (k,l)\neq (i,j)}} \big\{\frac{1}{n^2} M_{\theta}M^{(6)}\big\}^2\Bigg]^{1/2}\\
            =& \Bigg[n(n-1)\big\{\frac{1}{n^2}\big(M_{\theta}\sqrt{k}M^{(4)} + M^{(5)}+2\epsilon n^2\big)\big\}^2 \\
            +& \big(n(n-1) \times n(n-1) - n(n-1)\big)\big\{\frac{1}{n^2} M_{\theta}M^{(6)}\big\}^2\Bigg]^{1/2}\\
            \leq&\frac{\sqrt{n(n-1)}}{n^2}\big(\sqrt{k}M_{\theta}M^{(4)} + M^{(5)}+2\epsilon n^2\big)\\ +
            &\frac{\sqrt{n(n-1)\{n(n-1)-1\}}}{n^2}M_{\theta}M^{(6)}
        \end{align*}
        \end{enumerate}
      Putting all the results together, we have
        \begin{align*}
            ||\nabla^2 f_n^{\epsilon}(\theta, \mu)||_F =& 
            \bigg[||\nabla_{\theta}^2 f_n^{\epsilon}(\theta, \mu)||_F^2+2||\nabla_{\mu \theta} f_n^{\epsilon}(\theta, \mu)||_F^2+||\nabla_{\mu}^2 f_n^{\epsilon}(\theta, \mu)||_F^2\bigg]^{1/2}\\
            \leq& \bigg[\textbf{0} +\big(\frac{\sqrt{n(n-1)}}{n^2}\sqrt{k}M^{(3)}\big)^2+ \bigg\{\frac{\sqrt{n(n-1)}}{n^2}\big(\sqrt{k}M_{\theta}M^{(4)} + M^{(5)}+2\epsilon n^2\big)\\
            +& \frac{\sqrt{n(n-1)\{n(n-1)-1\}}}{n^2}M_{\theta}M^{(6)}\bigg\}^2\bigg]^{1/2}\\
            \leq&\frac{\sqrt{n(n-1)}}{n^2}\sqrt{k}M^{(3)} + \frac{\sqrt{n(n-1)}}{n^2}\big(\sqrt{k}M_{\theta}M^{(4)} + M^{(5)}+2\epsilon n^2\big)\\
            +& \frac{\sqrt{n(n-1)\{n(n-1)-1\}}}{n^2}M_{\theta}M^{(6)}\\
            =& L_{n,\epsilon}^{(4)}.
        \end{align*}
        
       Hence, we prove that the Hessian of $f_n^{\epsilon}$ , $\nabla^2 f_n^{\epsilon}$ is bounded by some positive constant $L_{n,\epsilon}^{(4)}>0$, thus $\nabla f_n^{\epsilon}$ is Lipschitz continuous, i.e., 
         \[||\nabla f_{n,1}^{\epsilon} - \nabla f_{n,2}^{\epsilon}|| \leq L_{n,\epsilon}^{(4)}||(\theta_1, vec(\mu_1)) - (\theta_2, vec(\mu_2)) ||.\]
           \end{enumerate}
    Let \[L_{n,\epsilon} = \max \{L_{n}^{(1)}, L_{n,\epsilon}^{(2)}, L_{n,\epsilon}^{(3)}, L_{n,\epsilon}^{(4)}\} = L_{n,\epsilon}^{(4)},\]
    because it depends on the bound for the norm of Hessian matrix $\nabla_{\mu \mu}^2 f_n{\theta, \mu}$. The Hessian matrix contains the Hessian matrix of entropy of $\mu$, which is $1/\mu (1-\mu)$. Since $\mu \in [\zeta, 1-\zeta]$, it can have large enough value when each $\mu_{ij}$ has either $\zeta$ or $1-\zeta$.

\end{enumerate}
    \item Assumption \hyperref[bound]{boundedness}. \\ 
  We want to show that there exists a positive constant $M_{n, \epsilon} >0$ such that
  \[|F_n|, ||\nabla F_n||, |f_n^{\epsilon}|, ||\nabla f_n^{\epsilon}|| \leq M_{n, \epsilon}. \]
  In fact, let $M_{n, \epsilon} := L_{n, \epsilon}^{(3)}>0$. Then we prove the \hyperref[bound]{boundedness} of the objective functions and their gradients. \qedsymbol
 \item Assumption \hyperref[ppl]{Projected PL inequality}. \\
 We want to show that the lower-level objective function $f_n^{\epsilon}(\theta, \cdot)$ satisfies the projected Polyak-Łojasiewicz (PL) inequality, i.e., 
 for any $(\theta, \mu) \in \Theta \times \mathcal{U}_{\zeta}$, there exists a positive constant $\kappa >0$ such that \[||G_{\alpha}^{\epsilon}(\mu;\theta)||_2^2 \geq \kappa(f_n^{\epsilon}(\theta, \mu) - f_n^{\epsilon}(\theta, \mu^{*}(\theta))).\]
  In fact, the $\ell_2$ regularization with regularization parameter $\epsilon$ greater than the minimum eigenvalue of the Hessian matrix of the lower-level objective function $\nabla_{\mu\mu}^2 f_n(\theta,\cdot)$ converts the lower-level objective function into a strongly convex function with parameter $\rho>0$. Using the proof in Appendix F of \cite{karimi2016linear}, the lower-level objective function satisfies the projected PL inequality. 
  
\end{enumerate}
\restoregeometry 
\doublespacing

\section*{Appendix B}\phantomsection\label{appendix:b}
\cite{mele2023approximate} suggest the likelihood function of ERGM with different specification for counting subgraphs such as the number of two-stars and the number of triangles. 
The log-likelihood function of ERGM using a variational mean-field approximation to the log-normalizing constant in their paper is 
\begin{align*}
    l_n^{MF}(\nu,\theta|g_n, \{X_i\}_{i=1}^n):=T_n(\nu,\theta;g_n, \{X_i\}_{i=1}^n)-\psi_n^{MF}(\nu,\theta),
\end{align*}
where 
\begin{align*}
    T_n(\nu, \theta, g_n, \{X_i\}_{i=1}^n)&=\frac{1}{n^2}\bigg[\underbrace{\sum_{i=1}^n\sum_{j= 1}^n \nu(X_i,X_j)g_{ij}}_{\text{Number of direct links}} +\frac{\beta }{2n}\underbrace{\sum_{i=1}^n\sum_{j=1}^n\sum_{k=1}^n g_{ij}g_{\underline{j}k}}_{\text{Number of path length 2}} +  \underbrace{\frac{2\gamma}{3n} \sum_{i=1}^n\sum_{j=1}^n\sum_{k=1}^n g_{ij}g_{jk}g_{ki}\bigg]}_{\text{Number of triangles with different scaling}}
\end{align*}
\begin{align*}
    \psi_n^{MF}(\nu, \theta) =& \sup_{\substack{\mu\in[0,1]^{n^2},\\\mu_{ij}=\mu_{ji},\forall i,j}} \frac{1}{n^2}\bigg\{\sum_{i=1}^n\sum_{j= 1}^n \nu_{ij}\mu_{ij} +\frac{\beta }{2n}\sum_{i=1}^n\sum_{j=1}^n\sum_{k=1}^n \mu_{ij}\mu_{\underline{j}k} +  \frac{2\gamma}{3n} \sum_{i=1}^n\sum_{j=1}^n\sum_{k=1}^n \mu_{ij}\mu_{jk}\mu_{ki}\\
    - &\frac{1}{2}\sum_{i,j}\big[\mu_{ij}\text{log}\mu_{ij} + (1-\mu_{ij})\text{log}(1-\mu_{ij})\big]\bigg\}.
\end{align*}
According to the authors, the first-order condition (FOC) of lower-level objective function $\psi_n^{MF}$ with respect to $\mu$ has a closed-form solution, which is:  
 \begin{align*}
        \mu_{ij}^{*} = 1/\bigg(1+\text{exp}\big(-2\alpha_{ij}-\beta n^{-1}\sum_{k=1}^n (\mu_{jk}^{*}+ \mu_{ki}^{*}) 
            -4 \gamma n^{-1} \sum_{k=1}^n \mu_{jk}^{*}\mu_{ki}^{*}\big)\bigg).\tag{FOC}\label{FOC}
    \end{align*}
In fact, the second and the third term are not the motifs to count the 2-stars and triangles in a given network, scaled by $1/n$. Using (\ref{melepotential}), we correct the above $T_n$ and  $\psi_n^{MF}$ to
\begin{align*}
     &T_n(\nu, \theta, g_n, \{X_i\}_{i=1}^n) =\frac{1}{n^2}\bigg[\underbrace{\sum_{i=1}^n\sum_{j= 1}^n\nu_{ij}g_{ij}}_{\text{Number of direct links}} +\underbrace{\frac{\beta}{n}\sum_{i=1}^n \sum_{j=1}^n \sum_{k=j+1}^n g_{ij}g_{ik}}_{\text{Number of two-stars}}+\underbrace{\frac{\gamma}{6n}  \sum_{i=1}^n\sum_{j=1}^n\sum_{k=1}^ng_{jk}g_{ki}\bigg]}_{\text{Number of triangles}}\\
     &\psi_n^{MF}(\nu, \theta) =\sup_{\substack{\mu\in[0,1]^{n^2},\\\mu_{ij}=\mu_{ji},\forall i,j}} \bigg\{\frac{1}{n^2}\big[\sum_{i=1}^n\sum_{j= 1}^n \nu_{ij}\mu_{ij} +\frac{\beta}{n}\sum_{i=1}^n \sum_{j=1}^n \sum_{k=j+1}^n \mu_{ij}\mu_{ik} +  \frac{\gamma}{6n} \sum_{i=1}^n\sum_{j=1}^n\sum_{k=1}^n \mu_{ij}\mu_{jk}\mu_{ki}\big]\\
     &\hspace{37mm}-\frac{1}{2n^2}\sum_{i,j}\big[\mu_{ij}\text{log}\mu_{ij} + (1-\mu_{ij})\text{log}(1-\mu_{ij})\big]\bigg\}.
\end{align*}
The first-order condition of $\psi_n^{MF}$ with respect $\mu$ will change accordingly from \ref{FOC} to the following: For $i\neq j\in [n],$
\[f(\theta,\mu\hspace{1mm}|\{X_i\}_{i=1}^n) = \frac{-1}{n^2}\bigg\{\nu_{ij} + \frac{\beta}{n}\big\{\sum_{k=1}^n \mu_{ik} - \mu_{ij} + \sum_{k=1}^n \mu_{jk} - \mu_{ji}\big\} + \frac{\gamma}{n} \sum_{k=1}^n \mu_{jk}\mu_{ki} - \text{log}\frac{\mu_{ij}}{1-\mu_{ij}}\bigg\} = 0\tag{FOC*}\label{FOC*}\]

If we rearrange the equation above, 
\begin{align*}
&f(\theta,\mu\hspace{1mm}|\{X_i\}_{i=1}^n) = \nu_1 + \nu_2 z_{ij} + \frac{\beta}{n}\big\{\sum_{k\neq j}^n \mu_{ik}  + \sum_{k\neq i}^n \mu_{jk}\big\} + \frac{\gamma}{n} \sum_{k=1}^n \mu_{jk}\mu_{ki} - \text{log}\frac{\mu_{ij}}{1-\mu_{ij}} \\
    &\text{exp}(\nu_1 + \nu_2 z_{ij} + \frac{\beta}{n}\big\{\sum_{k\neq j}^n \mu_{ik}  + \sum_{k\neq i}^n \mu_{jk}\big\} + \frac{\gamma}{n} \sum_{k=1}^n \mu_{jk}\mu_{ki}) = \frac{\mu_{ij}}{1-\mu_{ij}} \\
    & (1-\mu_{ij})\text{exp}(\nu_1 + \nu_2 z_{ij} + \frac{\beta}{n}\big\{\sum_{k\neq j}^n \mu_{ik}  + \sum_{k\neq i}^n \mu_{jk}\big\} + \frac{\gamma}{n} \sum_{k=1}^n \mu_{jk}\mu_{ki}) = \mu_{ij}\\
    & \mu_{ij} = \sigma (\nu_1 + \nu_2 z_{ij} + \frac{\beta}{n}\big\{\sum_{k\neq j}^n \mu_{ik}  + \sum_{k\neq i}^n \mu_{jk}\big\} + \frac{\gamma}{n} \sum_{k=1}^n \mu_{jk}\mu_{ki})
\end{align*}
where $\sigma(y) = 1/(1+\exp(-y))$.
The following algorithm is the corrected algorithm from their original algorithm:

    \begin{algorithm}
    \caption{Local optimization of mean-field approximation by \cite{mele2023approximate}} \label{algo:local-mf}
    \begin{algorithmic}[1]
    \Require Set the tolerance level $\varepsilon_{\text{tol}}$.
    \Require We provide a parameter $\theta = (\theta_1, \theta_2)$. 
    \State Set initial value of $\mu_0$ at $t = 0$. 
    \State Compute $\psi_{n,t}^{MF}$ via equation (\ref{lognormalizingmf}) and set $\text{diff} = 1$.
    \While {$\text{diff} > \epsilon$} 
        \State Given $\mu_{t}$, get $\mu_{t+1}$ via equation \begin{align*}
    \mu_{ij, t+1} =(1+ \exp(-(\theta_1 + \frac{\theta_2}{n} \sum_{k=1}^n \mu_{jk,t}\mu_{ki,t})))^{-1}
\end{align*}
        \State Compute $\psi_{n, t+1}^{MF}$ via equation (\ref{lognormalizingmf})
        \State $\text{diff} = \psi_{n, t+1}^{MF} - \psi_{n, t}^{MF}$
        \If{$\text{diff} < \varepsilon_{\text{tol}}$},
            \State \textbf{Break}
        \Else \State $\psi_{n, t}^{MF} = \psi_{n, t+1}^{MF}$
        \EndIf
    \EndWhile
    \end{algorithmic}
    \end{algorithm}

   \begin{algorithm}
    \caption{Multi-Start Algorithm} \label{algo:multi-start}
    \begin{algorithmic}[1]
    \Require Set the tolerance level $\varepsilon_{\text{tol}}$ and the number of initial values $K$.
    \Require We provide a parameter $\theta = (\theta_1, \theta_2)$. 
    \For{\texttt{$k=1$ to $K$}}
            \State Draw an initial value $\mu^{(k)} \in  U[0,1]^{n\times  n}$. 
            \State Given $\mu^{(k)}$ as an initial value, conduct mean-field approximation using Algorithm \ref{algo:local-mf}, where the optimum value is denoted by $\psi_n^{MF}(k; \theta)$ 
        \EndFor
    \State Step 3: Set $\bar{\psi}_n^{MF}(\theta) = \max_{k} \{\psi_n^{MF}(k; \theta)\}_{k=1}^{K}$
    \end{algorithmic}
    \end{algorithm}
    \begin{algorithm}
    \caption{Parameter Update}
    \begin{algorithmic}[1]
   
    \Require Set tuning parameters $(\varepsilon_{\text{tol}}, K)$ for mean-field approximation.  
    \Require The network data $g_n$. 
    \State Set initial parameter $\theta_0 = (\theta_{1,0}, \theta_{2,0})$ with $j = 0$.
    \State Find the mean-field approximation $\psi_n^{MF} (\theta_k)$ using multi-start algorithm (See Algorithm \ref{algo:multi-start}).
    \State Evaluate the loss function $\ell_n^{MF}(g_n, \theta_0) = T_n(g_n; \theta_k) - \bar{\psi}_n^{MF} (\theta_j)$
    \State Update $\theta_j \to \theta_{j+1}$ using BFGS, and set $j \to j + 1$.
    \end{algorithmic}
    \end{algorithm}

\FloatBarrier
\onehalfspacing
 \newgeometry{top=0.5in, bottom=0.5in, left=0.25in, right=0.25in}
\clearpage
\section*{Appendix C}\phantomsection\label{appendix:c}

In this appendix, I show the performance of each algorithm through 1000 Monte Carlo simulations. It includes the 5\% and 95\% quantiles of estimates, the sign recovery (1 if the sign of estimate matches the sign of the true parameter, 0 otherwise), and outliers, the number of extreme estimates beyond 1,000 in the absolute value during the simulations. 
I provide the estimation time of each algorithm. Note that the runtime results should be read cautiously, since the VRBEA and the algorithm by \cite{mele2023approximate} use GPU, whereas MCMC-MLE and MPLE use CPU. 
\subsection*{True parameter: [-1,1], positive transitivity}
\vspace{-0.5cm}
\begin{table}[H]
\centering
\footnotesize

\caption{Monte Carlo Simulation Results: Comparison of algorithms, True parameter: [-1,1]}
\begin{threeparttable}
\begin{tabular}{c||cc||cc||cc||cc}
  \hline  
  $n=50$ & \multicolumn{2}{c||}{M \& Z Mean-Field}  & \multicolumn{2}{c||}{VRBEA} & \multicolumn{2}{c||}{MCMC-MLE}  & \multicolumn{2}{c}{MPLE}\\ 
  \hline
   No perturb& $\theta_1$ & $\theta_2$  &$\theta_1$ & $\theta_2$ &$\theta_1$ & $\theta_2$  &$\theta_1$ & $\theta_2$\\ 
  \hline

   bias &  2.7650 & 10.5975  &0.0015 & 0.0005 

   & 0.0066 & 2.1821 &  0.0038 & 1.7953  \\ 
 
  mean &1.7650 & -9.5975  &-0.9985 & 1.0005
                         
  & -0.9934 & -1.1821 &  -0.9962 & -0.7953\\
                
  median & -1.9976 & 0.6600 & -0.9985 & 1.0004
                         
  & -0.9942 & -0.3290 &  -0.9960 & -0.1423\\  
                
  MAD & 7.3223 & 18.4530  &0.0003 & 0.0002
                         
  & 0.0571 & 7.1717 &  0.0594 & 7.4895  \\ 
                
  se & 32.4929 & 78.6824 & 0.0003 & 0.0002 
                         
   & 0.0723 & 9.0710 &  0.0750 & 9.4913 \\ 
                
  0.05 & -2.1463 & -12.8913  & -0.9990 & 1.0001 
                         
  & -1.1149 & -16.2759 &  -1.1221 & -16.5750\\
                
  0.95 & -1.7967 & 0.7060  & -0.9979 & 1.0009
                         
  & -0.8756 & 12.3422 &  -0.8741 & 13.4159\\
                
  sign recovery (\%) & 91.69 & 75.98 & 100.00 & 100.00
                         
  & 100.0000 & 48.1000 &  100.0000 & 49.7000  \\ 
                
  outliers & 36 & 36  & 0 & 0
                         
  & 0& 0 & 0 & 0   \\ 
  \hline  
              
  time (sec) & 2438.6715 & 2.4387  & 1632.4468 & 1.6324 
                         
  & 5457.1 & 5.4571 &  81.2 &  0.0812   \\ 
\hline  
\hline

   $n=100$  & $\theta_1$ & $\theta_2$  &$\theta_1$ & $\theta_2$ &$\theta_1$ & $\theta_2$  &$\theta_1$ & $\theta_2$\\ 
  \hline
                
   bias & 0.5063 & 0.5587 & 0.0019 & 0.0003 
                         
 & 0.0035 & 0.8574 &  0.0031 & 0.6830    \\ 
 
  mean & -0.4937 & 1.5587  & -0.9981 & 1.0003
                         
  & -0.9965 & 0.1426 &  -0.9969 & 0.3170\\
                
  median & -0.7943 & 1.2265 & -0.9981 & 1.0003
                         
  & -0.9978 & 0.4701 & -0.9975 & 0.5269\\ 
                
  MAD & 0.5071 & 0.5596  & 0.0001 & 0.0000
                         
  & 0.0380 & 4.8223 &  0.0387 & 4.9584\\

  se & 0.5089 & 0.5615  & 0.0001 & 0.0001 
                         
  & 0.0477 & 6.0110 & 0.0485 & 6.1584  \\ 
                
  0.05 & -1.0000 & 1.0000 &  -0.9982 & 1.0002  
                         
  & -1.0721 & -10.0989 & -1.0742 & -10.0057 \\
                
  0.95 & 0.0261 & 2.1326 &  -0.9980 & 1.0004
                         
  &  -0.9159 & 9.3575 &  -0.9142 & 9.8089\\
                
  sign recovery (\%) & 52.00 & 100.00 & 100.0000 & 100.0000
                         
  & 100.0000 & 53.1000 &   100.0000 & 53.5000 \\ 
                
  outliers & 0 & 0  & 0 & 0 
                         
  &  0 & 0 &  0 & 0   \\ 
  \hline  
              
  time & 105.2972 & 0.1053 & 1725.5416 & 1.7255 
                         
  &  7179.7395 & 7.1797 &  90.9196 & 0.0901   \\ 
\hline  
\hline
$n=200$  & $\theta_1$ & $\theta_2$  &$\theta_1$ & $\theta_2$ &$\theta_1$ & $\theta_2$  &$\theta_1$ & $\theta_2$\\ 
  \hline
                
   bias & 9.5106 & 5.8718  & 0.0019 & 0.0003
                         
   &  0.0002 & 0.0886 &  0.0003 & 0.0352   \\ 
 
  mean & 8.5106 & -4.8718  & -0.9981 & 1.0003
                         
  & 1.0002 & 0.9114 &  -1.0003 & 0.9648\\
                
  median & -1.0000 & 1.0000 & -0.9981 & 1.0003
                         
  & -0.9993 & 0.9918 &  -0.9992 & 1.0320\\ 
                
  MAD & 12.8778 & 12.7479 & 0.0000 & 0.0000
                         
  & 0.0253 & 3.3256 &   0.0255 & 3.3716\\ 
                
  se & 40.5049 & 43.5731  & 0.0000 & 0.0000 
                         
  & 0.0316 & 4.1665 &   0.0318 & 4.1966 \\ 
                
  0.05 & -1.0000 & -23.6600 &  -0.9981 & 1.0002
                         
  & -1.0535 & -6.3934 &   -1.0529 & -6.2827  \\
                
  0.95 & 30.8534 & 5.5802 & -0.9981 & 1.0003
                         
  & -0.9501 & 7.6101 &   -0.9490 & 7.8678\\
                
  sign recovery (\%) & 49.75 & 81.28  & 100.00 & 100.00 
                         
  & 100.00 & 59.70 &   100.00 & 59.50\\ 
                
  outliers & 53 & 53  & 0 & 0 
                         
  &  0 & 0 &  0 & 0   \\ 
\hline  
              
  time &1376.0382 & 0.1376 & 1972.1556 & 1.9722 
                         
  &  9609.0013 & 9.609 &  168.6215 & 0.1686 \\
\hline
\hline
\end{tabular}
\end{threeparttable}
\end{table}

\restoregeometry 
\onehalfspacing
\begin{table}[H]
\centering
\footnotesize

\caption{Monte Carlo Simulation Results: Comparison of algorithms, True parameter: [-1,1]}
\begin{threeparttable}
\begin{tabular}{c||cc||cc||cc||cc}
  \hline  
  $n=50$ & \multicolumn{2}{c||}{M \& Z Mean-Field}  & \multicolumn{2}{c||}{VRBEA} & \multicolumn{2}{c||}{MCMC-MLE}  & \multicolumn{2}{c}{MPLE}\\ 
  \hline
   Perturbed by 0.5& $\theta_1$ & $\theta_2$  &$\theta_1$ & $\theta_2$ &$\theta_1$ & $\theta_2$  &$\theta_1$ & $\theta_2$\\ 
  \hline
                
   bias & 1.7862 & 6.9915 & 0.0115 & 0.0050
                         
   & 0.0068 & 2.1813 & 0.0038 & 1.7953 \\ 
 
  mean & 0.7862 & -5.9915   & -0.9885 & 0.9950
                         
  &-0.9932 & -1.1813 & -0.9962 & -0.7953 \\
                
  median & -2.0074 & 0.4326 &  -0.9759 & 0.9916
                         
  &-0.9942 & -0.3251  & -0.9960 & -0.1423\\ 
                
  MAD & 5.4316 & 11.9016  & 0.2528 & 0.2532
                         
  &0.0570 & 7.1575  & 0.0594 & 7.4895 \\ 
                
  se & 24.9993 & 47.4835 & 0.2907 & 0.2923
                         
  & 0.0724 & 9.0599  & 0.0750 & 9.4913 \\ 
                
  0.05 & -2.1708 & -8.8405  &-1.4442 & 0.5545
                         
  & -1.1161 & -16.4009 & -1.1221 & -16.5750\\
                
  0.95 & -1.8201 & 0.9071  & -0.5549 & 1.4548
                         
  &-0.8745 & 12.4150 & -0.8741 & 13.4159\\
                
  sign recovery (\%) & 93.40 & 79.80 & 100.00 & 100.00
                         
  & 100.00 & 48.60 &  100.00 & 49.70 \\ 
                
   outliers & 39 & 39  & 0 & 0
                         
  & 0 & 0  & 0 & 0   \\ 
  \hline  
              
  time & 1772.7090 & 1.7727  & 1859.2716 & 1.8593 
                         
  & 4412.8055& 4.4128 &  64.2038 & 0.0642 \\ 
\hline  
\hline

   $n=100$  & $\theta_1$ & $\theta_2$  &$\theta_1$ & $\theta_2$ &$\theta_1$ & $\theta_2$  &$\theta_1$ & $\theta_2$\\ 
  \hline
                
   bias & 0.0826 & 0.1073  & 0.0189 & 0.0163
                         
   & 0.0030 & 0.8390  & 0.0031 & 0.6830 \\ 
 
  mean & -0.9174 & 1.1073  & -1.0189 & 1.0163
                         
  & -0.9970 & 0.1610 & -0.9969 & 0.3170 \\
                
  median & -0.9360 & 1.0958 & -1.0262 & 1.0077
                         
  & -0.9976 & 0.4409  & -0.9975 & 0.5269\\ 
                
  MAD & 0.3102 & 0.3035  & 0.2484 & 0.2453
                         
  & 0.0381 & 4.8365 & 0.0387 & 4.9584\\ 
                
  se & 0.3823 & 0.3904 & 0.2872 & 0.2845
                         
  &0.0479 & 6.0320 & 0.0485 & 6.1584\\ 
                
  0.05 & -1.4543 & 0.5481 &  -1.4602 & 0.5700
                         
  & -1.0738 & -10.0622  & -1.0742 & -10.0057\\
                
  0.95 & -0.0844 & 1.8596 &  -0.5551 & 1.4647
                         
  & -0.9175 & 9.4006  & -0.9142 & 9.8089\\
                
  sign recovery (\%) & 97.1 & 100.00 & 100.0000 & 100.0000
                         
  & 100.00 & 52.90 &   100.00 & 53.50\\ 
                
  outliers & 0 & 0  & 0 & 0
                         
  &  0 & 0 &  0 & 0   \\ 
  \hline  
              
  time & 85.1134 & 0.0851  & 1723.9980 & 1.7240
                         
  &  6203.4211 & 6.2034 &  76.2498 & 0.0762   \\ 
\hline  
\hline
$n=200$  & $\theta_1$ & $\theta_2$  &$\theta_1$ & $\theta_2$ &$\theta_1$ & $\theta_2$  &$\theta_1$ & $\theta_2$\\ 
  \hline
                
   bias & 1.8862 & 1.4339 & 0.0068 & 0.0082

   &0.0002 & 0.0901  & 0.0003 & 0.0352  \\

  mean & 0.8862 & 2.4339  & -1.0068 & 1.0082
                         
  & -1.0002 & 0.9099 & -1.0003 & 0.9648\\
                
  median & -0.8188 & 1.1985 & -1.0219 & 1.0143
                         
  & -0.9991 & 0.9854 & -0.9992 & 1.0320\\ 
                
  MAD & 2.7885 & 2.2094  & 0.2480 & 0.2465
                         
  &0.0251 & 3.3151 & 0.0255 & 3.3716\\ 
                
  se & 4.2184 & 3.4167  & 0.2884 & 0.2865
                         
  & 0.0314 & 4.1448 & 0.0318 & 4.1966 \\ 
                
  0.05 & -1.4440 & 0.5632 &  -1.4418 & 0.5554
                         
  & -1.0514 & -6.3231 & -1.0529 & -6.2827  \\
                
  0.95 & 6.9157 & 6.3213 & -0.5455 & 1.4459
                         
  &-0.9506 & 7.6110  & -0.9490 & 7.8678\\
                
  sign recovery (\%) & 72.40 & 99.50  & 100.0000 & 100.0000 
                         
  & 100.00 & 59.70 &  100.0000 & 59.50\\ 
                
   outliers & 2 & 2  & 0 & 0 
                         
  & 0 & 0 & 0 & 0   \\ 
  \hline  
              
  time & 119.3128 & 0.1193  & 1966.7521 & 1.9668
                         
  &  8196.1991 & 8.1962 &  132.1054 & 0.1321 \\ 
\hline  
\hline
\end{tabular}

\end{threeparttable}
\end{table}

\begin{table}[H]
\centering
\footnotesize

\caption{Monte Carlo Simulation Results: Comparison of algorithms, True parameter: [-1,1]}
\begin{threeparttable}
\begin{tabular}{c||cc||cc||cc||cc}
  \hline  
  $n=50$ & \multicolumn{2}{c||}{M \& Z Mean-Field}  & \multicolumn{2}{c||}{VRBEA} & \multicolumn{2}{c||}{MCMC-MLE}  & \multicolumn{2}{c}{MPLE}\\ 
  \hline
   Perturbed by 1& $\theta_1$ & $\theta_2$  &$\theta_1$ & $\theta_2$ &$\theta_1$ & $\theta_2$  &$\theta_1$ & $\theta_2$\\ 
  \hline
                
   bias &  2.1067 & 7.2073  & 0.0385 & 0.0031
                         
   & 0.0060 & 1.9934  & 0.0038 & 1.7953  \\ 
 
  mean & 1.1067 & -6.2073   & -0.9615 & 1.0031
                         
  &-0.9940 & -0.9934  & -0.9962 & -0.7953 \\
                
  median & -2.0086 & 0.4595 &  -0.9604 & 1.0006
                         
  & -0.9963 & -0.2583  & -0.9960 & -0.1423\\ 
                
  MAD & 6.0565 & 12.6061  & 0.4897 & 0.4998
                         
  & 0.0575 & 7.3042  & 0.0594 & 7.4895\\ 
                
  se & 27.6919 & 52.7056 & 0.5654 & 0.5776
                         
  & 0.0728 & 9.2747 & 0.0750 & 9.4913\\ 
                
  0.05 & -2.1753 & -6.3864  &-1.8435 & 0.1084
                         
  & -1.1162 & -16.3055 & -1.1221 & -16.5750\\
                
  0.95 & -1.8179 & 1.4540  & -0.1169 & 1.9145
                         
  & -0.8756 & 12.9877 & -0.8741 & 13.4159 \\
                
  sign recovery (\%) & 93.09 & 77.38 & 99.9 & 100.00
                         
  & 100.0000 & 49.00 &  100.0000 & 49.7000 \\ 
                
   outliers & 41 & 41  & 0 & 0
                         
  & 0 & 0  & 0 & 0   \\ 
  \hline  
              
  time & 1675.29 & 1.6753  &  1623.3243 & 1.6233
                         
  &  4998.1263 & 4.9981 &  77.7478 & 0.0777   \\ 
\hline  
\hline  

   $n=100$  & $\theta_1$ & $\theta_2$  &$\theta_1$ & $\theta_2$ &$\theta_1$ & $\theta_2$  &$\theta_1$ & $\theta_2$\\ 
  \hline
                
   bias & 0.2603 & 0.0565  &0.0419 & 0.0249
                         
   & 0.0037 & 0.8448 & 0.0031 & 0.6830 \\ 
 
  mean & -0.7397 & 0.9435  & -0.9581 & 0.9751
                         
  & -0.9963 & 0.1552 & -0.9969 & 0.3170\\
                
  median & -0.7648 & 1.0838 & -0.9475 & 0.9752
                         
  & -0.9969 & 0.3679  & -0.9975 & 0.5269\\ 
                
  MAD &0.5978 & 0.6852  & 0.4899 & 0.5069
                         
  & 0.0381 & 4.8258 & 0.0387 & 4.9584\\ 
                
  se & 1.6101 & 2.3273 & 0.5709 & 0.5874
                         
  &0.0477 & 6.0149  & 0.0485 & 6.1584 \\ 
                
  0.05 & -1.8721 & -0.0758 &  -1.8759 & 0.0803
                         
  & -1.0714 & -10.1526  & -1.0742 & -10.0057\\
                
  0.95 & 0.0109 & 1.9639  &  -0.0829 & 1.9040
                         
  & -0.9171 & 9.5193 & -0.9142 & 9.8089 \\
                
  sign recovery (\%) & 98.30 & 100.00 & 100.0000 & 100.0000
                         
  & 100.00 & 52.80 &   100.0000 & 53.5000\\ 
                
  outliers & 5 & 5  & 0 & 0
                         
  &  0 & 0 &  0 & 0   \\ 
  \hline  
              
  time & 275.6156 & 0.2756  & 1726.3974 & 1.7264
                         
  &  6370.4030 & 6.3704 &  93.4851 & 0.0935   \\ 
\hline  
\hline
$n=200$  & $\theta_1$ & $\theta_2$  &$\theta_1$ & $\theta_2$ &$\theta_1$ & $\theta_2$  &$\theta_1$ & $\theta_2$\\ 
  \hline
                
   bias &  4.8291 & 0.6404 & 0.0268 & 0.0076
                         
   & 0.0002 & 0.1170 & 0.0003 & 0.0352   \\ 
 
  mean & 3.8291 & 0.3596 & -1.0268 & 0.9924
                         
  & -0.9998 & 0.8830 & -1.0003 & 0.9648 \\
                
  median & -0.6152 & 1.3457 & -1.0420 & 0.9897
                         
  & -0.9995 & 1.0330 & -0.9992 & 1.0320\\ 
                
  MAD &  7.3059 & 4.2824  &  0.5047 & 0.5086
                         
  & 0.0252 & 3.3156  & 0.0255 & 3.3716 \\ 
                
  se & 29.1295 & 27.8631  & 0.5829 & 0.5828
                         
  & 0.0315 & 4.1531  & 0.0318 & 4.1966 \\ 
                
  0.05 & -1.8271 & 0.0224&  -1.9202 & 0.0967
                         
  & -1.0520 & -6.1530 & -1.0529 & -6.2827   \\
                
  0.95 & 17.7823 & 11.9454 & -0.1032 & 1.8976
                         
  & -0.9492 & 7.5958 & -0.9490 & 7.8678\\
                
  sign recovery (\%) & 72.70 & 93.10  & 99.8 & 100.0000 
                         
  & 100.00 & 59.20 &  100.00 & 59.50\\ 
                
   outliers & 30 & 30  & 0 & 0 
                         
  &   0 & 0 &  0 & 0    \\ 
  \hline  
              
  time & 362.1119 & 0.3621  & 1951.9874 & 1.9520
                         
   &9354.3148 & 9.3543 &  161.5817 & 0.1616  \\ 
\hline  
\hline
\end{tabular}

\end{threeparttable}
\end{table}
\subsection*{True parameter: [-1,-1], negative transitivity}

\begin{table}[H]
\centering
\footnotesize

\caption{Monte Carlo Simulation Results: Comparison of algorithms, True parameter: [-1,-1]}
\begin{threeparttable}
\begin{tabular}{c||cc||cc||cc||cc}
  \hline  
  $n=50$ & \multicolumn{2}{c||}{M \& Z Mean-Field}  & \multicolumn{2}{c||}{VRBEA} & \multicolumn{2}{c||}{MCMC-MLE}  & \multicolumn{2}{c}{MPLE}\\ 
  \hline
   No perturb& $\theta_1$ & $\theta_2$  &$\theta_1$ & $\theta_2$ &$\theta_1$ & $\theta_2$  &$\theta_1$ & $\theta_2$\\ 
  \hline
                
   bias &   0.2607 & 1.8613 & 0.0014 & 0.0005
                         
   & 0.0049 & 1.6441 & 0.0022 & 1.2939  \\ 
 
  mean & -1.2607 & -2.8613  &-0.9986 & -0.9995
                         
  & -0.9951 & -2.6441  & -0.9978 & -2.2939\\
                
  median & -1.9988 & -1.1906 & -0.9986 & -0.9996
                         
  & -0.9971 & -1.8372 & -0.9996 & -1.5659 \\  
                
  MAD & 1.5114 & 3.1491  & 0.0002 & 0.0002
                         
  & 0.0588 & 7.3441  & 0.0602 & 7.6100 \\ 
                
  se & 13.3173 & 20.7453 & 0.0003 & 0.0002
                         
   & 0.0737 & 9.2480  & 0.0755 & 9.6231 \\ 
                
  0.05 & -2.1659 & -1.8807  & -0.9990 & -0.9998
                         
  & -1.1176 & -19.0549  & -1.1233 & -19.1106\\
                
  0.95 & -1.8540 & -1.1505  & -0.9981 & -0.9992
                         
  & -0.8732 & 11.4488 & -0.8718 & 12.3911\\
                
  sign recovery (\%) & 98.40 & 100.00 & 100.00 & 100.00
                         
  & 100.00 & 58.60 &  100.00 & 57.20  \\ 
                
  outliers & 6 & 6  & 0 & 0
                         
  & 0& 0 & 0 & 0   \\ 
  \hline  
              
  time (sec) & 537.3482 & 0.5373  & 3282.3476 & 3.2823
                         
  & 6461.5935  & 6.4616 &  89.7022 &  0.0897   \\ 
\hline  
\hline  

   $n=100$  & $\theta_1$ & $\theta_2$  &$\theta_1$ & $\theta_2$ &$\theta_1$ & $\theta_2$  &$\theta_1$ & $\theta_2$\\ 
  \hline
                
   bias & 0.8126 & 0.6575 & 0.0018 & 0.0004
                         
 & 0.0026 & 0.7963  & 0.0022 & 0.6067   \\ 
                 
  mean & -0.1874 & -0.3425  & -0.9982 & -0.9996
                         
  & -0.9974 & -1.7963  & -0.9978 & -1.6067\\
                
  median & -0.1871 & -0.2661 & -0.9982 & -0.9996
                         
  & -0.9987 & -1.2951  & -0.9988 & -1.2063\\ 
                
  MAD & 0.1476 & 0.1653  & 0.0001 & 0.0000
                         
  & 0.0385 & 5.0953 & 0.0392 & 5.2549\\

  se & 0.7420 & 0.9466  & 0.0001 & 0.0001 
                         
  & 0.0484 & 6.4332 & 0.0494 & 6.6120  \\ 
                
  0.05 & -0.4666 & -0.7268 &  -0.9983 & -0.9997  
                         
  & -1.0740 & -13.0886  & -1.0760 & -13.2354 \\
                
  0.95 & -0.0502 & -0.1441 &  -0.9981 & -0.9995
                         
  &  -0.9158 & 8.3385 & -0.9146 & 8.7305\\
                
  sign recovery (\%) & 97.20 & 99.00 & 100.0000 & 100.0000
                         
  & 100.0000 & 59.3000  & 100.0000 & 57.3000  \\ 
                
  outliers & 9 & 9  & 0 & 0 
                         
  &  0 & 0 &  0 & 0   \\ 
  \hline  
              
  time & 610.1112 & 0.6101 & 3227.2959 & 3.2273
                         
  &  7069.3556 & 7.0694 &  86.8743 & 0.0869  \\ 
\hline  
\hline
$n=200$  & $\theta_1$ & $\theta_2$  &$\theta_1$ & $\theta_2$ &$\theta_1$ & $\theta_2$  &$\theta_1$ & $\theta_2$\\ 
  \hline
                
   bias & 4.2656 & 3.0211  & 0.0019 & 0.0003
                         
   &  0.0019 & 0.4558  & 0.0019 & 0.4217   \\ 
                 
  mean & 3.2656 & 2.0211   & -0.9981 & -0.9997
                         
  &  -0.9981 & -1.4558  & -0.9981 & -1.4217\\
                
  median & -1.0000 & -1.0000 & -0.9981 & -0.9997 
                         
  & -0.9979 & -1.5357  & -0.9976 & -1.6201\\ 
                
  MAD &  4.3394 & 3.0869 & 0.0000 & 0.0000
                         
  & 0.0253 & 3.3256 &   0.0255 & 3.3716\\ 
                
  se & 4.6135 & 3.9613 & 0.0000 & 0.0000 
                         
  & 0.0317 & 4.3343  & 0.0320 & 4.3924  \\ 
                
  0.05 & -1.0000 & -1.0000 & -0.9982 & -0.9997
                         
  & -1.0519 & -8.5547  & -1.0499 & -8.6112  \\
                
  0.95 & 10.9393 & 9.9199 & -0.9981 & -0.9997
                         
  & -0.9460 & 5.6673  & -0.9450 & 5.7377\\
                
  sign recovery (\%) & 50.00 & 51.80  & 100.00 & 100.00 
                         
  & 100.0000 & 63.3000  & 100.0000 & 61.9000\\ 
                
  outliers & 17 & 17  & 0 & 0 
                         
  &  0 & 0 &  0 & 0   \\ 
\hline  
              
  time &117.1705 & 0.1172 & 3227.3765 & 3.2274
                         
  & 9444.6754 & 9.4447 &  154.9559 & 0.1550   \\ 
\hline
\hline
\end{tabular}

\end{threeparttable}
\end{table}

\begin{table}[H]
\centering
\footnotesize

\caption{Monte Carlo Simulation Results: Comparison of algorithms, True parameter: [-1,-1]}
\begin{threeparttable}
\begin{tabular}{c||cc||cc||cc||cc}
  \hline  
  $n=50$ & \multicolumn{2}{c||}{M \& Z Mean-Field}  & \multicolumn{2}{c||}{VRBEA} & \multicolumn{2}{c||}{MCMC-MLE}  & \multicolumn{2}{c}{MPLE}\\ 
  \hline
   Perturbed by 0.5 & $\theta_1$ & $\theta_2$  &$\theta_1$ & $\theta_2$ &$\theta_1$ & $\theta_2$  &$\theta_1$ & $\theta_2$\\ 
  \hline
                
   bias &   0.1039 & 2.5243  & 0.0053 & 0.0181
                         
   & 0.0049 & 1.6369 &  0.0022 & 1.2939   \\ 
                 
  mean & -0.8961 & -3.5243 & -0.9947 & -1.0181
                         
  & -0.9951 & -2.6369  & -0.9978 & -2.2939 \\
                
  median & -2.0158 & -1.748 & -0.9920 & -1.0255
                         
  & -0.9971 & -1.8372 & -0.9996 & -1.5659 \\  
                
  MAD & 2.1993 & 3.4000 & 0.2522 & 0.2406
                         
  & 0.0588 & 7.3441  & 0.0602 & 7.6100 \\ 
                
  se & 14.6296 & 18.2255 & 0.2901 & 0.2792
                         
   & 0.0736 & 9.2837  & 0.0755 & 9.6231 \\ 
                
  0.05 & -2.1613 & -3.3540  & -1.4388 & -1.4445
                         
  & -1.1162 & -19.2972  & -1.1233 & -19.1106 \\
                
  0.95 & -1.8402 & -0.8105  & -0.5450 & -0.5595
                         
  & -0.8732 & 11.4488 & -0.8718 & 12.3911\\
                
  sign recovery (\%) & 96.70 & 100.00 & 100.00 & 100.00
                         
  & 100.00 & 57.90 & 100.0000 & 57.20 \\ 
                
  outliers & 17 & 17  & 0 & 0
                         
  & 0& 0 & 0 & 0   \\ 
  \hline  
              
  time (sec) & 1023.3097 & 1.0233  & 1685.2355 & 1.6582
                         
  &  5088.1109   & 5.0881 &  80.1254 &  0.0801   \\ 
\hline  
\hline

   $n=100$  & $\theta_1$ & $\theta_2$  &$\theta_1$ & $\theta_2$ &$\theta_1$ & $\theta_2$  &$\theta_1$ & $\theta_2$\\ 
  \hline
                
   bias &  0.2134 & 0.2036 & 0.0125 & 0.0001
                         
 & 0.0023 & 0.8045 & 0.0022 & 0.6067  \\ 
                 
  mean & -0.7866 & -0.7964  & -0.9875 & -1.0001
                         
  & -0.9977 & -1.8045 & -0.9978 & -1.6067 \\
                
  median & -0.7507 & -0.8094 & -0.9885 & -1.0059
                         
  & -0.9991 & -1.3375  & -0.9988 & -1.2063\\ 
                
  MAD & 0.3326 & 0.3498  & 0.2503 & 0.2508
                         
  &0.0385 & 5.0849  & 0.0392 & 5.2549\\

  se & 0.4035 & 0.4341  & 0.2889 & 0.2891
                         
  & 0.0485 & 6.4246 & 0.0494 & 6.6120 \\ 
                
  0.05 & -1.4350 & -1.4243 &  -1.4315 & -1.4638
                         
  & -1.0739 & -12.8481  & -1.0760 & -13.2354 \\
                
  0.95 & -0.1220 & -0.0363  & -0.9146 & 8.7305
                         
  &  -0.9158 & 8.3385 & -0.9146 & 8.7305\\
                
  sign recovery (\%) & 98.20 & 95.80 & 100.0000 & 100.0000
                         
  & 100.0000 & 60.3000  & 100.0000 & 57.3000  \\ 
                
  outliers & 9 & 9  & 0 & 0 
                         
  &  0 & 0 &  0 & 0   \\ 
  \hline  
              
  time & 87.6936  & 0.0877 & 1787.2565& 1.7873
                         
  &  7168.9404 & 7.1689 &  93.3880 & 0.09338  \\ 
\hline  
\hline
$n=200$  & $\theta_1$ & $\theta_2$  &$\theta_1$ & $\theta_2$ &$\theta_1$ & $\theta_2$  &$\theta_1$ & $\theta_2$\\ 
  \hline
                
   bias & 7.0402 & 1.1410 & 0.0023 & 0.0041
                         
   &  0.0020 & 0.4649 & 0.0019 & 0.4217   \\ 
                 
  mean & 6.0402 & 0.1410   & -1.0023 & -0.9959
                         
  &  -0.9980 & -1.4649 & -0.9981 & -1.4217\\
                
  median & -0.5410 & -0.6925  & -1.0031 & -0.9992
                         
  & -0.9976 & -1.5875 & -0.9976 & -1.6201\\ 
                
  MAD &   9.5270 & 4.5960 & 0.2424 & 0.2469
                         
  & 0.0257 & 3.4845 & 0.0259 & 3.5513\\ 
                
  se & 43.7809 & 28.5481 & 0.2826 & 0.2844 
                         
  & 0.0317 & 4.3304 & 0.0320 & 4.3924 \\ 
                
  0.05 & -1.3887 & -1.4676 & -1.4384 & -1.4442
                         
  & -1.0507 & -8.4858 & -1.0499 & -8.6112   \\
                
  0.95 & 14.7753 & 12.8828 & -0.5426 & -0.5492
                         
  & -0.9456 & 5.6501 & -0.9450 & 5.7377\\
                
  sign recovery (\%) & 62.60 & 72.90  & 100.00 & 100.00 
                         
  & 100.00 & 62.90  & 100.00 & 61.9000\\ 
                
  outliers & 11 & 11  & 0 & 0 
                         
  &  0 & 0 &  0 & 0   \\ 
\hline  
              
  time &224.7921 & 0.2248 & 2009.4478 & 2.0094
                         
  & 9725.2553 & 9.7253 &  173.7222 & 0.1737   \\ 
\hline
\hline
\end{tabular}

\end{threeparttable}
\end{table}

\begin{table}[H]
\centering
\footnotesize

\caption{Monte Carlo Simulation Results: Comparison of algorithms, True parameter: [-1,-1]}
\begin{threeparttable}
\begin{tabular}{c||cc||cc||cc||cc}
  \hline  
  $n=50$ & \multicolumn{2}{c||}{M \& Z Mean-Field}  & \multicolumn{2}{c||}{VRBEA} & \multicolumn{2}{c||}{MCMC-MLE}  & \multicolumn{2}{c}{MPLE}\\ 
  \hline
   Perturbed by 1 & $\theta_1$ & $\theta_2$  &$\theta_1$ & $\theta_2$ &$\theta_1$ & $\theta_2$  &$\theta_1$ & $\theta_2$\\ 
  \hline
                
   bias &   0.8930 & 4.6491  & 0.0089 & 0.0074
                         
   & 0.0043 & 1.4630  & 0.0022 & 1.2939 \\ 
                 
  mean & -0.1070 & -5.6491 & -1.0089 & -0.9926
                         
  & -0.9957 & -2.4630 & -0.9978 & -2.2939 \\
                
  median & -2.0131 & -1.9466 & -1.0322 & -0.9916
                         
  & -0.9982 & -1.8682 & -0.9996 & -1.5659\\  
                
  MAD & 3.7484 & 7.2314 & 0.4892 & 0.5057 
                         
  & 0.0591 & 7.4670 & 0.0602 & 7.6100\\ 
                
  se & 18.8527 & 31.7752 & 0.5679 & 0.5826
                         
   & 0.0740 & 9.4575  & 0.0755 & 9.6231 \\ 
                
  0.05 & -2.2012 & -5.1480   &  -1.8889 & -1.8888
                         
  & -1.1178 & -18.9249  & -1.1233 & -19.1106 \\
                
  0.95 & -1.8435 & -0.2024  & -0.1090 & -0.1105
                         
  & -0.8726 & 11.9230 & -0.8718 & 12.3911\\
                
  sign recovery (\%) & 95.39 & 97.89 & 100.00 & 100.00
                         
  & 100.00 & 57.60 & 100.00 & 57.20 \\ 
                
  outliers & 23 & 23  & 0 & 0
                         
  & 0& 0 & 0 & 0   \\ 
  \hline  
              
  time (sec) & 1361.1460 & 1.3611  & 2063.9508 & 2.0640
                         
  &   4952.0582  & 4.9521 &  93.1160 &  0.0931
   \\ 
\hline  
\hline

   $n=100$  & $\theta_1$ & $\theta_2$  &$\theta_1$ & $\theta_2$ &$\theta_1$ & $\theta_2$  &$\theta_1$ & $\theta_2$\\ 
  \hline
                
   bias &  0.8186 & 0.4070 & 0.0142 & 0.0081
                         
 & 0.0030 & 0.8067 & 0.0022 & 0.6067 \\ 
                 
  mean & -0.1814 & -1.4070  & -0.9858 & -1.0081
                         
  & -0.9970 & -1.8067 & -0.9978 & -1.6067 \\
                
  median & -0.6032 & -0.8094 & -0.9769 & -1.0186
                         
  & -0.9983 & -1.3150  & -0.9988 & -1.2063 \\ 
                
  MAD & 1.0916 & 1.2500  & 0.4987 & 0.4855
                         
  &0.0385 & 5.0800 & 0.0392 & 5.2549\\

  se & 16.4357 & 17.4157  & 0.5786 & 0.5700
                         
  & 0.0484 & 6.4268  & 0.0494 & 6.6120 \\ 
                
  0.05 & -1.8278 & -1.8615 & -1.0760 & -13.2354
                         
  & -1.0739 & -12.8481  & -1.0760 & -13.2354 \\
                
  0.95 & -0.0147 & 0.1341   & -0.0753 & -0.0879 
                         
  &  -0.9154 & 8.3957 & -0.9146 & 8.7305\\
                
  sign recovery (\%) & 96.00 & 93.30 & 99.80 & 100.00
                         
  & 100.0000 & 59.90  & 100.0000 & 57.3000  \\ 
                
  outliers & 9 & 9  & 0 & 0 
                         
  &  0 & 0 &  0 & 0   \\ 
  \hline  
              
  time & 135.5748  & 0.1356 & 1791.0188& 1.791
                         
  &  6398.8733 & 6.3989 &  97.0823 & 0.09708  \\ 
\hline  
\hline
$n=200$  & $\theta_1$ & $\theta_2$  &$\theta_1$ & $\theta_2$ &$\theta_1$ & $\theta_2$  &$\theta_1$ & $\theta_2$\\ 
  \hline

   bias & 8.5863 & 3.7572 & 0.0012 & 0.0058

   &  0.0022 & 0.4729 & 0.0019 & 0.4217  \\ 

  mean & 7.5863 & 2.7572   & -0.9988 & -1.0058

  &  -0.9980 & -1.4649 & -0.9981 & -1.4217\\

  median & -0.2584 & -0.6899  & -0.9771 & -1.0178

  & -0.9977 & -1.5800  & -0.9976 & -1.6201\\ 

  MAD &   10.8776 & 6.3450 & 0.4946 & 0.5020

  & 0.0256 & 3.4968 & 0.0259 & 3.5513\\ 

  se & 40.0066 & 23.8867 & 0.5738 & 0.5798

  & 0.0316 & 4.3348  & 0.0320 & 4.3924 \\ 

  0.05 & -1.6453 & -1.9103 & -1.9089 & -1.8999

  & -1.0510 & -8.6583 & -1.0499 & -8.6112   \\

  0.95 & 18.9772 & 12.8353 & -0.1013 & -0.0945

  & -0.9463 & 5.6139 & -0.9450 & 5.7377 \\

  sign recovery (\%) & 57.40 & 69.40  & 100.00 & 100.00 

  & 100.00 & 63.00  & 100.00 & 61.9000\\ 

  outliers & 35 & 35  & 0 & 0 

  &  0 & 0 &  0 & 0   \\ 
\hline  

  time &237.7299 & 0.2377 & 2029.4932 & 2.0294

  & 9483.3453 & 9.4383 &  162.610 & 0.1626   \\ 
\hline
\hline
\end{tabular}

\end{threeparttable}
\end{table}

\restoregeometry 
\doublespacing

\clearpage
\section*{Appendix D}\phantomsection\label{appendix:D}
\cite{snijders2002markov} illustrates a Markov Chain Monte Carlo Maximum Likelihood Estimation (MCMC-MLE) using the stochastic iteration algorithm proposed by \cite{robbins1951stochastic}. I describe it here in detail. First, I briefly summarize the likelihood function, log-likelihood function, score function and Hessian function of ERGM.
\begin{align}
    &\pi_n(\theta; g_{\text{obs}}) = \frac{\text{exp}\big(\langle \theta, T(g_{\text{obs}}) \rangle \big)}{\sum_{w\in \mathcal{G}_n}\text{exp}\big( \langle\theta, T(w)\rangle\big)}\tag{Likelihood}\\
    &\ell(\theta) = \theta^{\top} T(g_{\text{obs}}) - \text{log}\bigg(\sum_{w\in\mathcal{G}_n}\theta^{\top} T(w)\bigg)\tag{Log-likelihood}\\
    &s(\theta) = \nabla_{\theta}\ell(\theta) =T(g_{\text{obs}}) - E_{\mathbb{P_{\theta}}}[T(W)]\tag{Score}\\
    &H(\theta) = \frac{d}{d \theta} s(\theta) = \frac{d^2}{d\theta d\theta^{\top}} \ell(\theta) \tag{Hessian}
\end{align}
Since the second term of score function is intractable, \cite{geyer1991markov} proposes a method to approximate the expectation of sufficient statistics over ERGM using the Markov chain Monte Carlo (MCMC). That is, the sample counterpart of the second term can be computed by generating network samples $\{W_m\}_{m=1}^M$ by the MCMC for fixed $\theta$, $E_{\mathbb{P_{\theta}}}[T(W)] \approx \frac{1}{M}\sum_{m=1}^M T(w_m)$. \cite{snijders2002markov} improves it using the stochastic iterative algorithm by \cite{robbins1951stochastic}. The following algorithm illustrates his algorithm.
\begin{algorithm}

    \caption{MCMC-MLE} \label{algo:mcmc-mle}
    \begin{algorithmic}
    \Require Set an initial value $\theta^{(0)}$ and tuning parameters: 
 Tolerance level $\varepsilon_{\text{tol}}$, Burn-in parameter $B$, thining parameter $K$, the number of samples $M$. 
    \While{$||\theta^{(t+1)}-\theta^{(t)}|| \geq \varepsilon_{\text{tol}}$}
    \State Step 1. Run MCMC using $\theta^{(t)}$. Collect $M$ networks for every $K$th generated network \par
    \hskip  \algorithmicindent \hskip  \algorithmicindent after $B$ burn-in.
    \State Step 2. Compute the score function $s(\theta^{(t)})$ and the Hessian function $H(\theta^{(t)})$ of \par
    \hskip  \algorithmicindent \hskip  \algorithmicindent log-likelihood function  of ERGM.
    \State Step 3: Use the Newton-Raphson method to update $\theta^{(t)}$\par
    \hskip  \algorithmicindent  \hskip\algorithmicindent $\theta^{(t+1)} = \theta^{(t)} + \alpha H(\theta^{(t)})^{-1} s(\theta^{(t)}) $
    \State Step 4: \textbf{If} $||\theta^{(t+1)}-\theta^{(t)}|| \leq \varepsilon_{\text{tol}}$  \textbf{Break} \par \hskip  \algorithmicindent \hskip  \algorithmicindent \textbf{Else} $\theta^{(t)} = \theta^{(t+1)}$
 
    \EndWhile
    \end{algorithmic}

    \end{algorithm}
\clearpage
Maximum Pseudo-Likelihood Estimation (MPLE) was first proposed by \cite{besag1974spatial}, further developed by \cite{strauss1990pseudolikelihood}, \cite{wasserman1996logit}. They construct a log-likelihood function using the conditional probability of forming a link between unit $i$ and $j$ given any pair of unit $l$ and $k$ other than $i$ and $j$, that is, 
\begin{align*}
    \ell_{\text{pseudo}}(\theta) = \sum_{i=1}^n \sum_{j=i+1} \text{log}\big(\text{Pr}_{\theta}(G_{ij}= g_{ij} \hspace{1mm}|\hspace{1mm} G_{lk} = g_{lk}\hspace{2mm}\text{for}\hspace{2mm}(l,k) \neq (i,j), \hspace{2mm} i,j,l,k \in [n])\big)
\end{align*}
\section*{Appendix E}\phantomsection\label{appendix:E}
\subsection*{Sigmoid Saturation}
In fact, using the mean value theorem,  
\begin{align*}
    |\mu_{ij, k+1} - \mu_{ij, k}| = &|\sigma (h(\mu_k)) - \sigma (h(\mu_{k-1}))|\\ 
    =& |\langle \frac{d}{d h} h(\bar{\mu}) \frac{\partial}{\partial \mu} h (\bar{\mu}), \mu_{k}- \mu_{k-1}\rangle | \\
    \leq & | \frac{d}{d h}|  h(\bar{\mu}) | \|\frac{\partial}{\partial \mu} h (\bar{\mu})\| \| \mu_{k}- \mu_{k-1}\|\\
    =&|\sigma(h (\bar{\mu})) (1- \sigma(h (\bar{\mu})))| \|\frac{\partial}{\partial \mu} h (\bar{\mu})\| \| \mu_{k}- \mu_{k-1}\|
\end{align*}
The sigmoid function reaches close to either 0 or 1 when its argument in absolute value exceeds 4. In other words, if $|h (\bar{\mu})| \geq 4$, then $\sigma(h (\bar{\mu})) \approx 0$ or $1$. This is called the sigmoid saturation. It is a well-known phenomenon in machine learning. Thus, the change in each element of $\mu$ will shrink due to the sigmoid saturation through the insensitivity of the sigmoid function. The magnitude of $h(\bar{\mu})$ can easily exceed 4 because $h(\bar{\mu})$ contains the derivatives of complex dependence terms such as $k-$stars or triangles.
\end{document}